\documentclass[my_style]{revtex4}
\usepackage{graphicx}
\usepackage{fancyhdr}
\usepackage{epsfig}
\renewcommand{\baselinestretch}{1.2}

\begin{document}
\begin{titlepage}

\thispagestyle{empty}
\def\thefootnote{\fnsymbol{footnote}}       % symbols for footnotes

\begin{center}
\mbox{ }

\end{center}
\begin{flushright}
\Large
\mbox{\hspace{10.2cm} hep-ph/0602136} \\
\end{flushright}
\begin{center}
\vskip 2cm
{\Huge\bf
MSSM Higgs Boson Searches at LEP}

\vskip 1cm
{\LARGE\bf Andr\'e~Sopczak}
\\
\smallskip
\Large 
Lancaster University
\end{center}

\vskip 2.7cm
\centerline{\Large \bf Abstract}

\vskip 2.7cm
\hspace*{-0.5cm}
\begin{picture}(0.001,0.001)(0,0)
\put(,0){
\begin{minipage}{\textwidth}
\Large
\renewcommand{\baselinestretch} {1.2}
Final results from the MSSM Higgs boson searches from the LEP experiments ALEPH, DELPHI,
L3 and OPAL are presented. 
The results are statistically combined and the statistical significance for signal and 
background hypotheses are given.
Upper bounds on the production cross sections are set for several Higgs-like topologies.
Interpretations for six benchmark scenarios in the MSSM, both for CP-conserving and 
CP-violating scenarios, are given. 
Limits on the $\tan\beta$ parameter, and in some scenarios upper limits on the neutral Higgs 
boson masses are set.
\renewcommand{\baselinestretch} {1.}

\normalsize
\vspace{5cm}
\begin{center}
{\sl \large
Presented at the 13th International Conference on Supersymmetry and Unification of Fundamental Interactions (SUSY'05), Durham, UK, July 18-23, 2005
\vspace{-6cm}
}
\end{center}
\end{minipage}
}
\end{picture}
\vfill

\end{titlepage}
\newpage
\thispagestyle{empty}
\mbox{ }
\newpage
\setcounter{page}{0}

%Title of paper
\title{{\small{International Conference on Supersymmetry -- SUSY'05, Durham, UK 
}}\\ %% Please keep this conference title here
\vspace{12pt}
MSSM Higgs Boson Searches at LEP\footnote{Presented at the 13th International Conference on Supersymmetry and Unification of Fundamental Interactions (SUSY'05), Durham, UK, July 18-23, 2005.}} %% Paper title goes here

\author{Andr\'e~Sopczak\footnote[7]{Email: andre.sopczak@cern.ch}\\
{\small (On behalf of the ALEPH, DELPHI, L3, OPAL Collaborations, and 
      the LEP Working Group for Higgs Boson Searches)}}
\affiliation{Lancaster University}

\begin{abstract}
Final results from the MSSM Higgs boson searches from the LEP experiments ALEPH, DELPHI,
L3 and OPAL are presented. 
The results are statistically combined and the statistical significance for signal and 
background hypotheses are given.
Upper bounds on the production cross sections are set for several Higgs-like topologies.
Interpretations for six benchmark scenarios in the MSSM, both for CP-conserving and 
CP-violating scenarios, are given. 
Limits on the $\tan\beta$ parameter, and in some scenarios upper limits on the neutral Higgs 
boson masses are set.
\end{abstract}

\maketitle

\pagestyle{plain}

\section{Introduction} % Section title should be in all capitals.
\vspace*{-1mm}

The LEP data-taking ended November 3, 2000, although some data excess were observed.
Very successful data-taking at the LEP accelerator was completed. 
The expectations were exceeded regarding luminosity and center-of-mass energy.
Limits on the SM Higgs boson were reported previously 
$m_{\rm H}^{\rm SM} > 114.4$ GeV/$c^2$~\cite{smlimit}.
Many searches beyond the SM have led to stringent limit, for a review see Ref.~\cite{beyond}.
This report focuses on new results in the MSSM for CP-conserving and CP-violating MSSM benchmarks,
based on statistical combinations of searches by LEP collaborations at LEP-1 ($\approx 91$ GeV) 
and LEP-2 ($\le 209$ GeV) energies~\cite{newmssm}.
No significant excess indicative for Higgs boson production has been observed in the data and
the searches set bounds on topological cross sections and MSSM parameters.
The report is structured as follows: 
benchmark parameters,
experimental searches,
limits on topological cross sections,
CP-conserving and CP-violating interpretations.

\section{Benchmark Parameters}
CP-conserving and CP-violating MSSM benchmarks are defined and listed in Table~\ref{tab:benchmarks}.
The benchmarks are designed to include 
$\rm h\to c\bar c,gg, W^+W^-$ decays for the large-$\mu$ benchmark, 
reduced gluon fusion at LHC for the gluophobic benchmark,
reduced $\rm h\to b\bar b, \tau^+\tau^-$ decays by a
$\sin \alpha_{\rm eff}/\cos\beta$ coupling factor for the small $\alpha_{\rm eff}$ benchmark,
and CP-even/odd mixing $\propto m^4_{\rm t} {\rm Im}(\mu A)/M^2_{\rm SUSY}$ in the CPX benchmark
scenario.

\begin{table}
\vspace*{-0.4cm}
\caption{Benchmark parameter sets for CP-conserving and CP-violating MSSM scenarios. 
}
\label{tab:benchmarks}
\begin{center}
\vspace*{-0.3cm}
\begin{tabular}{l|cccccc}
           &  (1)         & (2)      & (3)        &   (4)    &  (5)                    &(6) \\
           &$m_{\rm h}$-max& no-mixing& large-$\mu$&gluophobic&small-$\alpha_{\rm eff}$ &CPX \\ \hline  
           & \multicolumn{6}{c}{Parameters varied in the scan} \\   \hline  
$\tan\beta$& 0.4-40       &0.4-40    &0.7-50        &0.4-40    &0.4-40                   &0.6-40 \\
$m_{\rm A}$ (GeV/$c^2$)&0.1-1000&0.1-1000  &0.1-400   &0.1-1000&0.1-1000                 &  --  \\
$m_{\rm H^\pm}$ (GeV/$c^2$) &-- &--    &-- &--                    &--                     & 4-1000  \\ \hline  
           & \multicolumn{6}{c}{Fixed parameters}\\       \hline  
$M_{\rm SUSY}$ (GeV)& 1000       & 1000 &400 &350 &800 &500  \\
$M_2$ (GeV)         &  200       &  200&400 &300&500 &200  \\
$\mu$ (GeV)        &$-200$        & $-200$&1000&300&2000&2000 \\ 
$m_{\tilde g}$ (GeV/$c^2$) & 800 &  800&200 &500&500 &1000 \\ \hline
$X_{\rm t}$ (GeV) &2$M_{\rm SUSY}$ & 0 &$-300$ &$-750$&$-1100$&$A-\mu\cot\beta$ \\  \hline  
$A$ (GeV)   &~~$X_{\rm t}+\mu\cot\beta$~~&~~$X_{\rm t}+\mu\cot\beta$~~&~~$X_{\rm t}+\mu\cot\beta$~~&~~$X_{\rm t}+\mu\cot\beta$~~&~~$X_{\rm t}+\mu\cot\beta$&1000 \\
arg ($A$) = arg ($m_{\rm \tilde g})$ & -- & --&--&--&--&$90^\circ$
\end{tabular}
\end{center}
\vspace*{-1cm}
\end{table}

\clearpage
\section{Experimental Searches}
The search topologies are:

\begin{minipage}{0.49\textwidth}
\vspace*{3mm}
Higgsstrahlung: $\rm e^+e^- \to H_1Z$
\begin{itemize}
\item Four-jet:       $\rm (H_1\to b\bar b)(Z\to q\bar q)$
\item Missing energy: $\rm (H_1\to b\bar b,\tau^+\tau^-)(Z\to \nu\bar\nu)$
\item Leptonic:       $\rm (H_1\to b\bar b,q\bar q)(Z\to e^+e^-,\mu^+\mu^-)$
\item Tau-leptons:    $\rm (H_1\to \tau^+\tau^-)(Z\to q\bar q)$ and \\ 
		      $\rm (H_1\to b\bar b,\tau^+\tau^-)(Z\to \tau^+\tau^-)$
\item Cascade:        $\rm H_2Z\to (H_1H_1)Z$
\end{itemize}
\end{minipage} \hfill
\begin{minipage}{0.49\textwidth}
Pairproduction: $\rm e^+e^- \to H_2H_1$
\begin{itemize}
\item Four-b: $\rm (H_2\to b\bar b)(H_1\to b\bar b)$
\item Mixed: $\rm (H_2\to \tau^+\tau^-)(H_1\to b\bar b)$ and\\ 
             $\rm (H_2\to b\bar b)(H_1\to \tau^+\tau^-)$
\item Four-tau: $\rm (H_2\to \tau^+\tau^-)(H_1\to \tau^+\tau^-)$
\item Cascade: $\rm H_2H_1\to (H_1H_1)H_1$
\item[] \mbox{ }
\end{itemize}
\end{minipage}

\vspace*{3mm}
Additional experimental constraints are applied:
upper bound at 95\% CL on the possible additional Z boson decay width $\Delta\Gamma_{\rm Z} < 2$ MeV, 
decay mode independent $\rm H_1Z$ searches, and 
searches for the Yukawa processes $\rm b\bar b H_1$ and $\rm b\bar b H_2$.
Further details are given in Ref.~\cite{newmssm}.

\section{Limits on Topological Cross Sections}

\begin{figure}[h!]
\begin{center}
\includegraphics[width=0.32\textwidth]{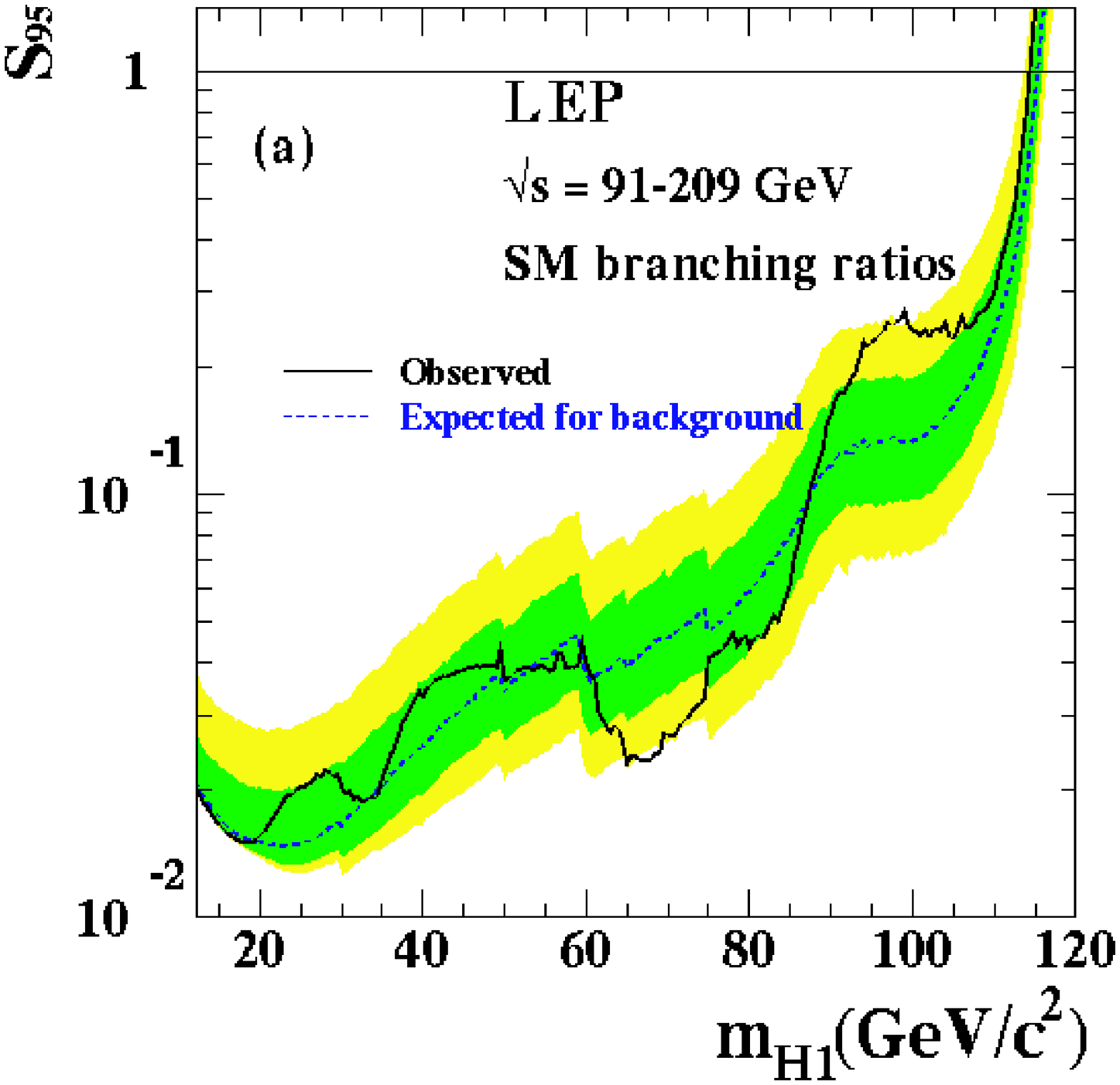} \hfill
\includegraphics[width=0.32\textwidth]{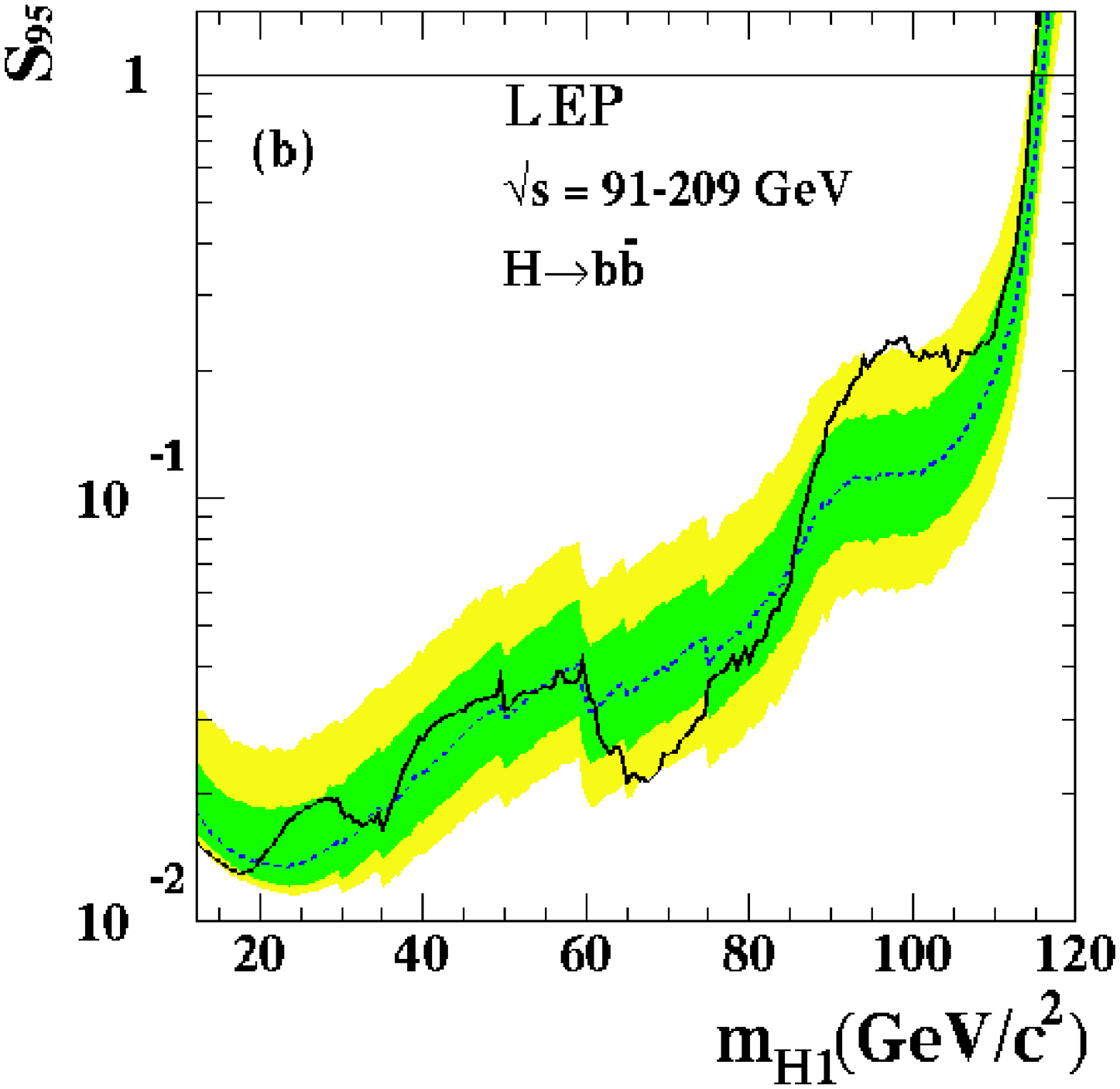} \hfill
\includegraphics[width=0.32\textwidth]{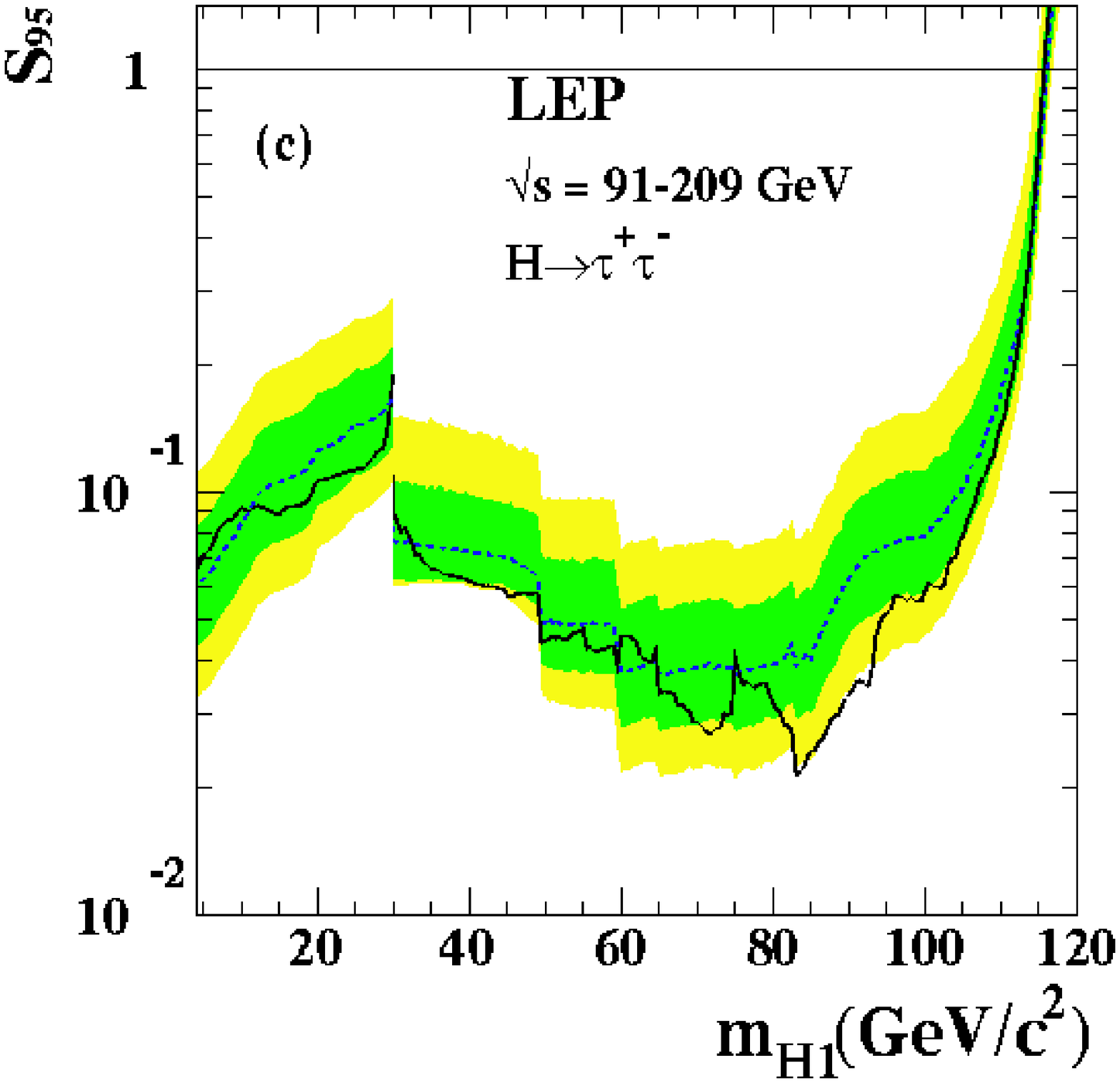}
\end{center}
\vspace*{-8mm}
\caption{Limits on  $S_{\rm 95} = \xi^2 = (g_{\rm HZZ}/g_{\rm HZZ}^{\rm SM})^2$.
Left:   SM Higgs boson decay. 
Center: $\rm H\to b\bar b$.
Right: $\rm H\to \tau^+\tau^-$.
}
\label{fig:xi}
\vspace*{-5mm}
\end{figure}

\begin{figure}[h!]
\begin{center}
\includegraphics[width=0.32\textwidth]{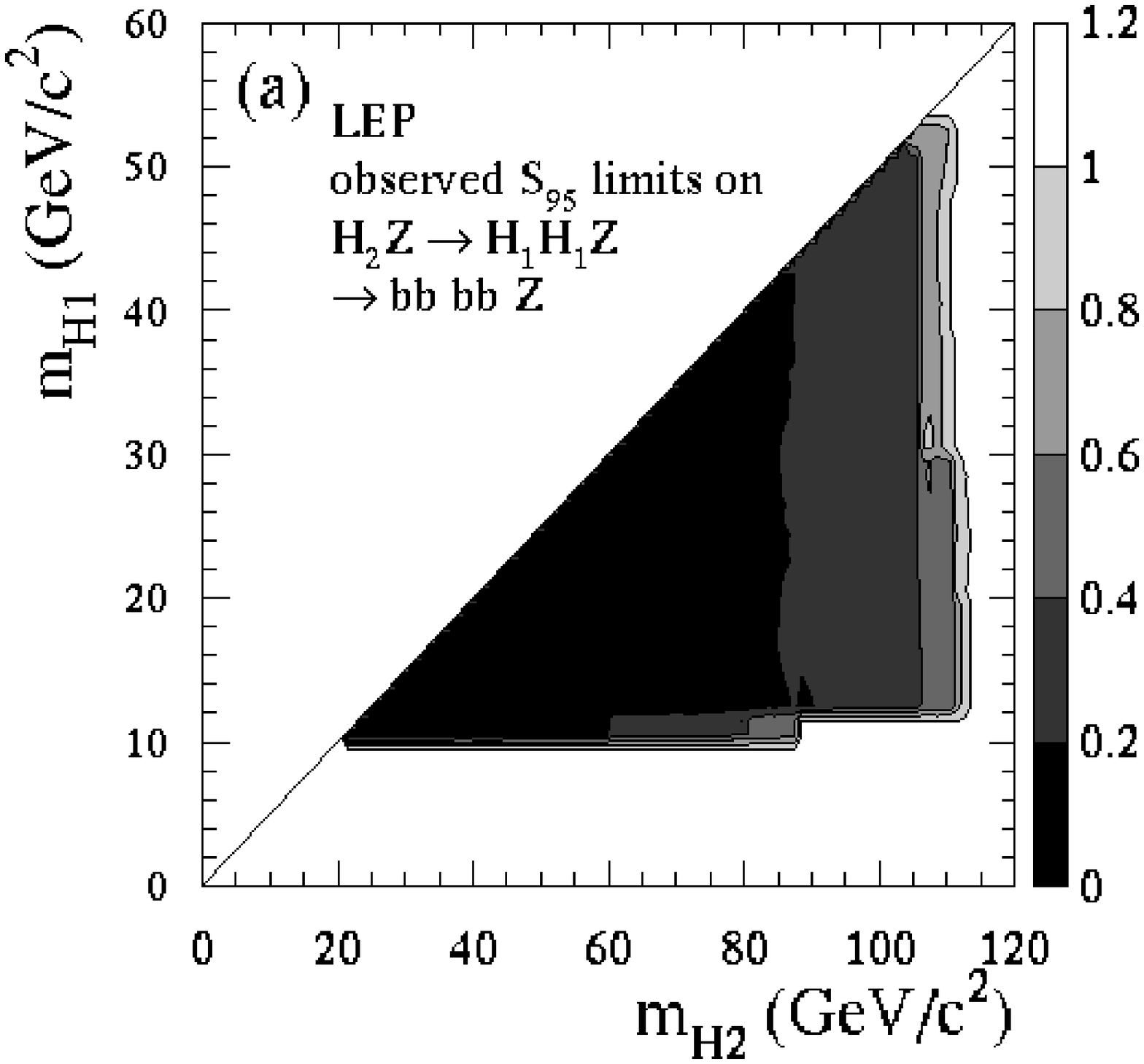} \hfill
\includegraphics[width=0.32\textwidth]{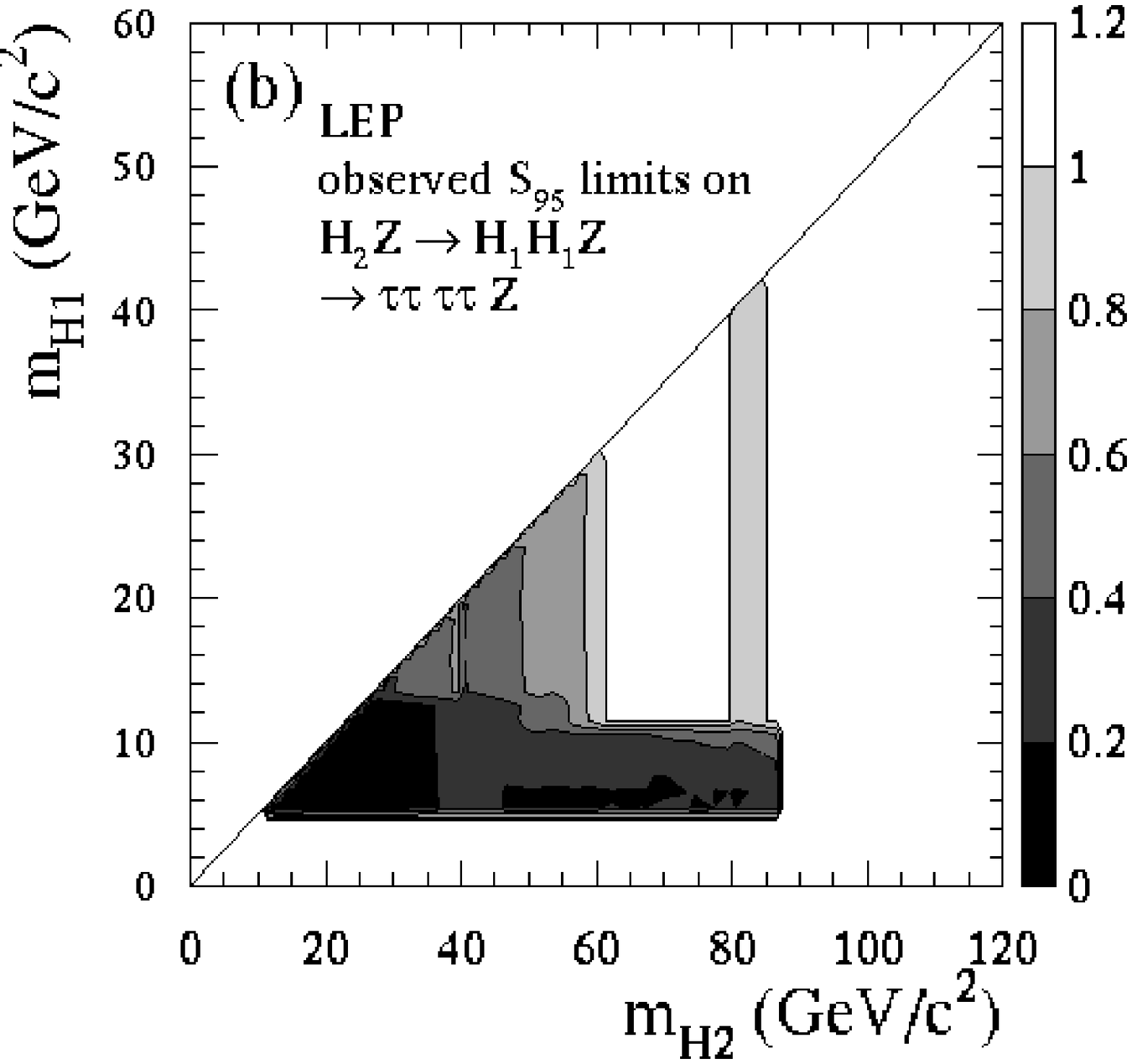} \hfill
\includegraphics[width=0.32\textwidth]{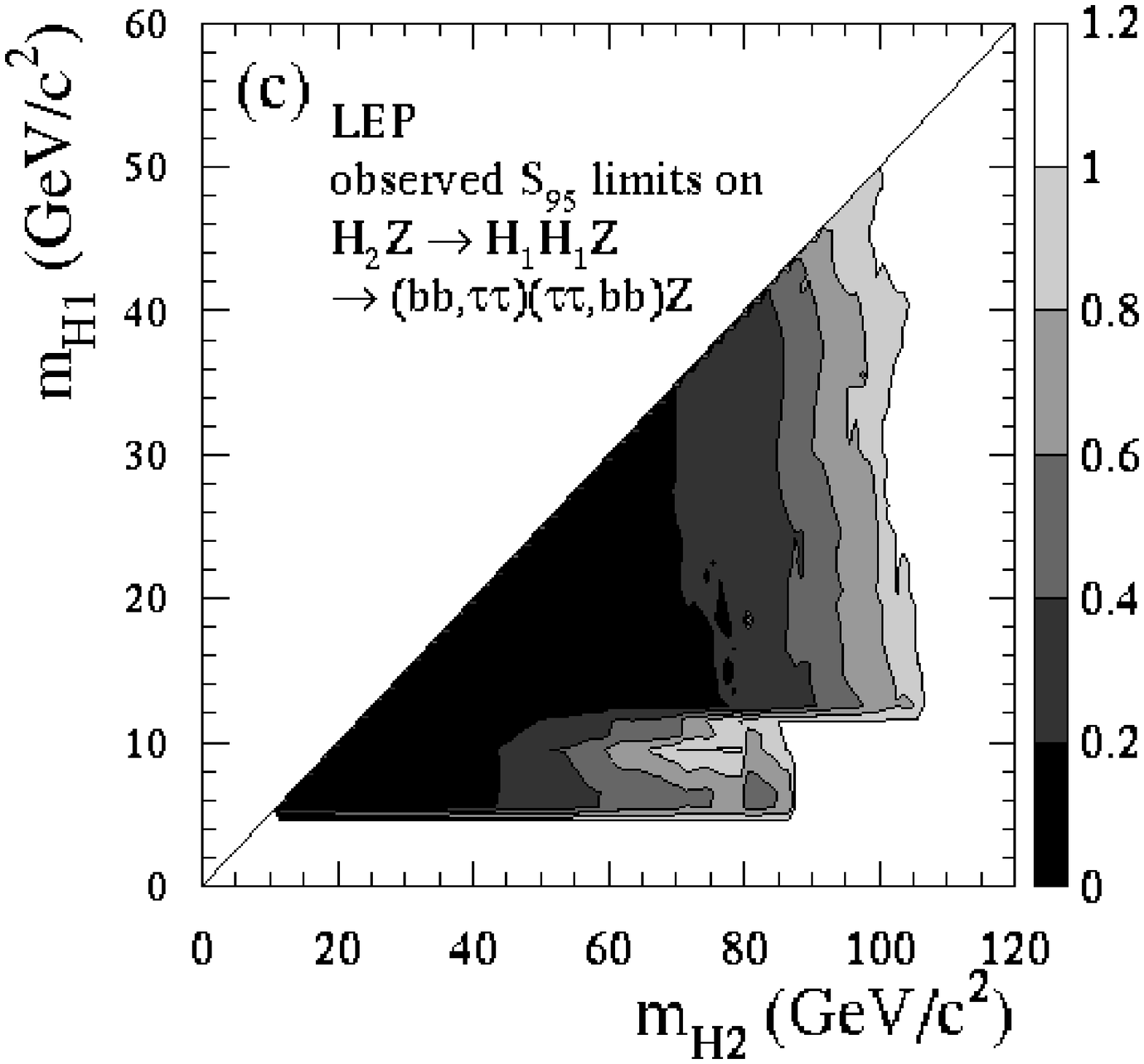}
\end{center}
\vspace*{-8mm}
\caption{Limits on $ S_{\rm 95}$ for the $\rm H_2Z \to H_1H_1Z$ process in different decay modes,
as indicated in the plots.
}
\label{fig:hhh}
\vspace*{-5mm}
\end{figure}

\begin{figure}[thp]
\vspace*{-2mm}
\begin{center}
\includegraphics[width=0.24\textwidth]{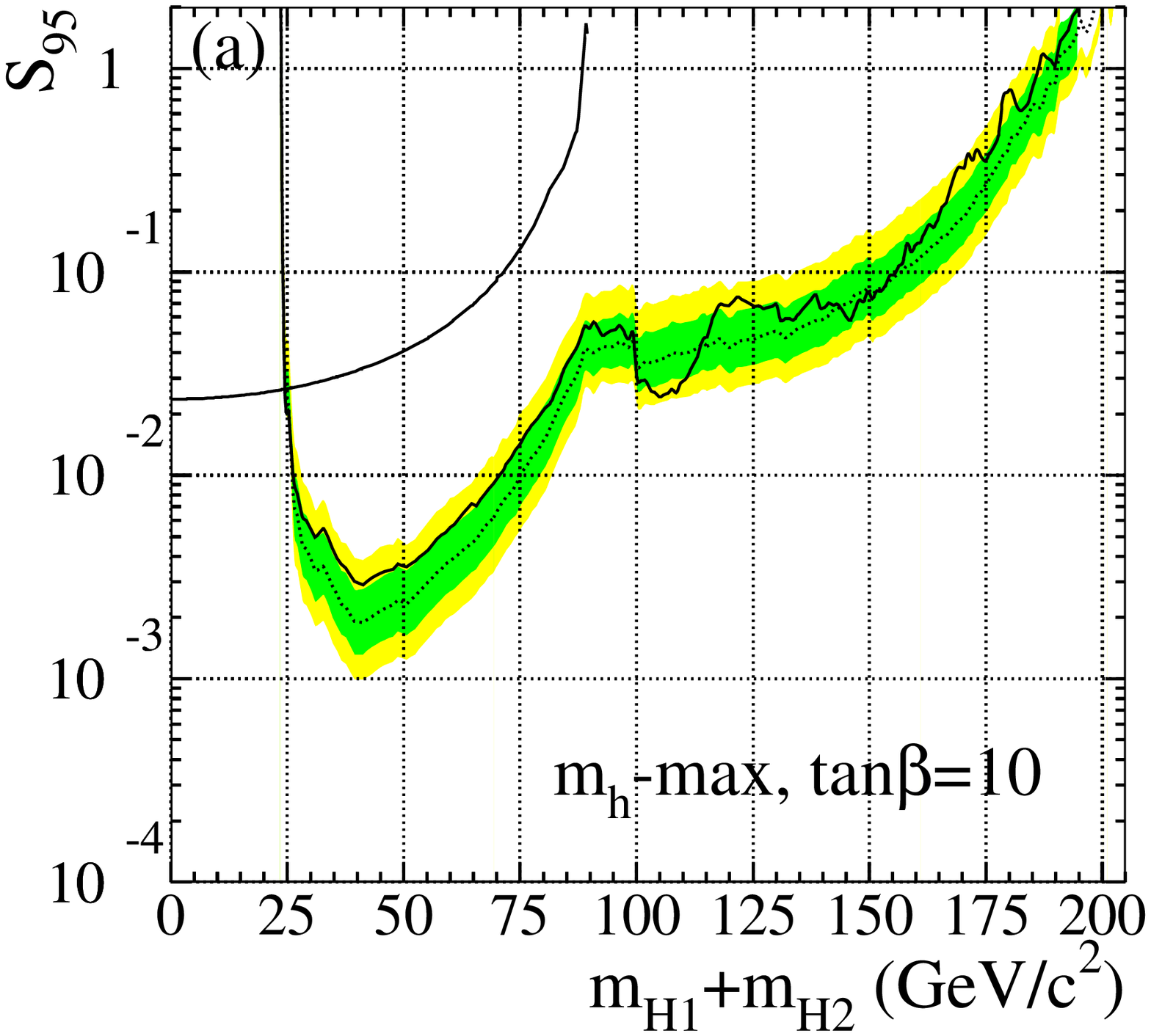} \hfill 
\includegraphics[width=0.24\textwidth]{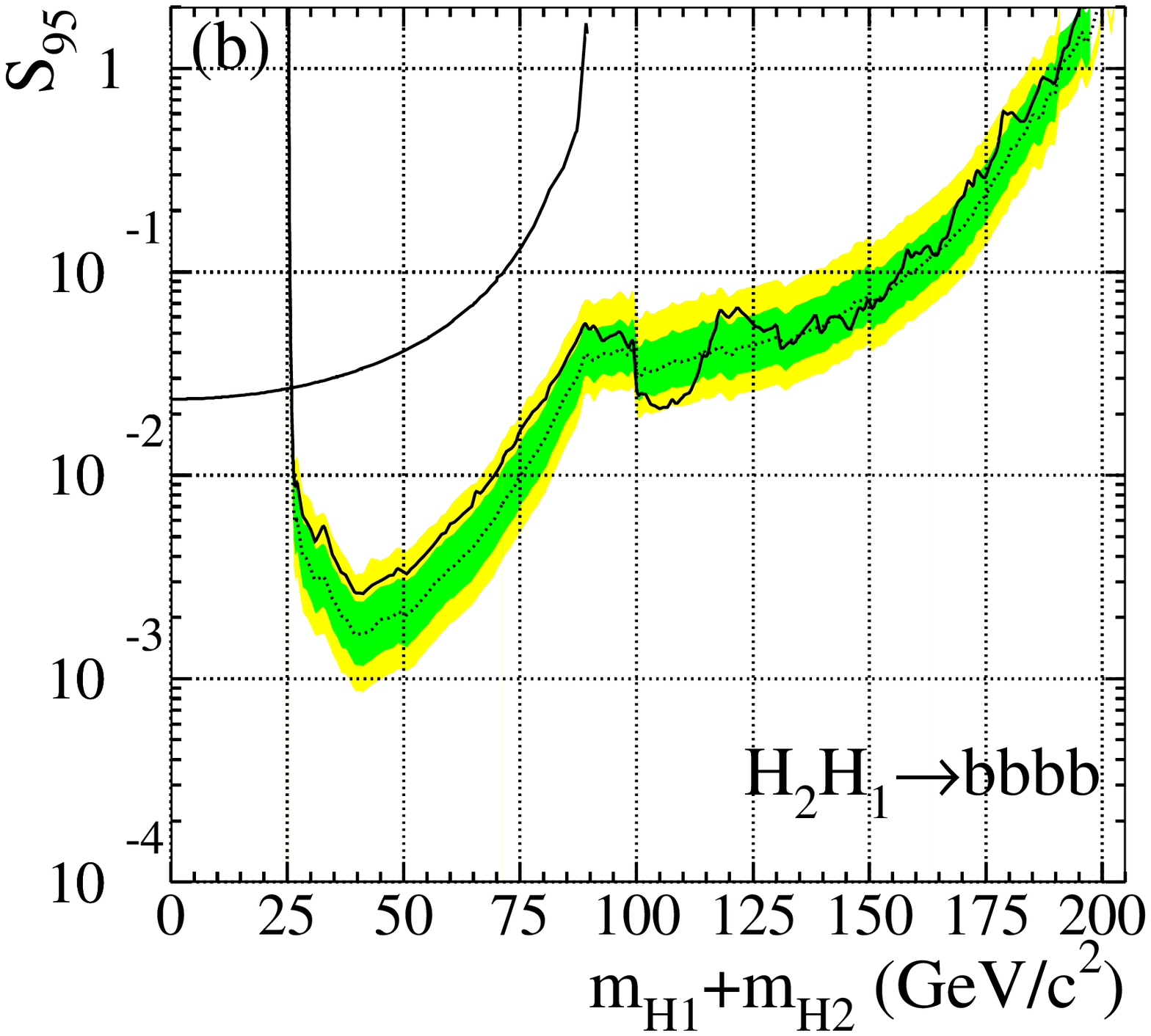} \hfill
\includegraphics[width=0.24\textwidth]{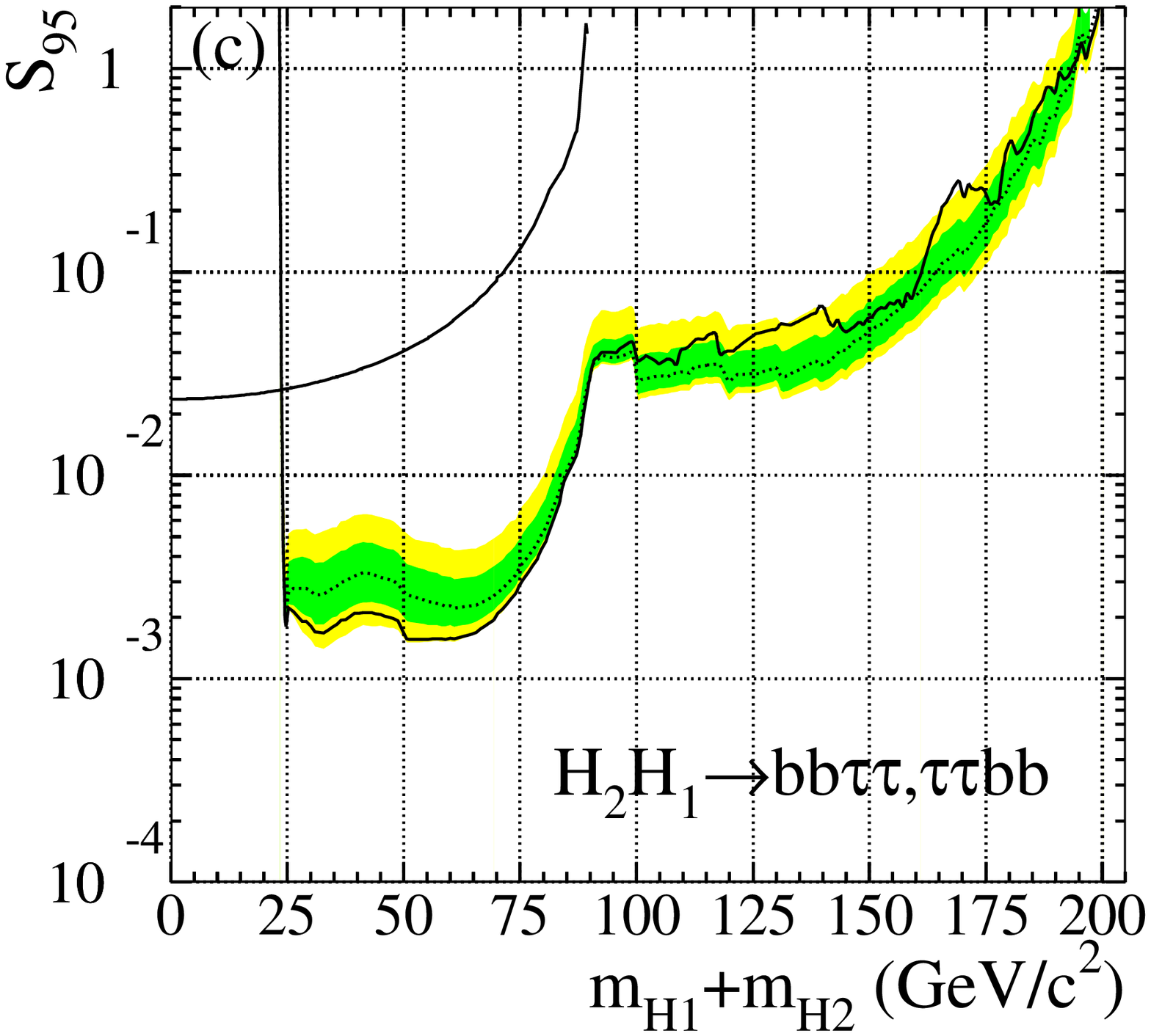} \hfill
\includegraphics[width=0.24\textwidth]{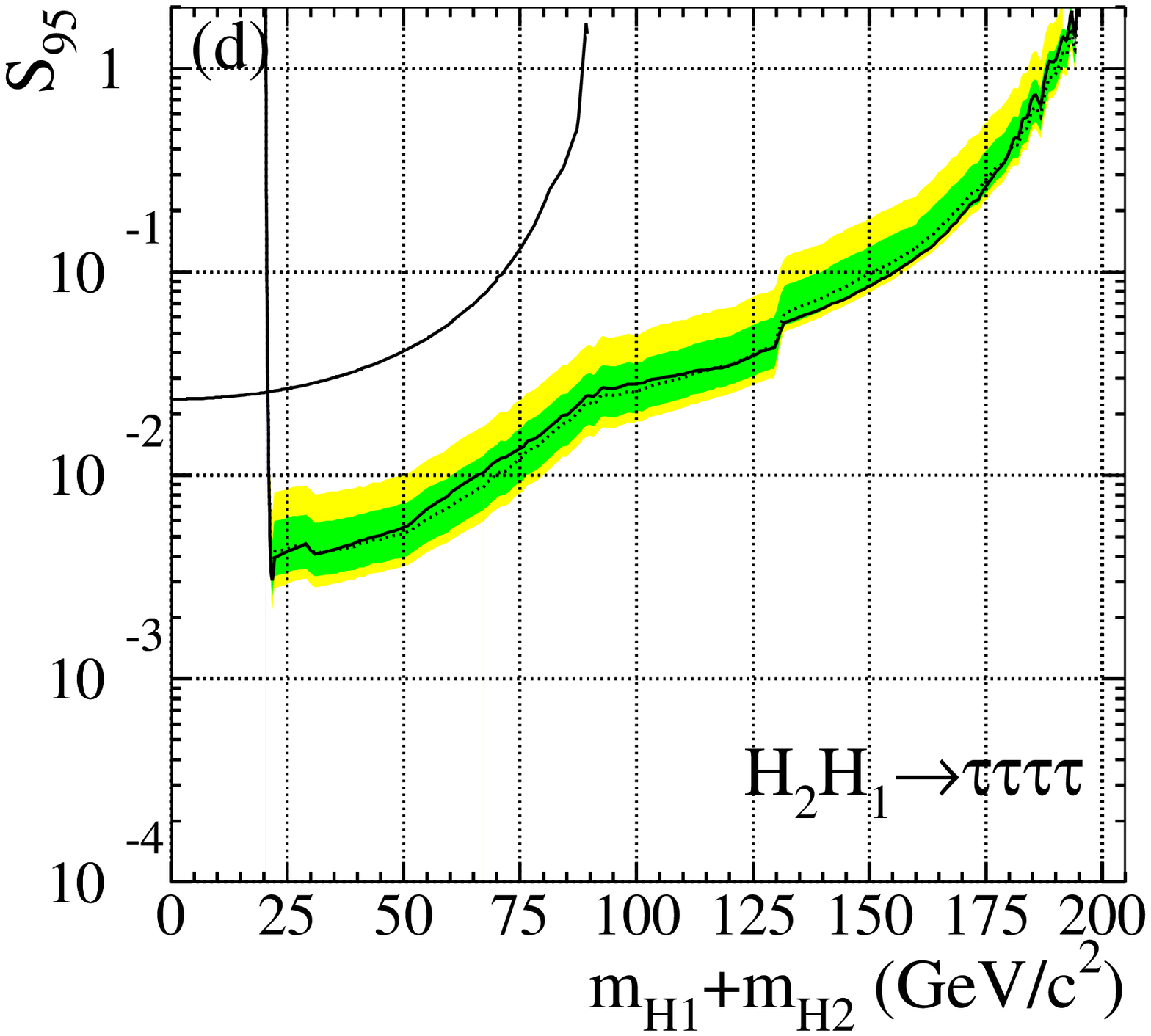}
\end{center}
\vspace*{-7mm}
\caption{Limits on $S_{\rm 95}$ for the $\rm H_1H_2$ process with different decays modes.
         Regions at low masses are excluded by the $\rm \Delta \Gamma_Z < 2$ MeV constraint.
Left: assuming branching ratios of the h-max scenario.
Center left:  $\rm b\bar bb\bar b$ decays.
Center right: $\rm b\bar b\tau^+\tau^-$ decays.
Right: $\tau^+\tau^-\tau^+\tau^-$ decays.
}
\label{fig:hh}
\vspace*{-7.5mm}
\end{figure}

\begin{figure}[hp]
\begin{center}
\includegraphics[width=0.24\textwidth]{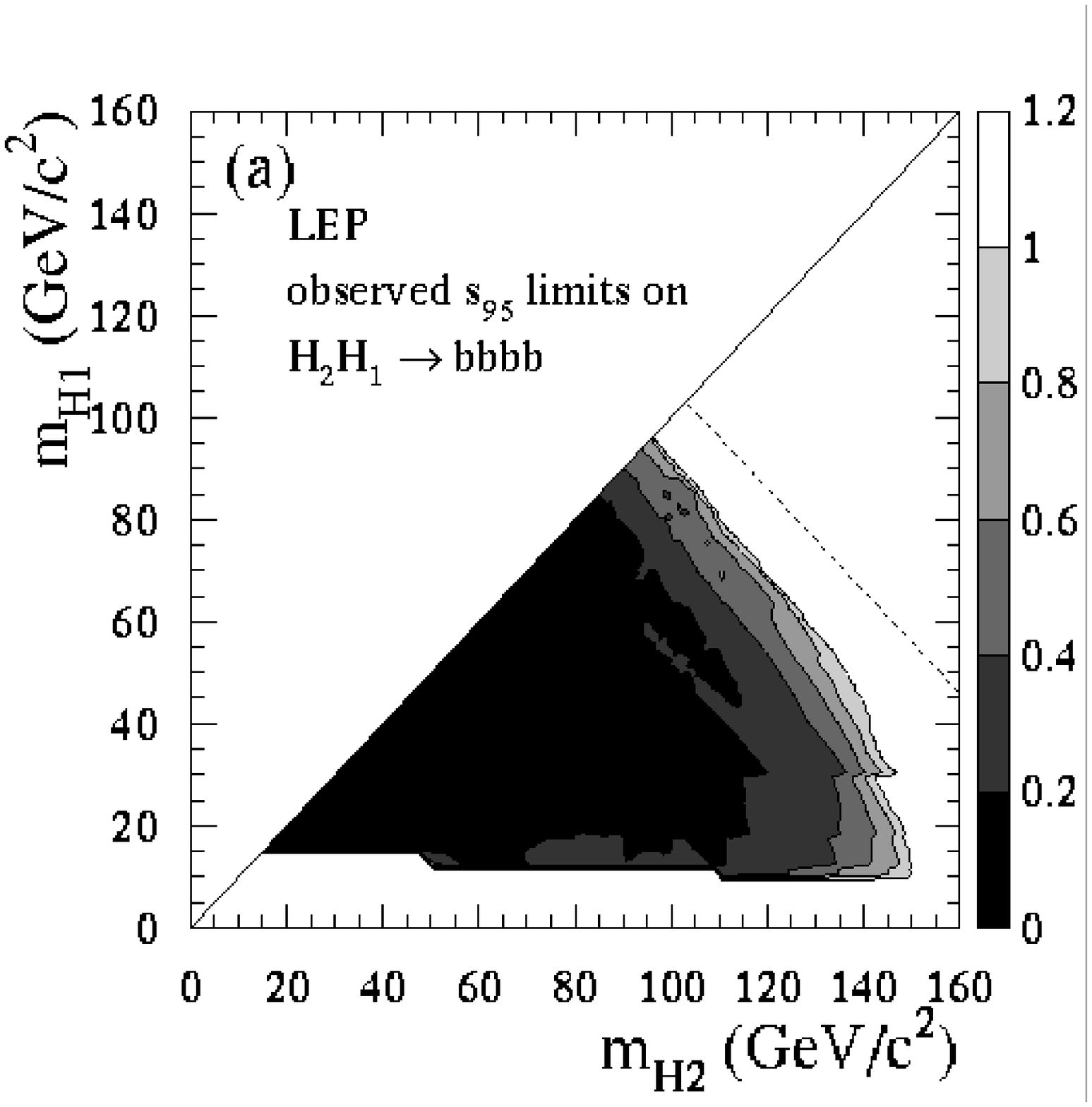} \hfill
\includegraphics[width=0.24\textwidth]{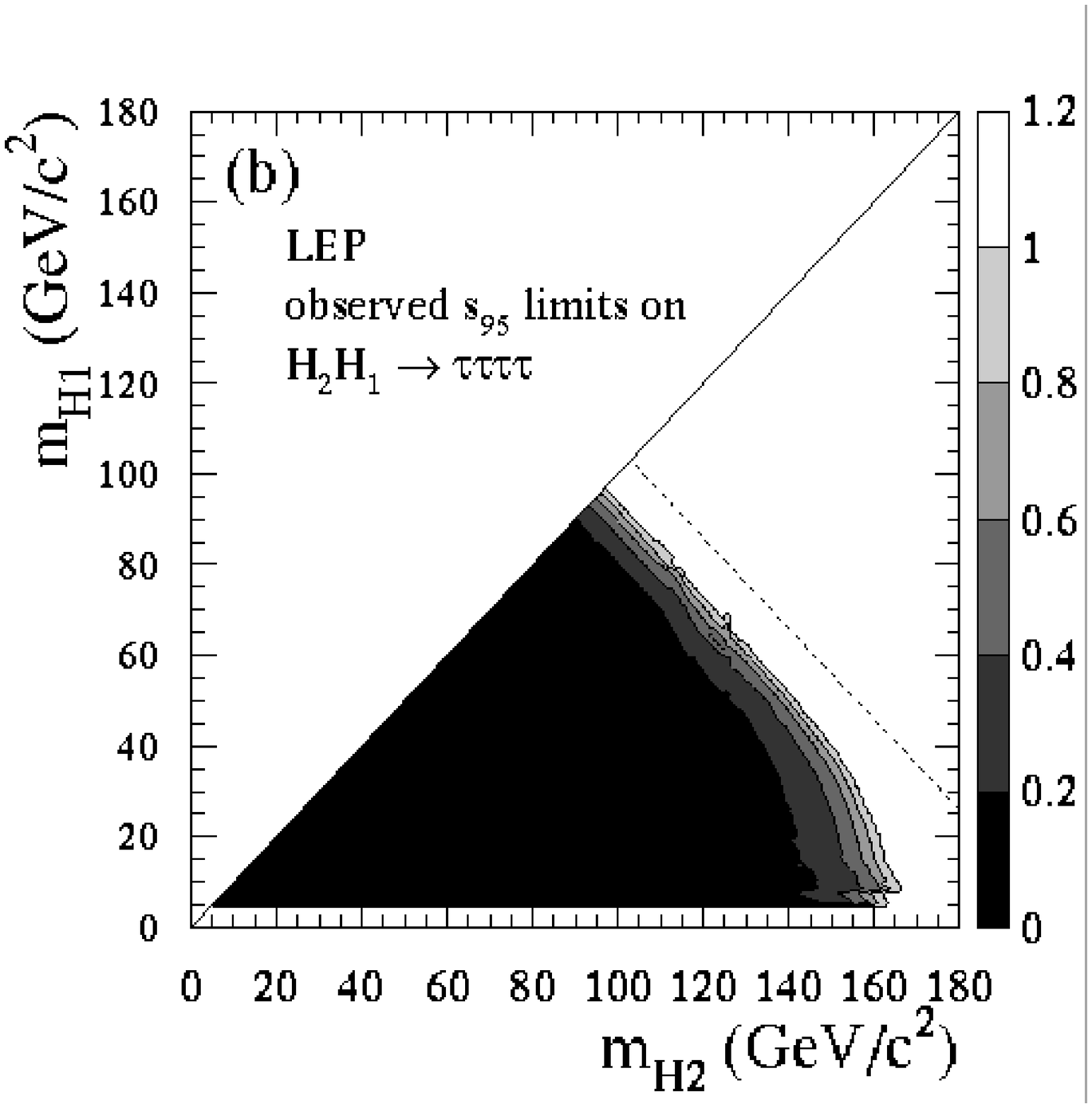} \hfill
\includegraphics[width=0.24\textwidth]{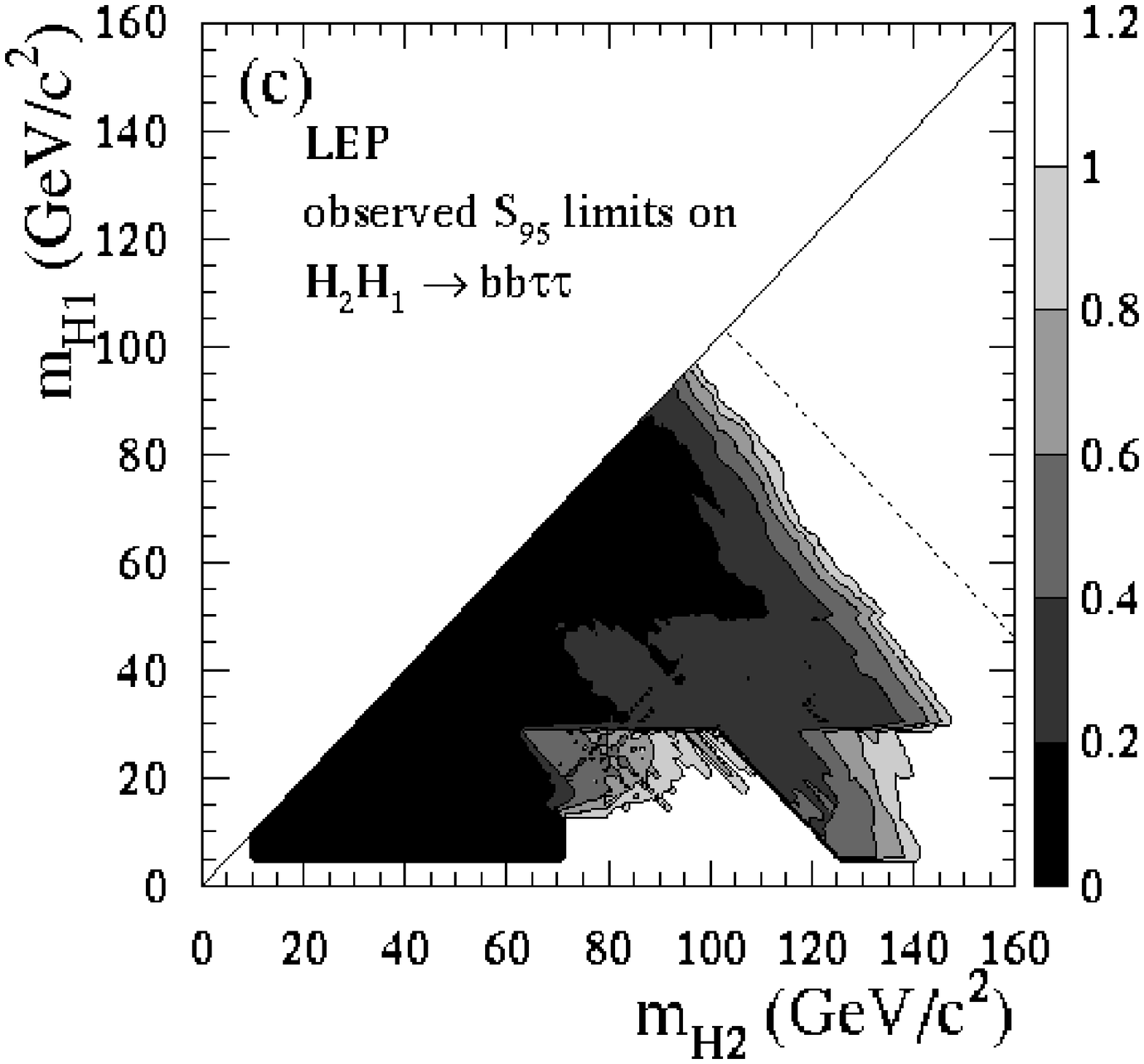} \hfill
\includegraphics[width=0.24\textwidth]{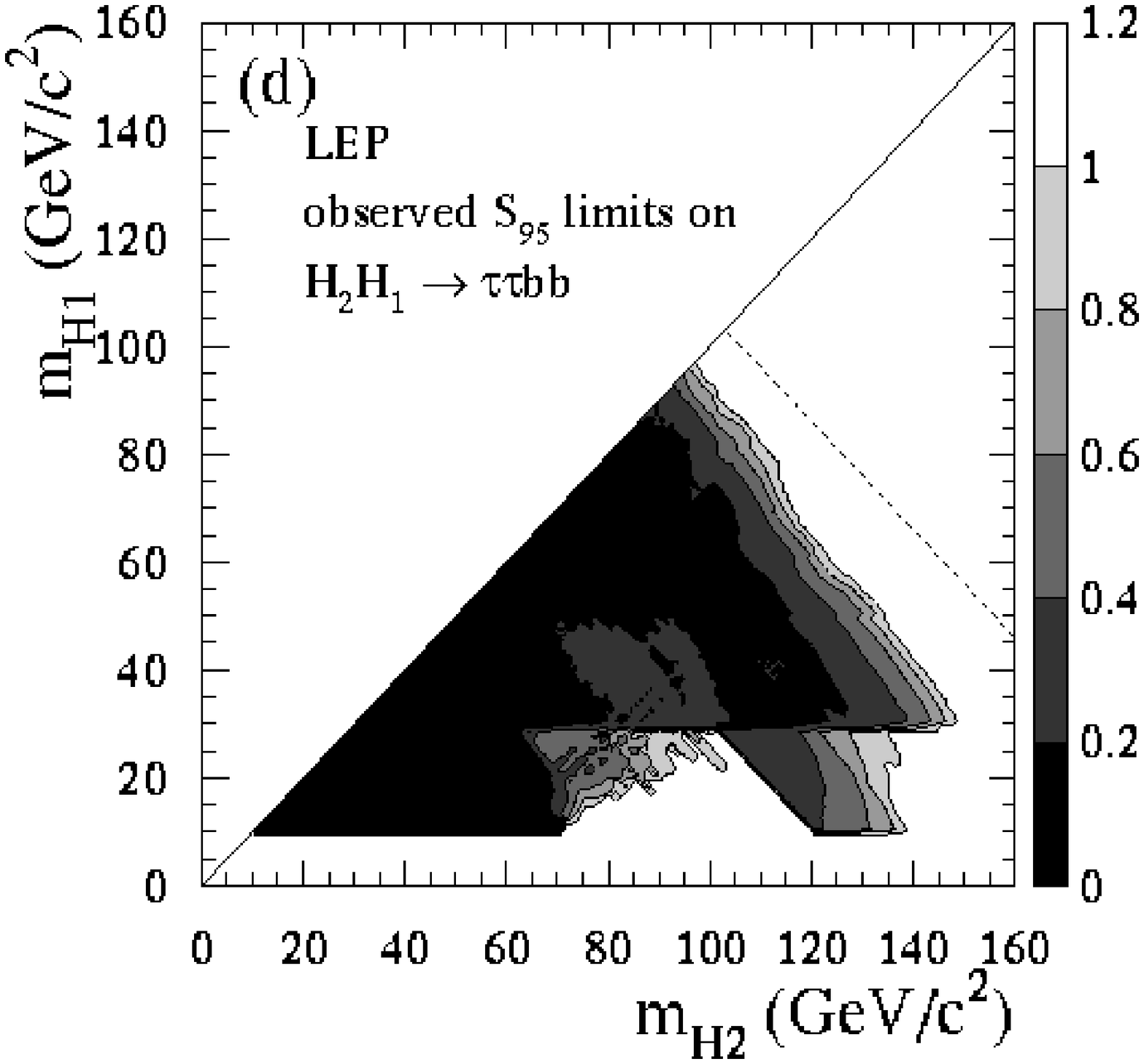}
\end{center}
\vspace*{-8mm}
\caption{Limits on $S_{\rm 95}$ for the $\rm H_1H_2$ process with different decays modes,
         as indicated in the plots.
}
\label{fig:hh_masses}
\vspace*{-7.5mm}
\end{figure}

\begin{figure}[h!]
\begin{center}
\includegraphics[width=0.24\textwidth]{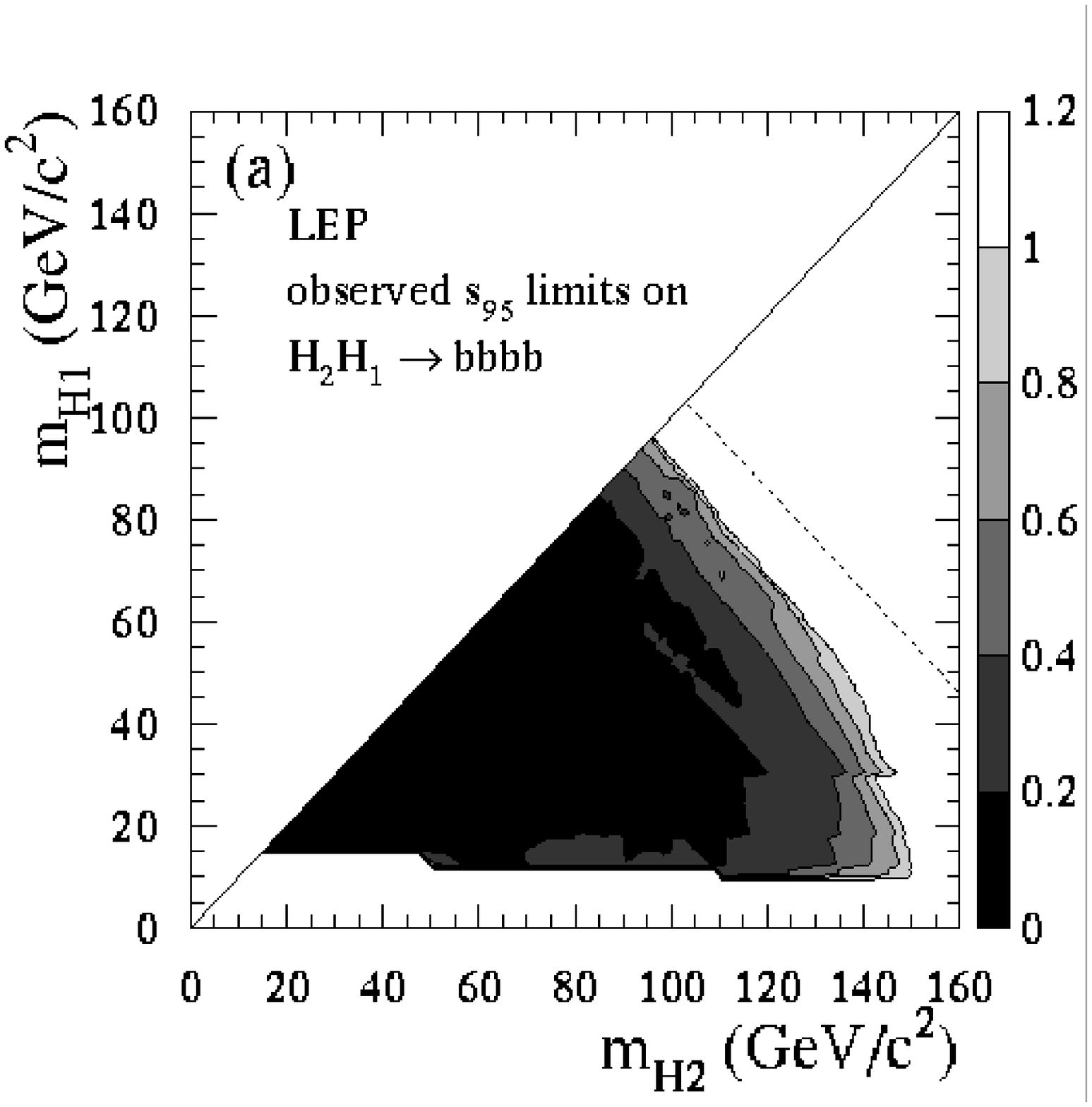} \hfill
\includegraphics[width=0.24\textwidth]{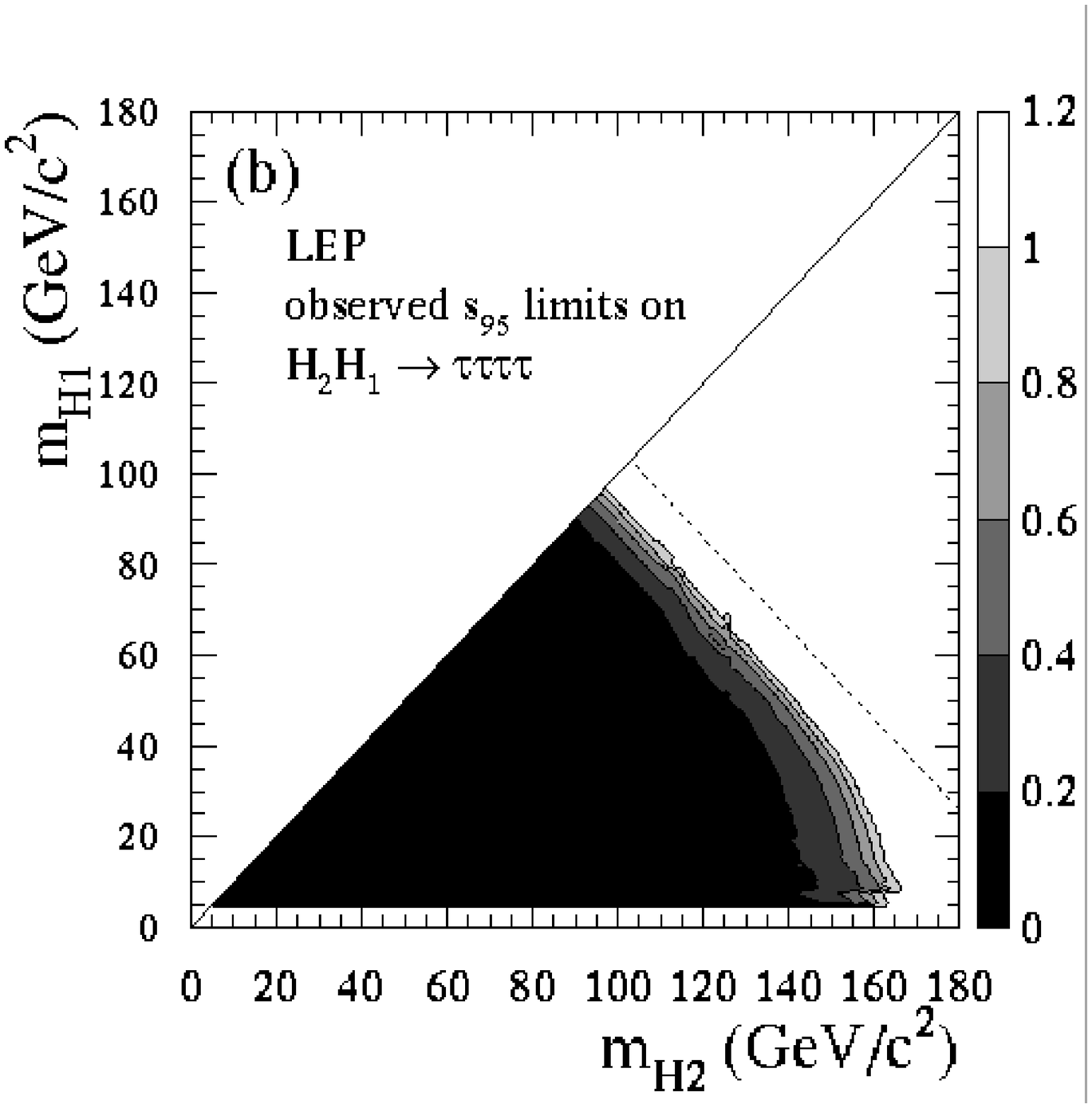} \hfill
\includegraphics[width=0.24\textwidth]{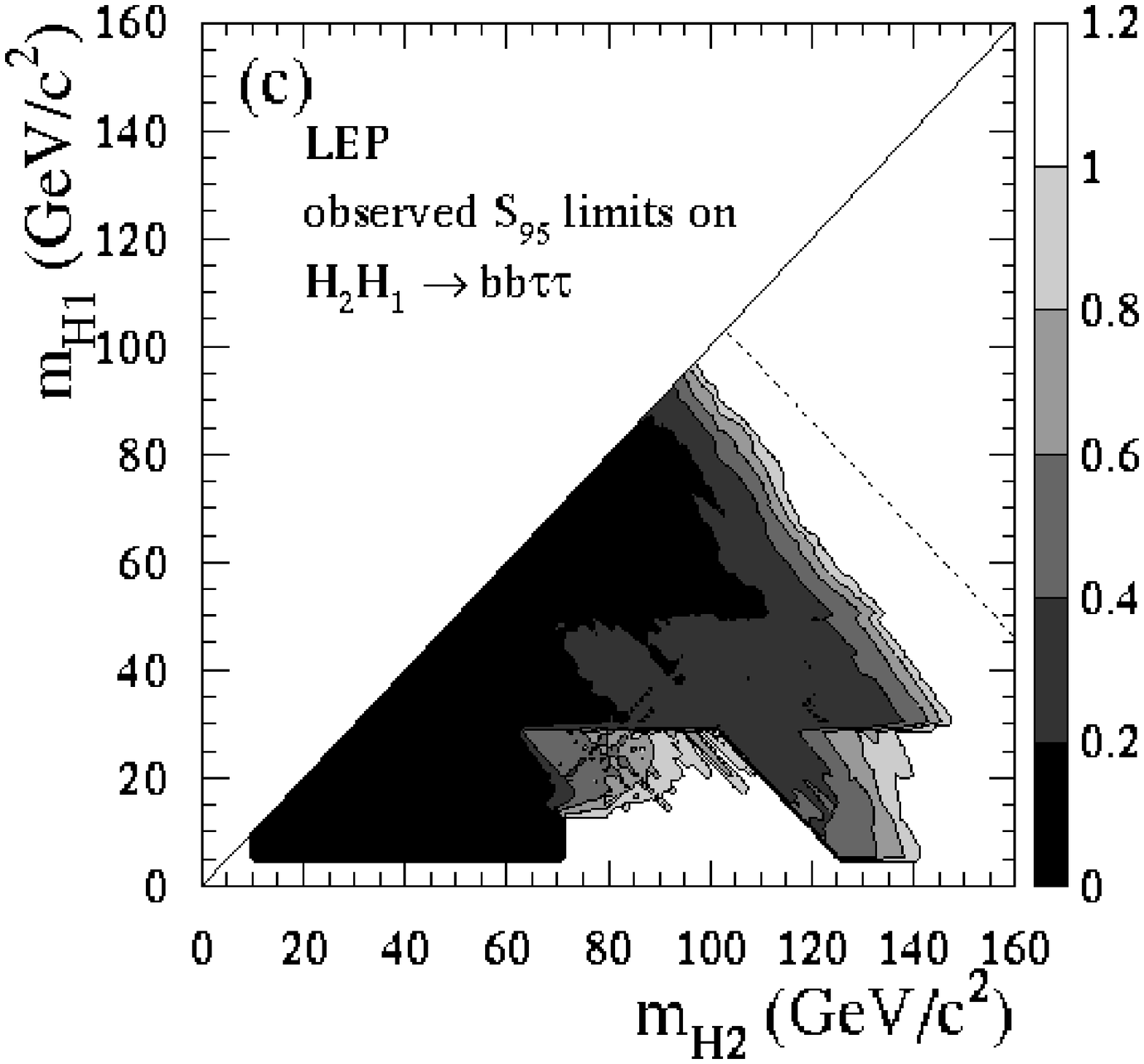}
\end{center}
\vspace*{-8mm}
\caption{Limits on $S_{\rm 95}$ for the $\rm H_1H_2\to H_1H_1H_1$ process with different decays 
modes, as given in the plots.
}
\label{fig:hhh_masses}
\vspace*{-6mm}
\end{figure}

In CP-conserving scenarios $\rm H_1$ is the scalar Higgs boson h, 
and $\rm H_2$ the pseudo-scalar A.
Limits on the production cross section are set for 
$S_{\rm 95} = \sigma_{\max} / \sigma_{\rm ref}$, where $\sigma_{\rm ref}$ is a reference 
cross section.
Figure~\ref{fig:xi} shows the limits on $S_{\rm 95} = \xi^2 = (g_{\rm HZZ}/g_{\rm HZZ}^{\rm SM})^2$.
The limits on $\rm H_2Z \to H_1H_1Z$ are shown in Fig.~\ref{fig:hhh}.
Limits are given in Fig.~\ref{fig:hh} for the $\rm H_1H_2$ process with 
`h-max', $\rm b\bar b$ and $\tau^+\tau^-$ decay modes, and as a function of scalar and 
pseudoscalar masses in in Fig.~\ref{fig:hh_masses}. The limits for the process
$\rm H_1H_2\to H_1H_1H_1$ are given in Fig.~\ref{fig:hhh_masses}.

\clearpage

\section{CP-Conserving Interpretations}
\begin{figure}[h!]
\vspace*{-1mm}
\begin{center}
\includegraphics[width=0.3\textwidth]{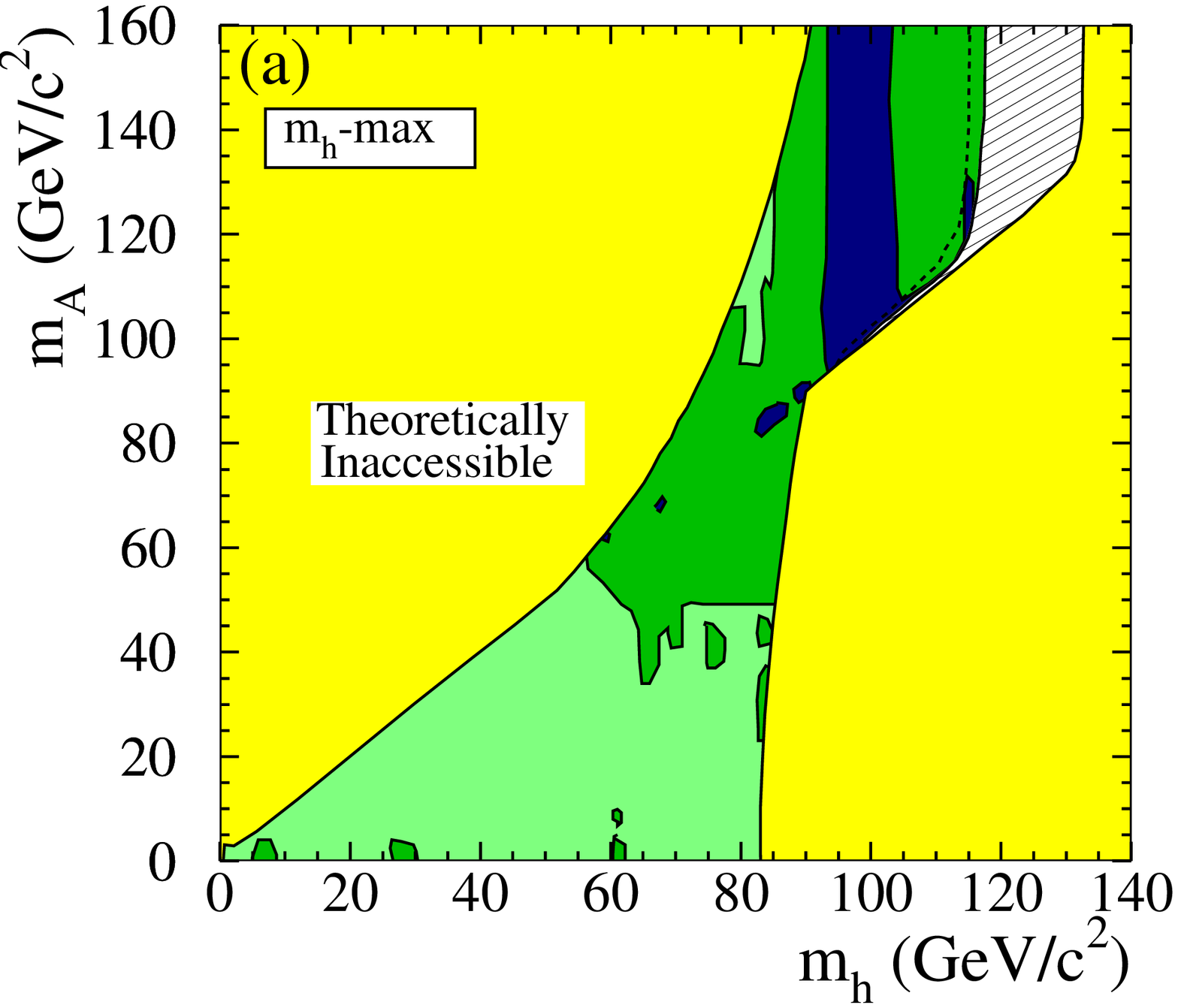} \hfill
\includegraphics[width=0.3\textwidth]{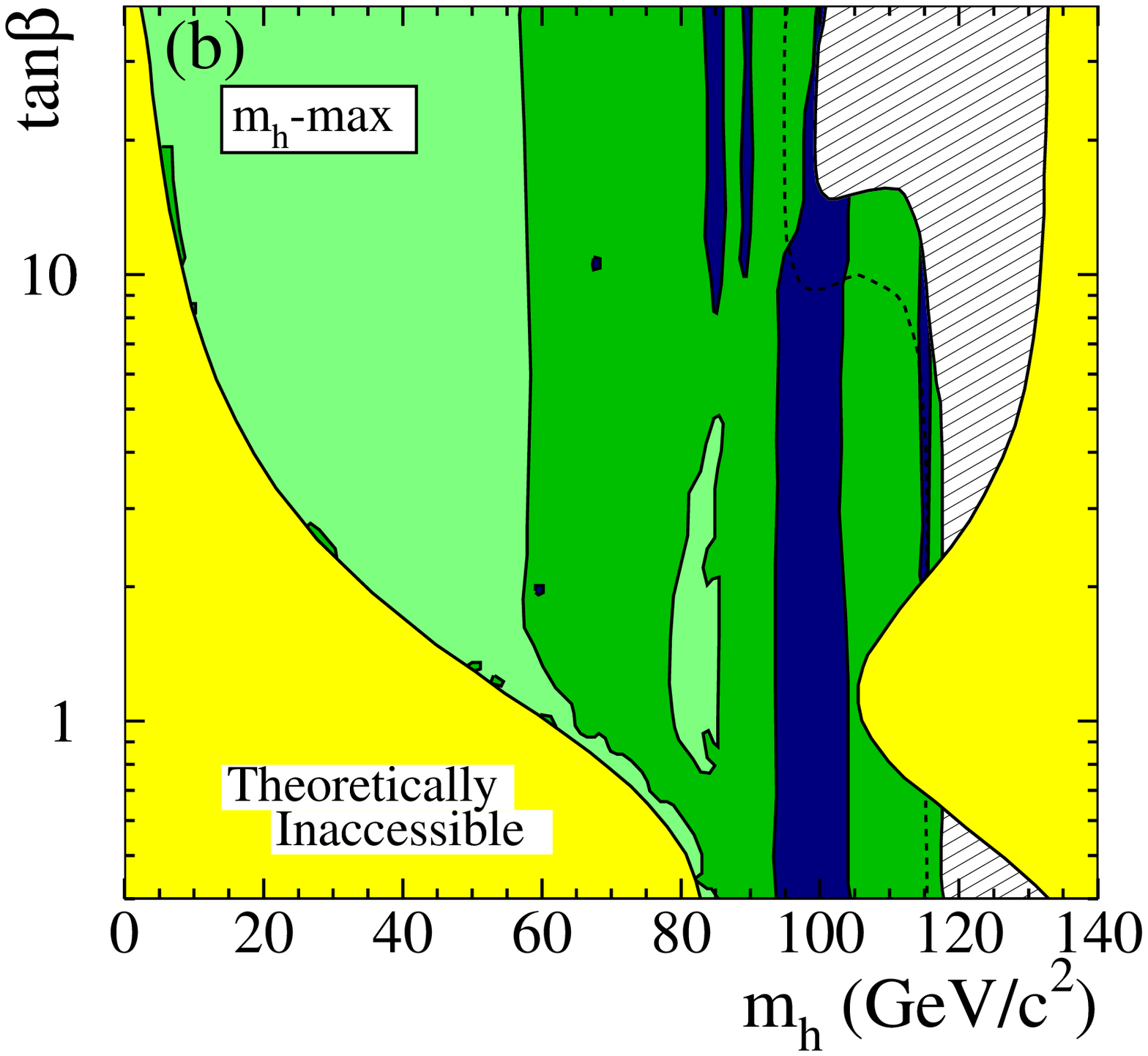} \hfill
\includegraphics[width=0.3\textwidth]{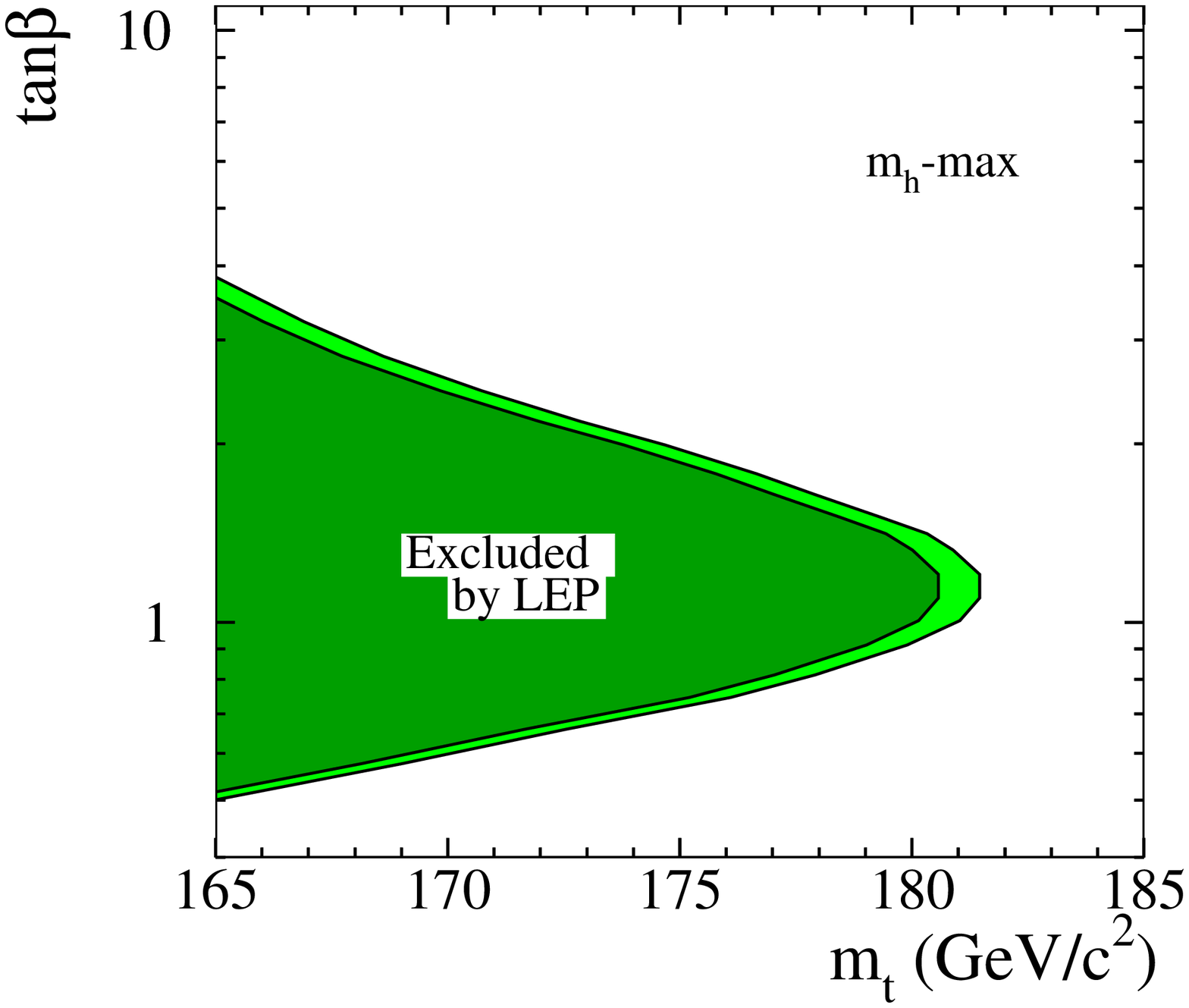}
\end{center}
\vspace*{-8mm}
\caption{Left and center: Contour regions of $1-CL_{\rm b}$ in the h-max scenario.
Light-green: $1\sigma$, dark-green: 1-2$\sigma$, blue: $>$$2\sigma$.
Expected limits in the absence of a signal are indicated with a 
dotted line. In the hatched region appearent excess would not be significant.
Right: resulting limit on $\tan\beta$ as a function of the top quark mass. 
}
\label{fig:clb_hmax}
\vspace*{-9.5mm}
\end{figure}

\begin{figure}[h!]
\vspace*{-1mm}
\begin{center}
\includegraphics[width=0.24\textwidth]{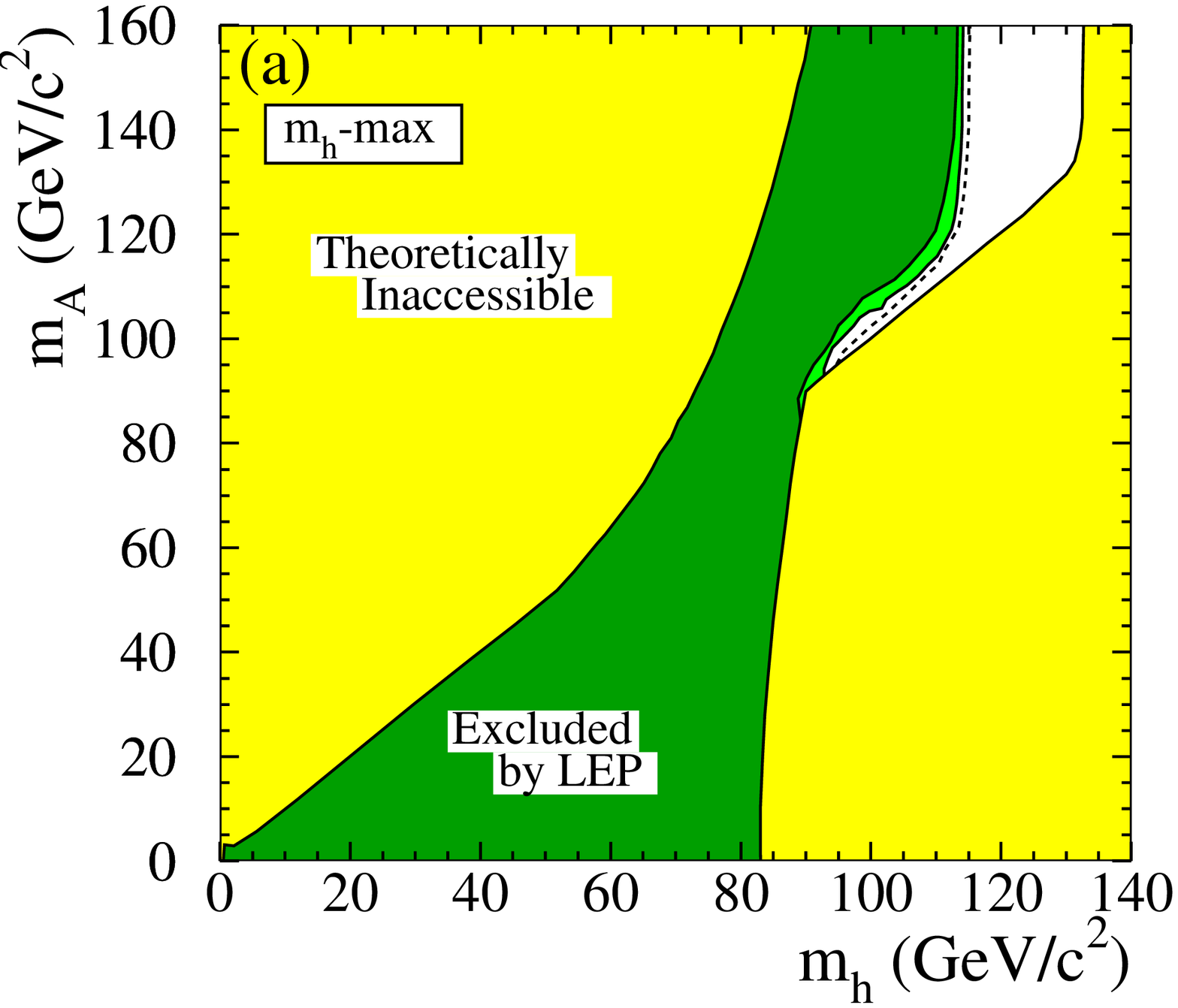} \hfill
\includegraphics[width=0.24\textwidth]{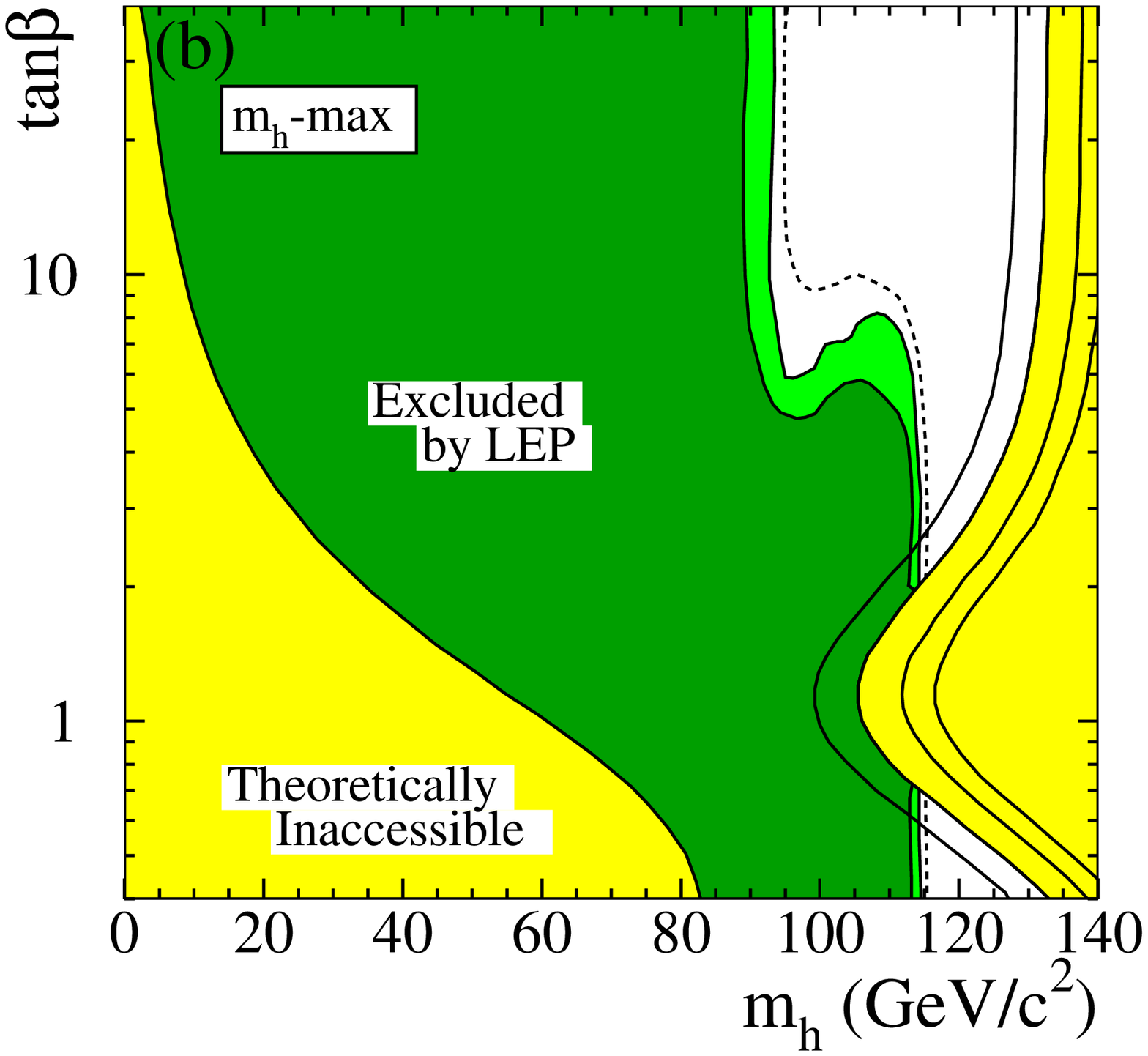} \hfill
\includegraphics[width=0.24\textwidth]{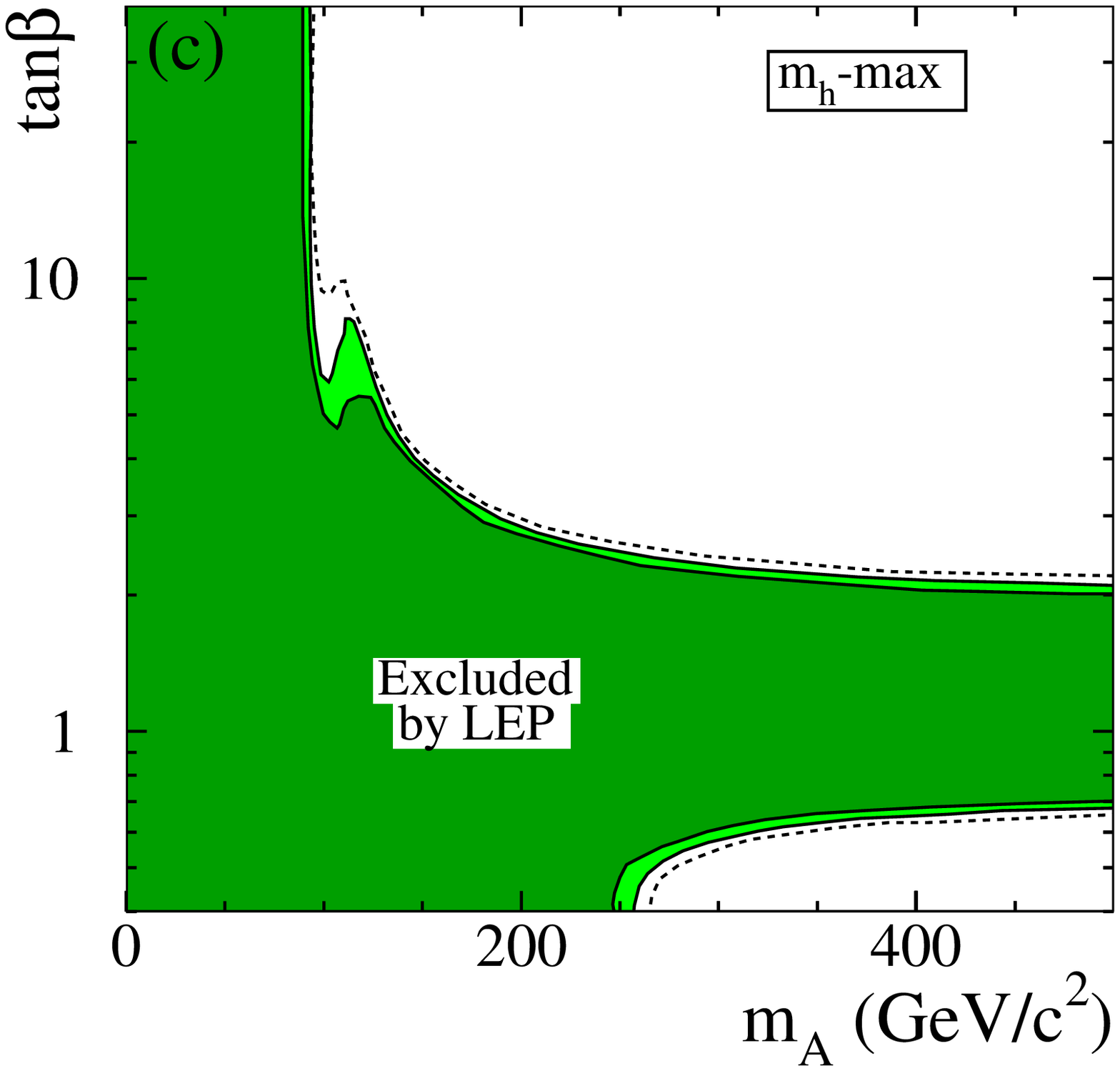} \hfill
\includegraphics[width=0.24\textwidth]{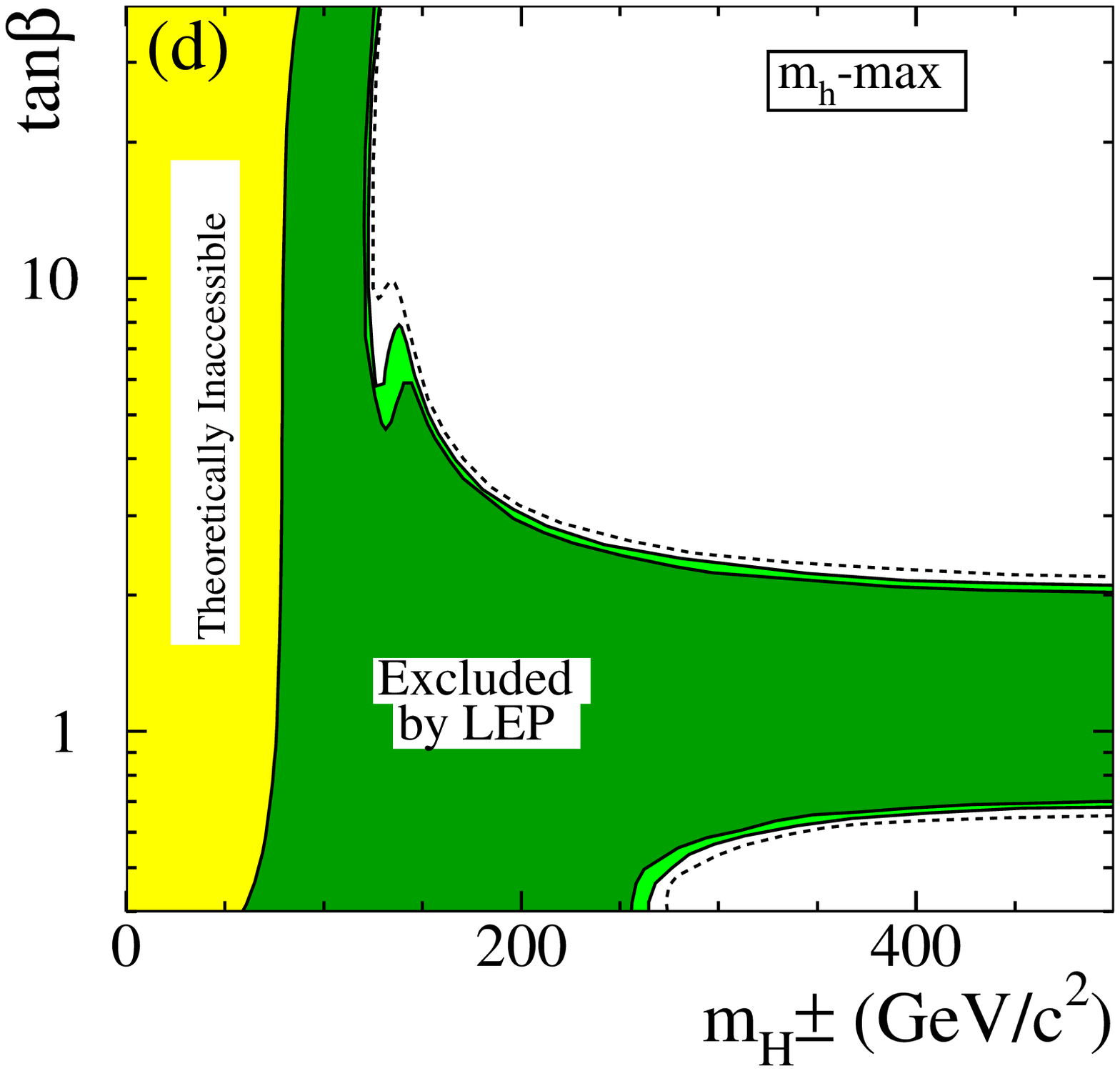}
\end{center}
\vspace*{-8mm}
\caption{Excluded regions in the h-max scenario at 95\% CL
(light-green) and at 99.7\% CL (dark-green).
Expected limits in the absence of a signal are indicated with a 
dotted line.
}
\label{fig:cls_hmax}
\vspace*{-4mm}
\end{figure}

\begin{figure}[h!]
\begin{center}
\includegraphics[width=0.24\textwidth]{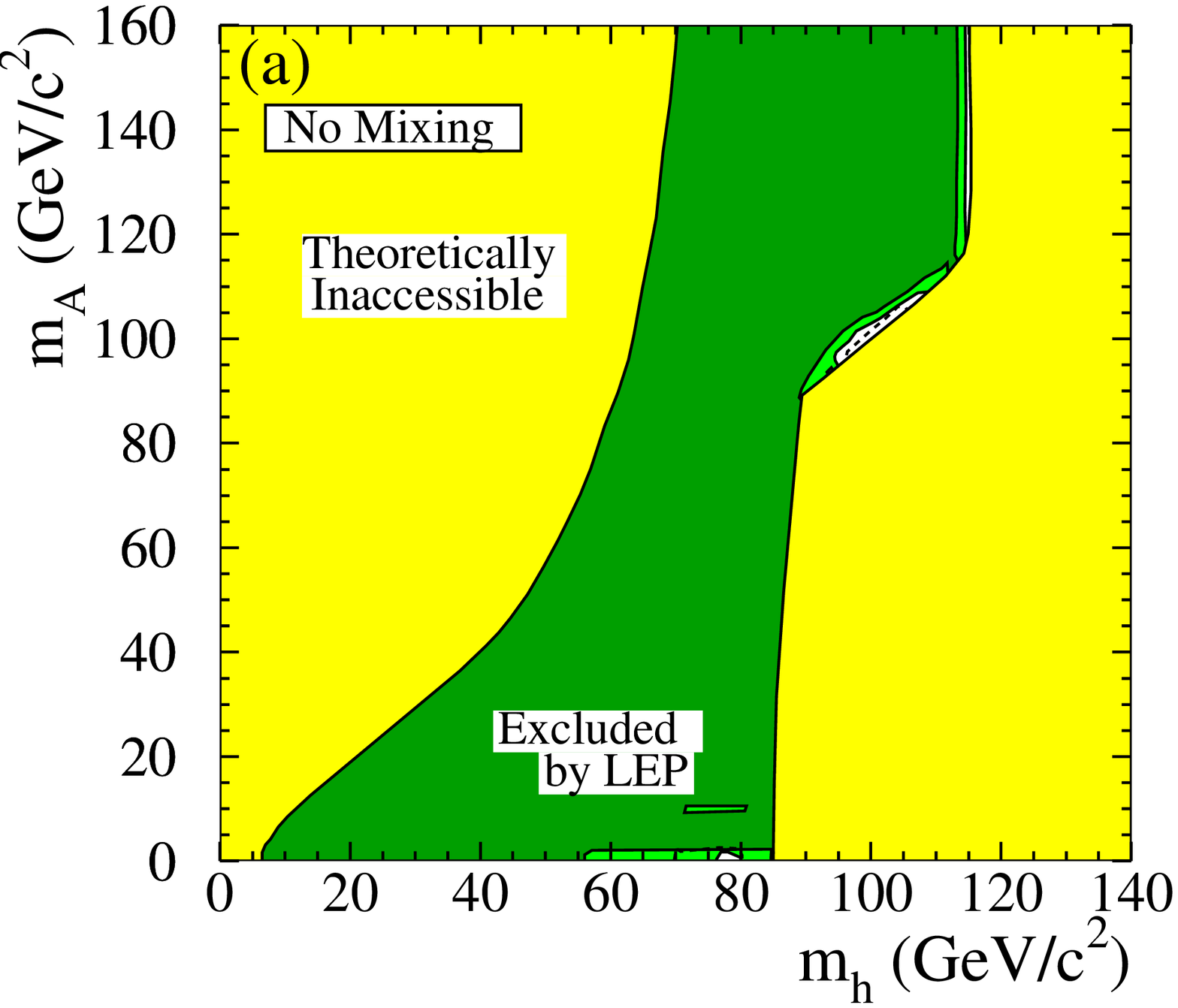} \hfill
\includegraphics[width=0.24\textwidth]{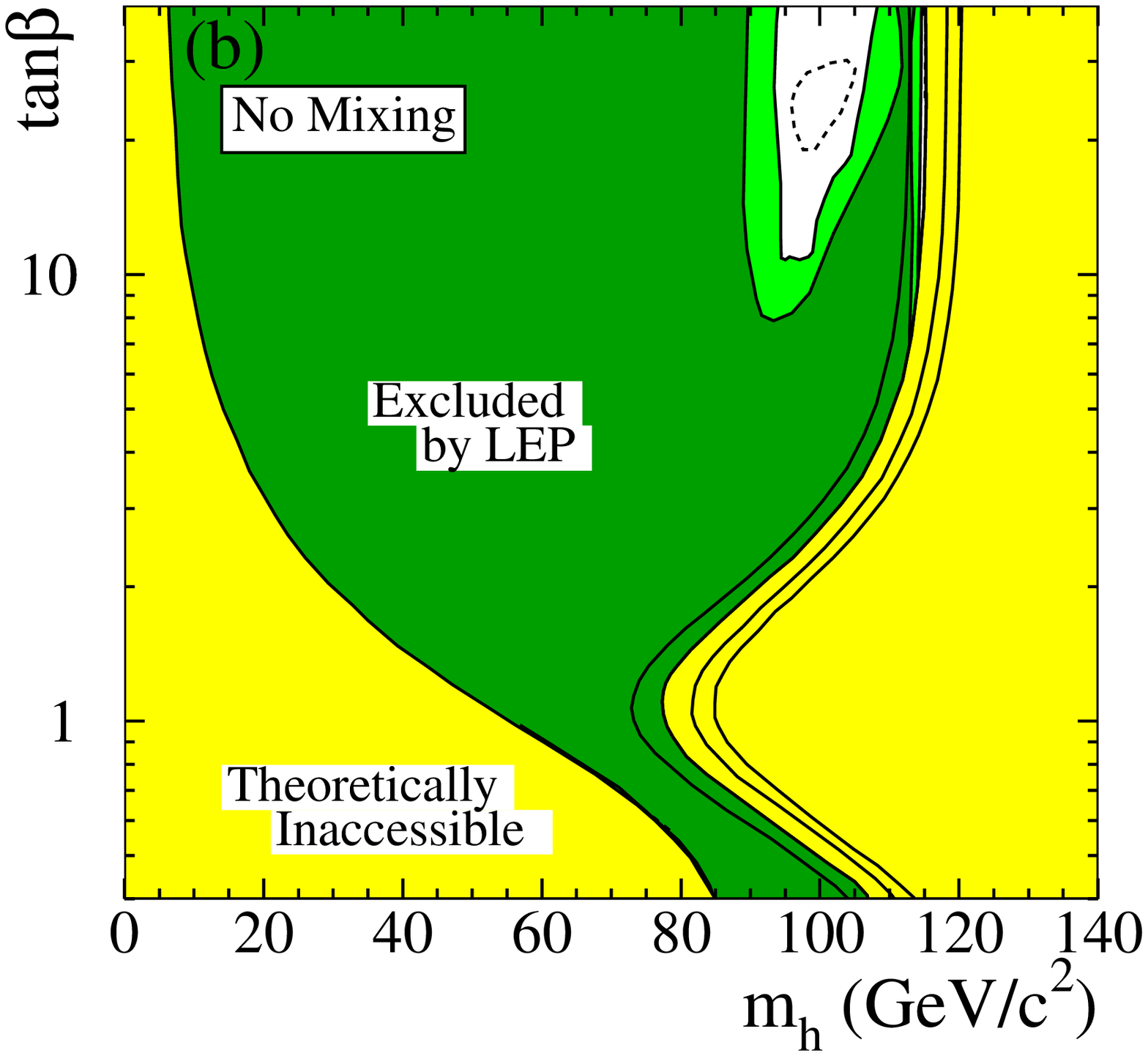} \hfill
\includegraphics[width=0.24\textwidth]{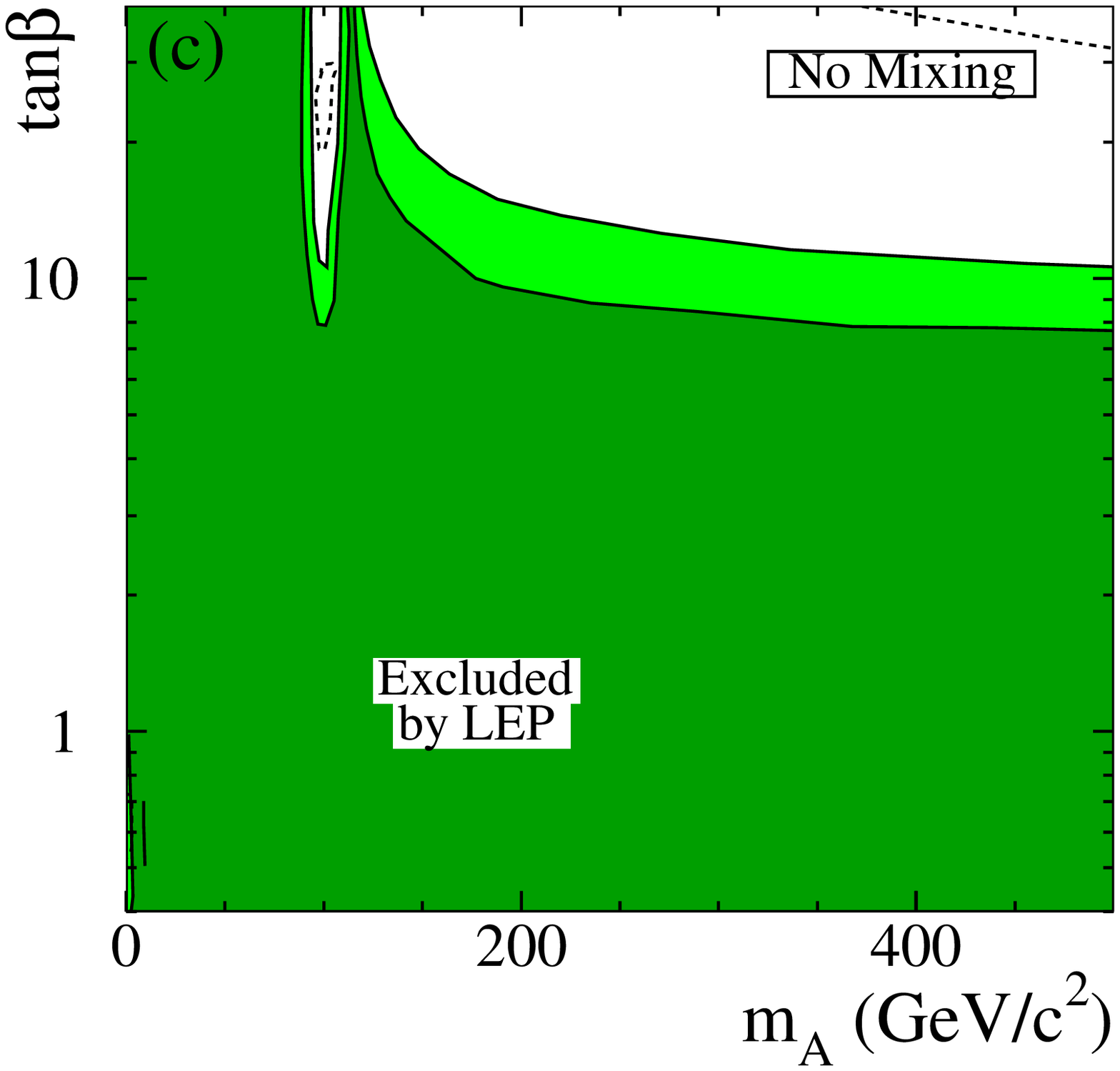} \hfill
\includegraphics[width=0.24\textwidth]{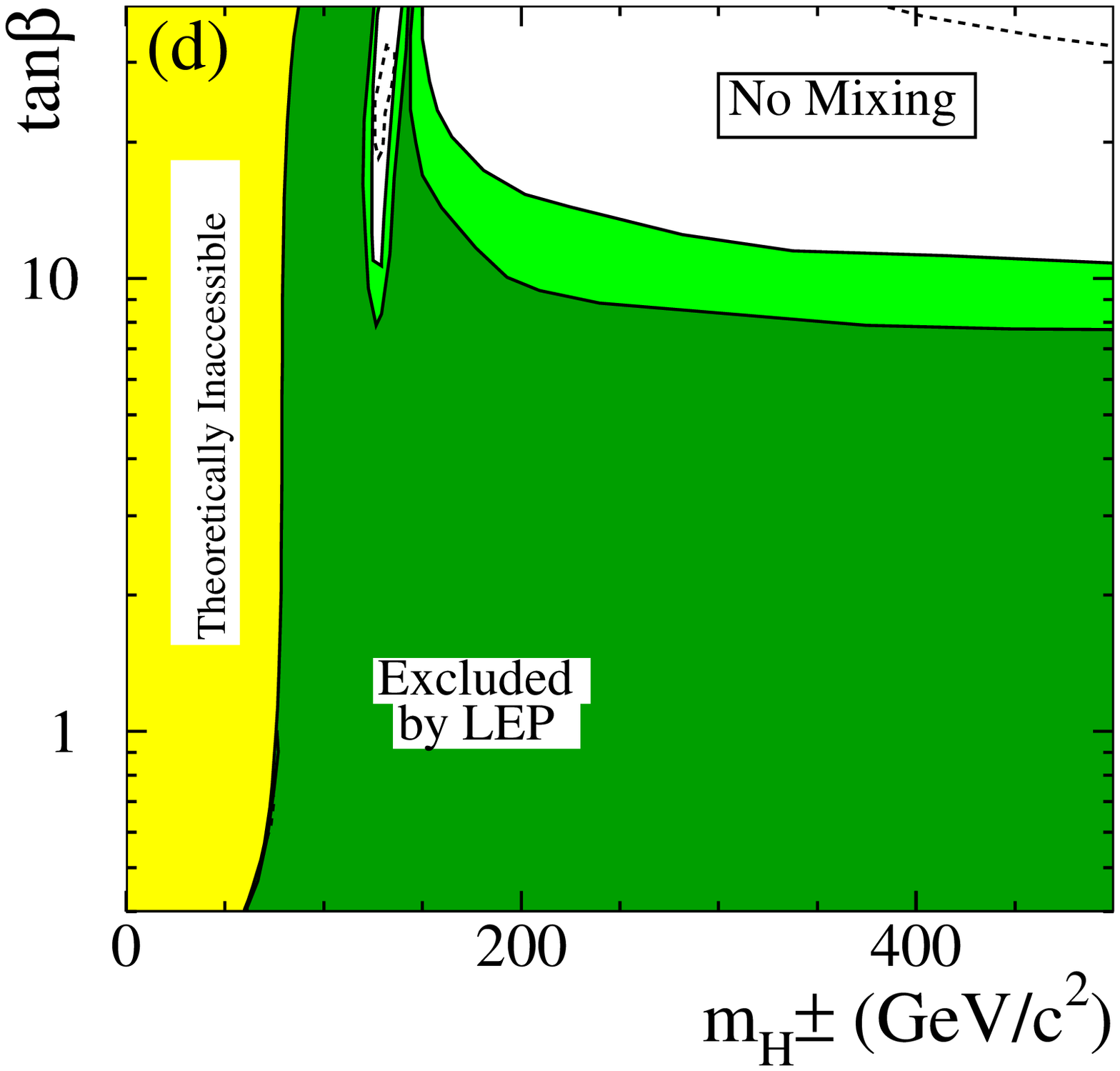}
\end{center}
\vspace*{-8mm}
\caption{Excluded regions in the no-mixing scenario at 95\% CL
(light-green) and at 99.7\% CL (dark-green).
Expected limits in the absence of a signal are indicated with a 
dotted line.
}
\label{fig:cls_hmin}
\vspace*{-4mm}
\end{figure}

\begin{figure}[h!]
\begin{center}
\includegraphics[width=0.24\textwidth]{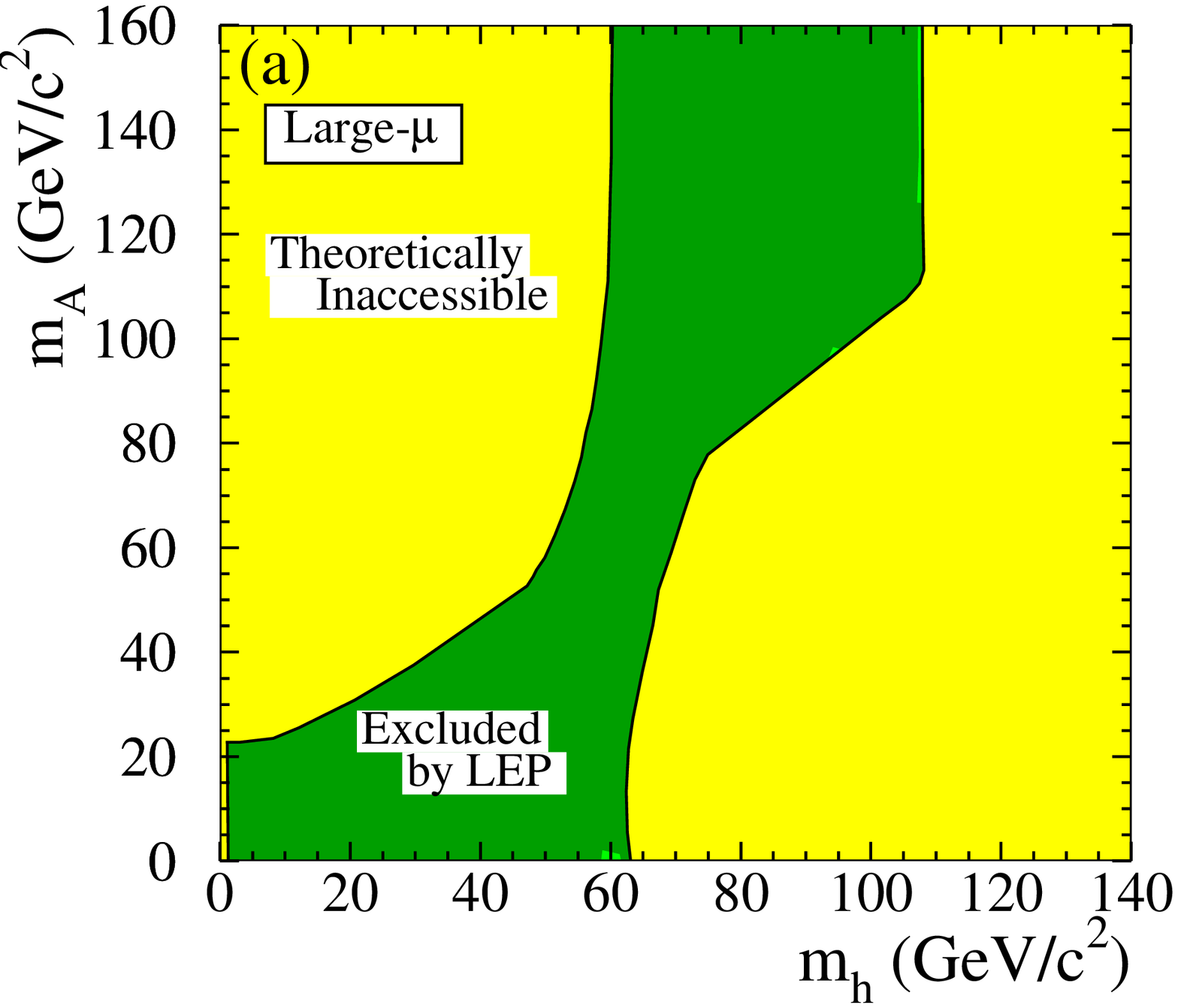} \hfill
\includegraphics[width=0.24\textwidth]{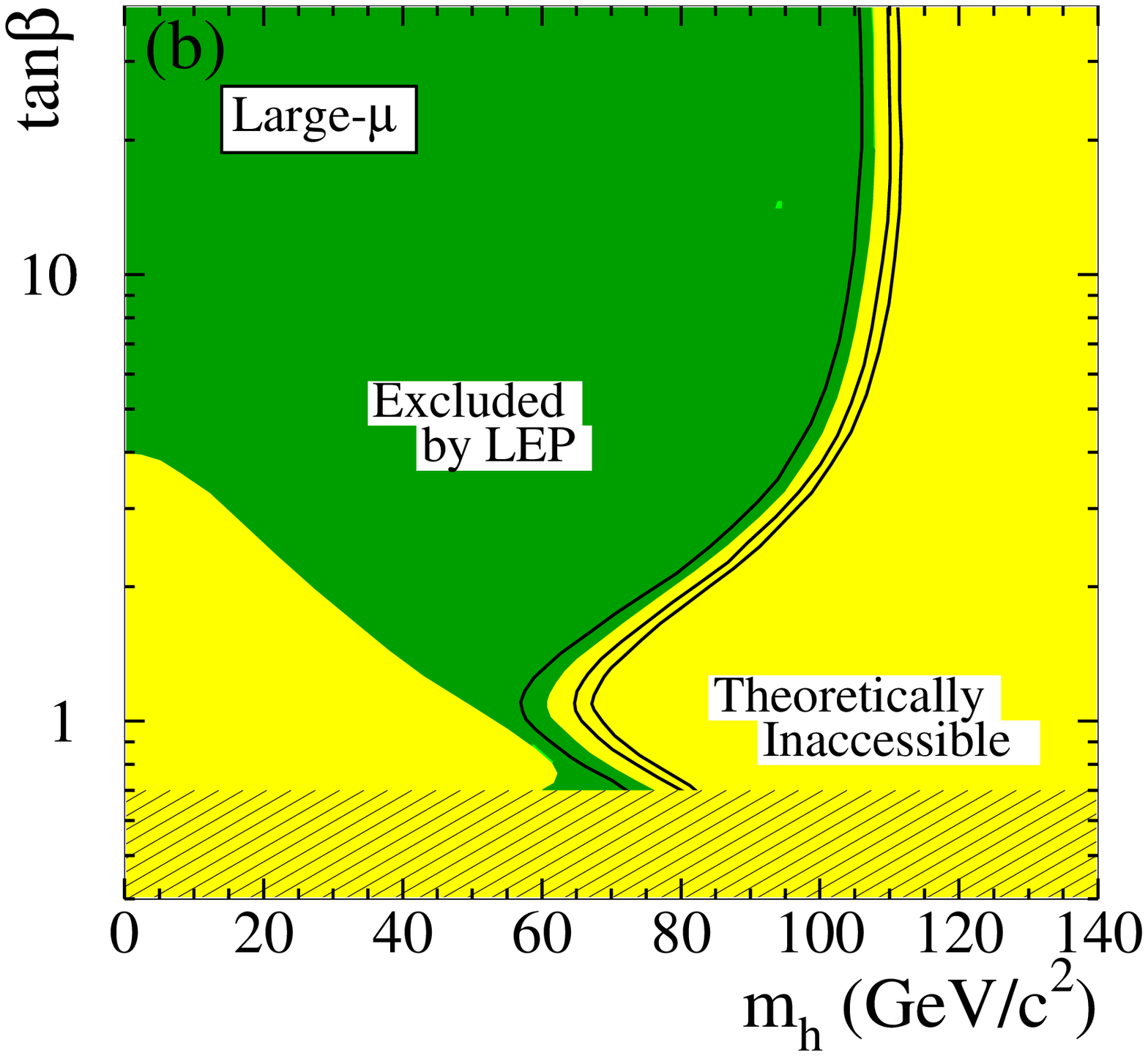} \hfill
\includegraphics[width=0.24\textwidth]{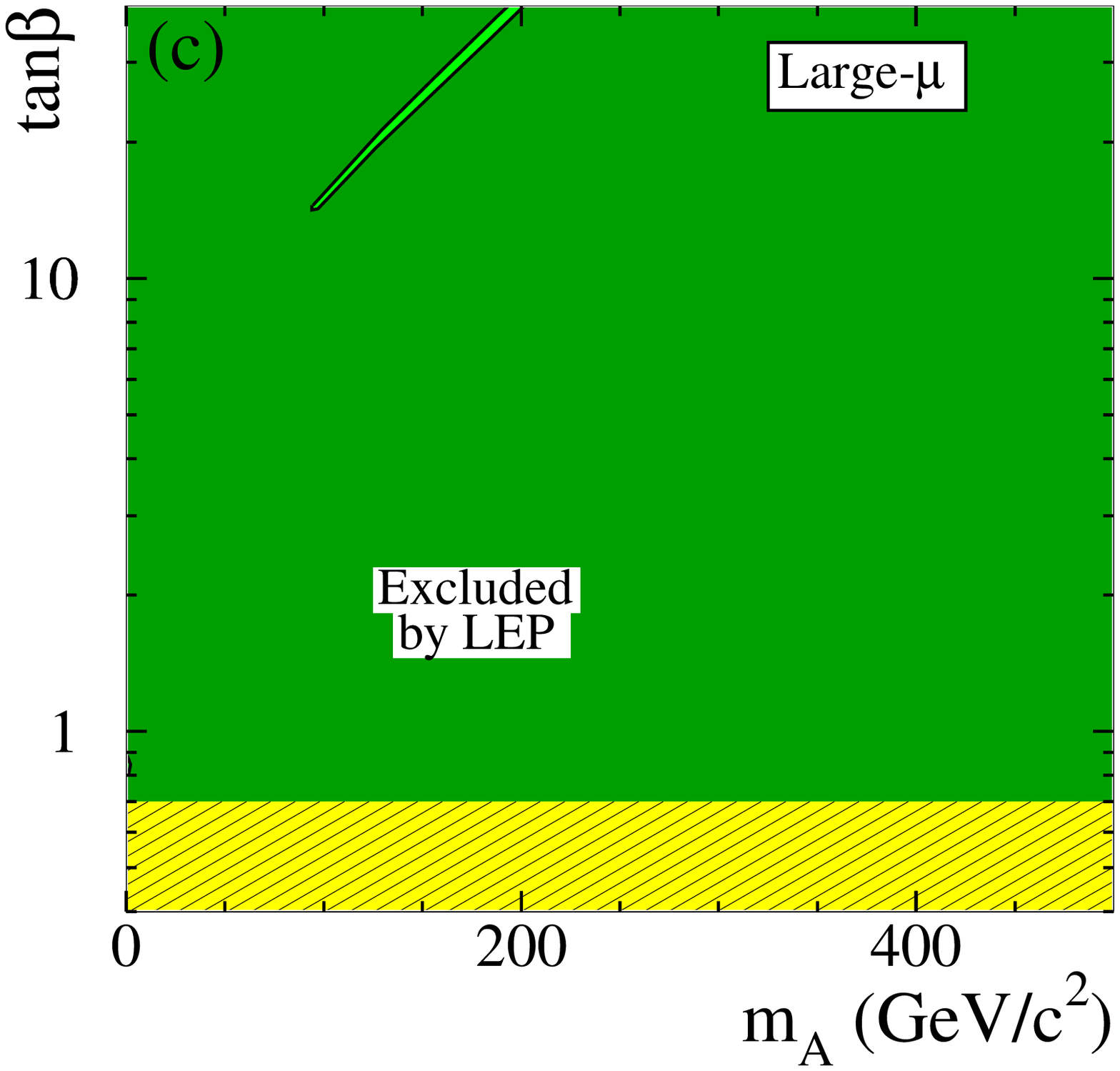} \hfill
\includegraphics[width=0.24\textwidth]{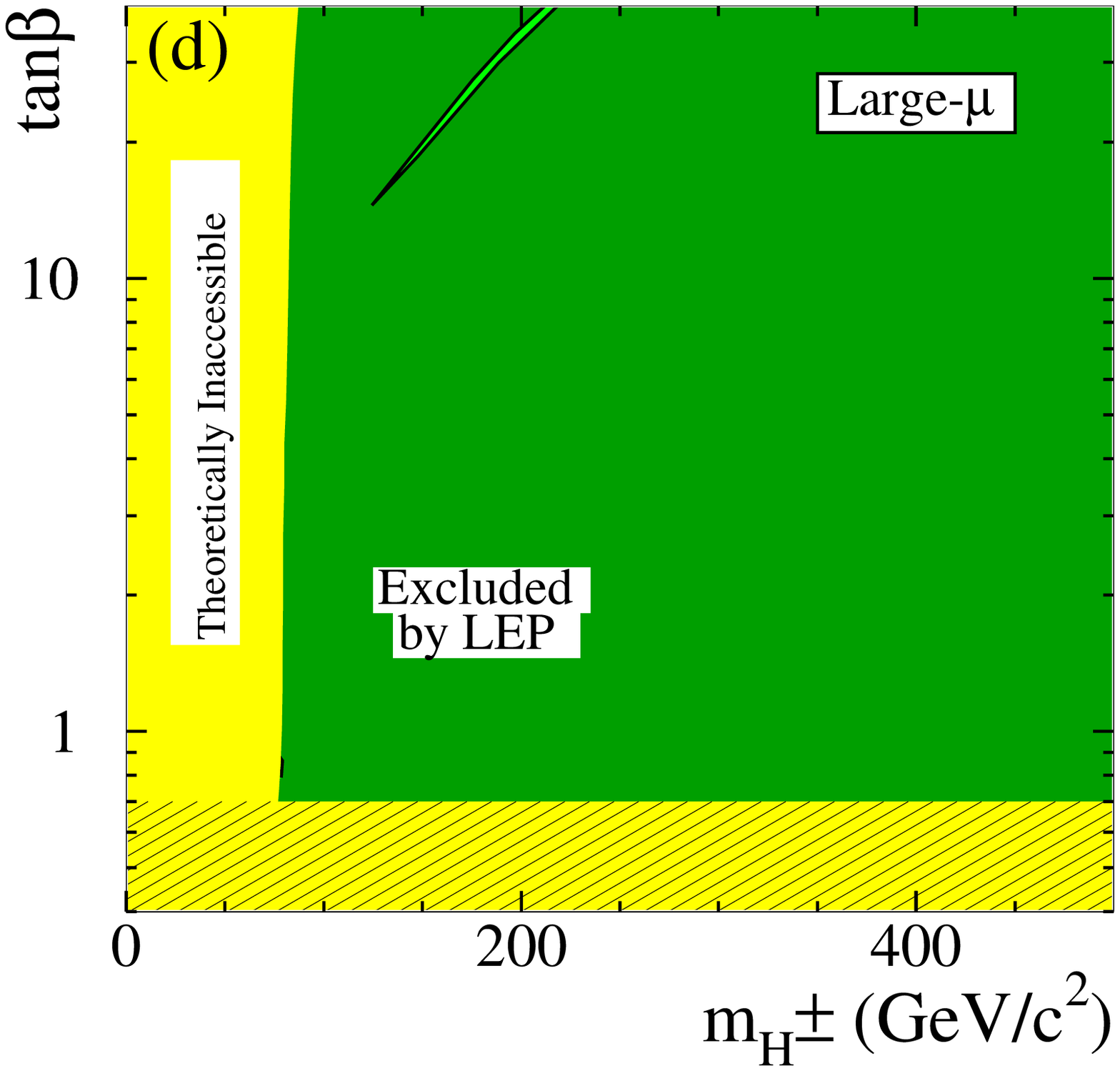}
\end{center}
\vspace*{-8mm}
\caption{Excluded regions in the large-$\mu$ scenario at 95\% CL
(light-green) and at 99.7\% CL (dark-green).
Expected limits in the absence of a signal are indicated with a 
dotted line.
}
\label{fig:large_mu}
\vspace*{-4mm}
\end{figure}

\begin{figure}[thp]
\begin{center}
\includegraphics[width=0.24\textwidth]{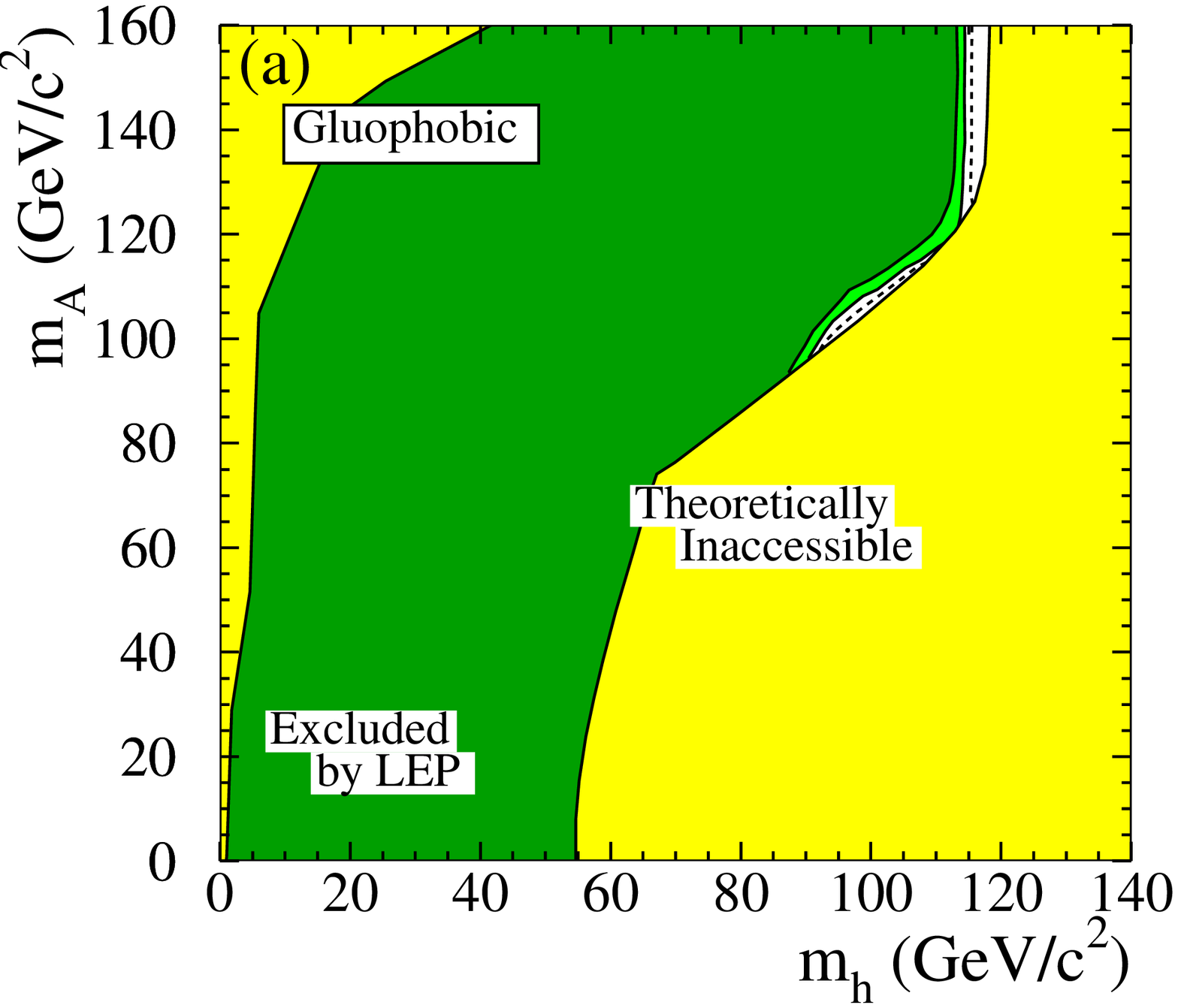} \hfill
\includegraphics[width=0.24\textwidth]{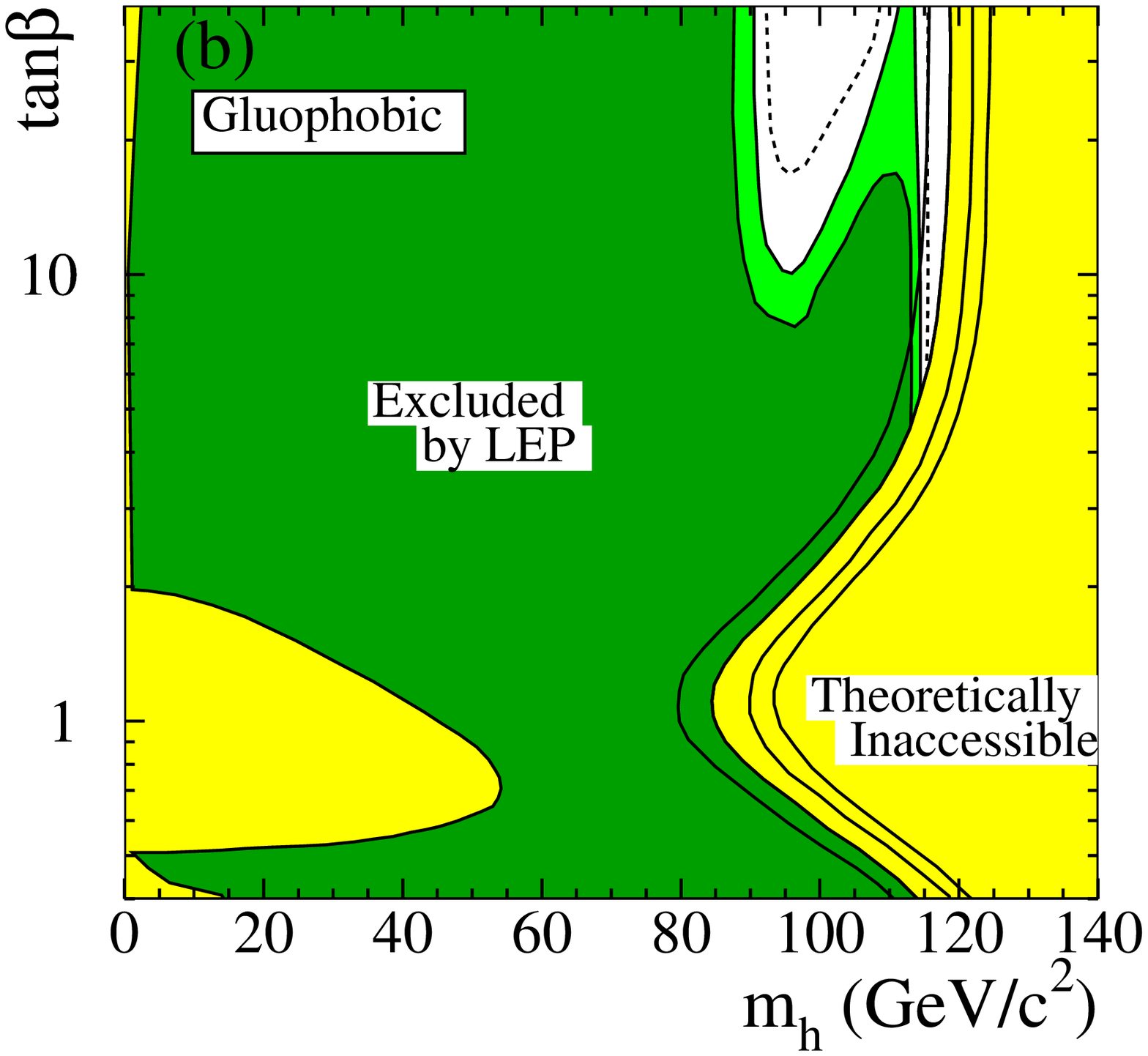} \hfill
\includegraphics[width=0.24\textwidth]{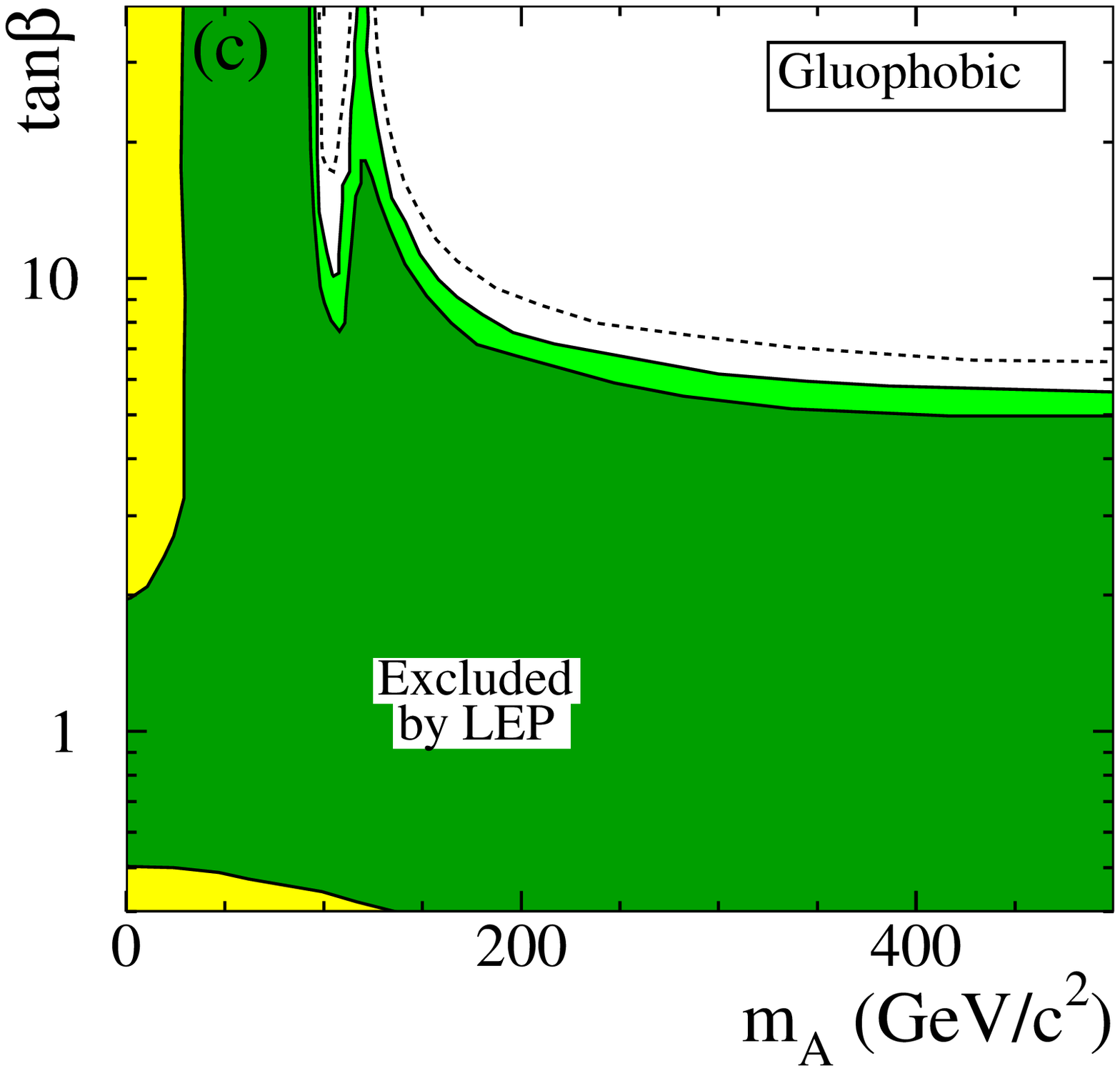} \hfill
\includegraphics[width=0.24\textwidth]{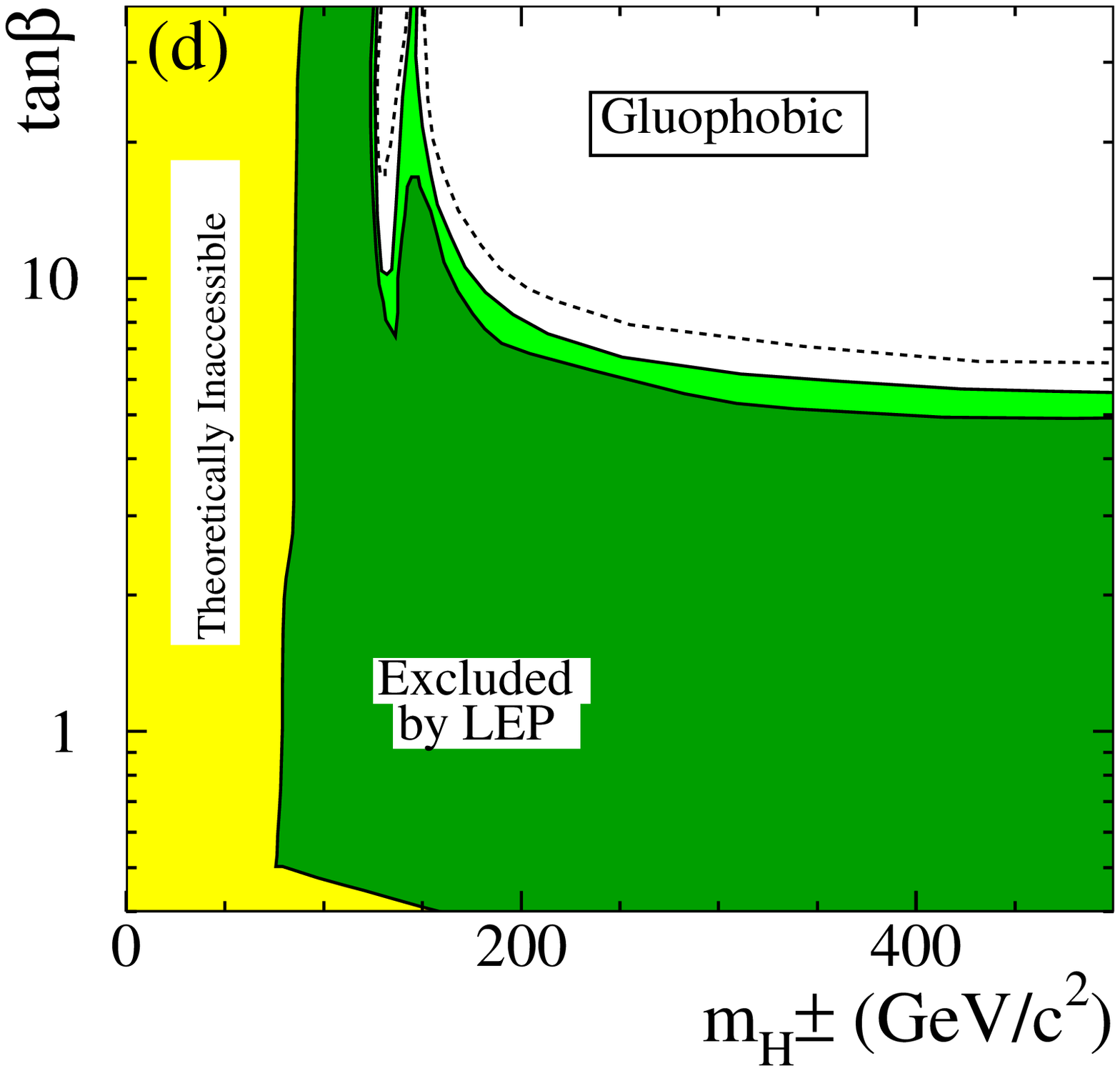}
\end{center}
\vspace*{-8mm}
\caption{Excluded regions in the gluophobic scenario at 95\% CL
(light-green) and at 99.7\% CL (dark-green).
Expected limits in the absence of a signal are indicated with a 
dotted line.
}
\label{fig:gluophobic}
\vspace*{-4mm}
\end{figure}

\begin{figure}[h!]
\begin{center}
\includegraphics[width=0.24\textwidth]{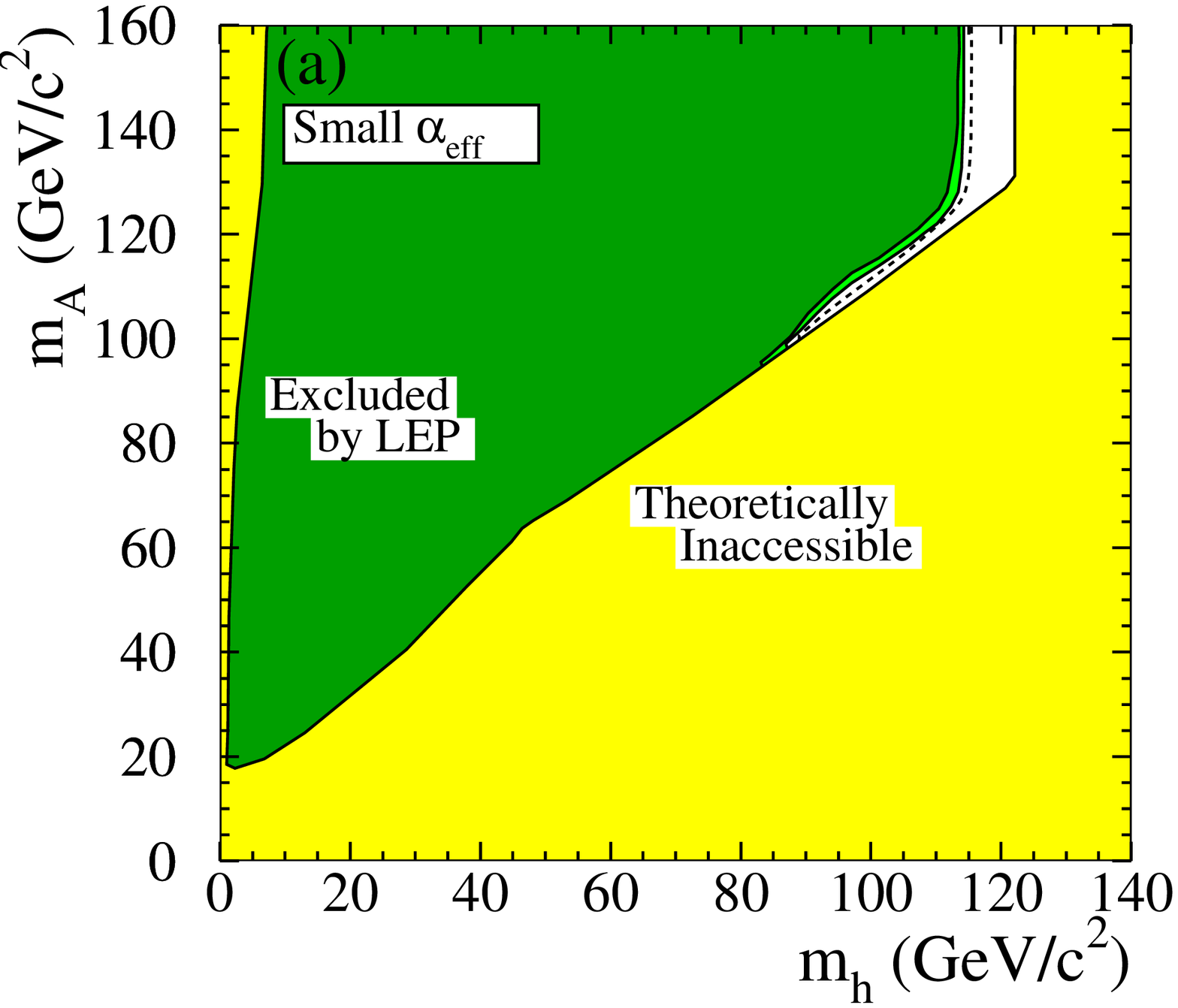} \hfill
\includegraphics[width=0.24\textwidth]{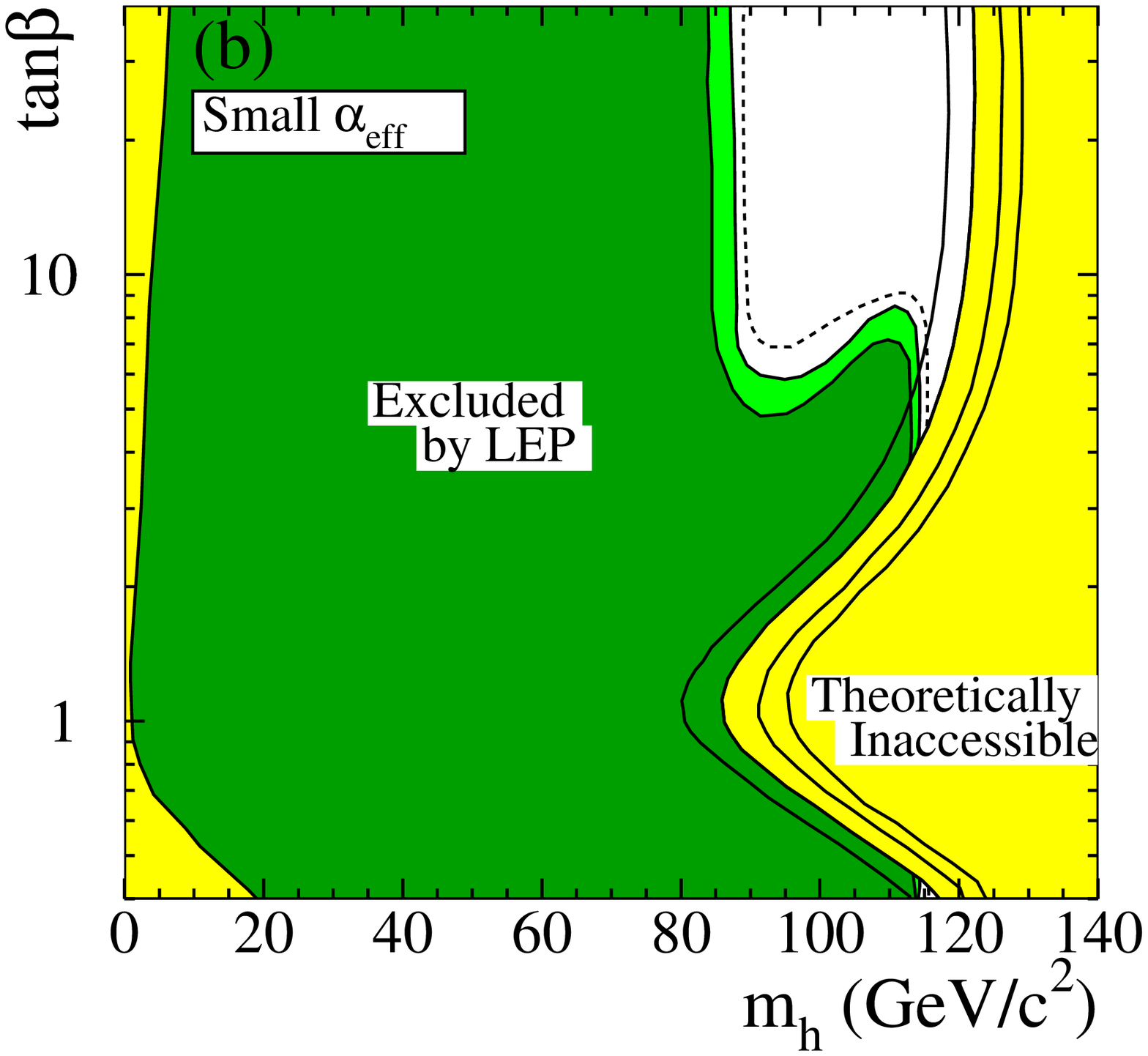} \hfill
\includegraphics[width=0.24\textwidth]{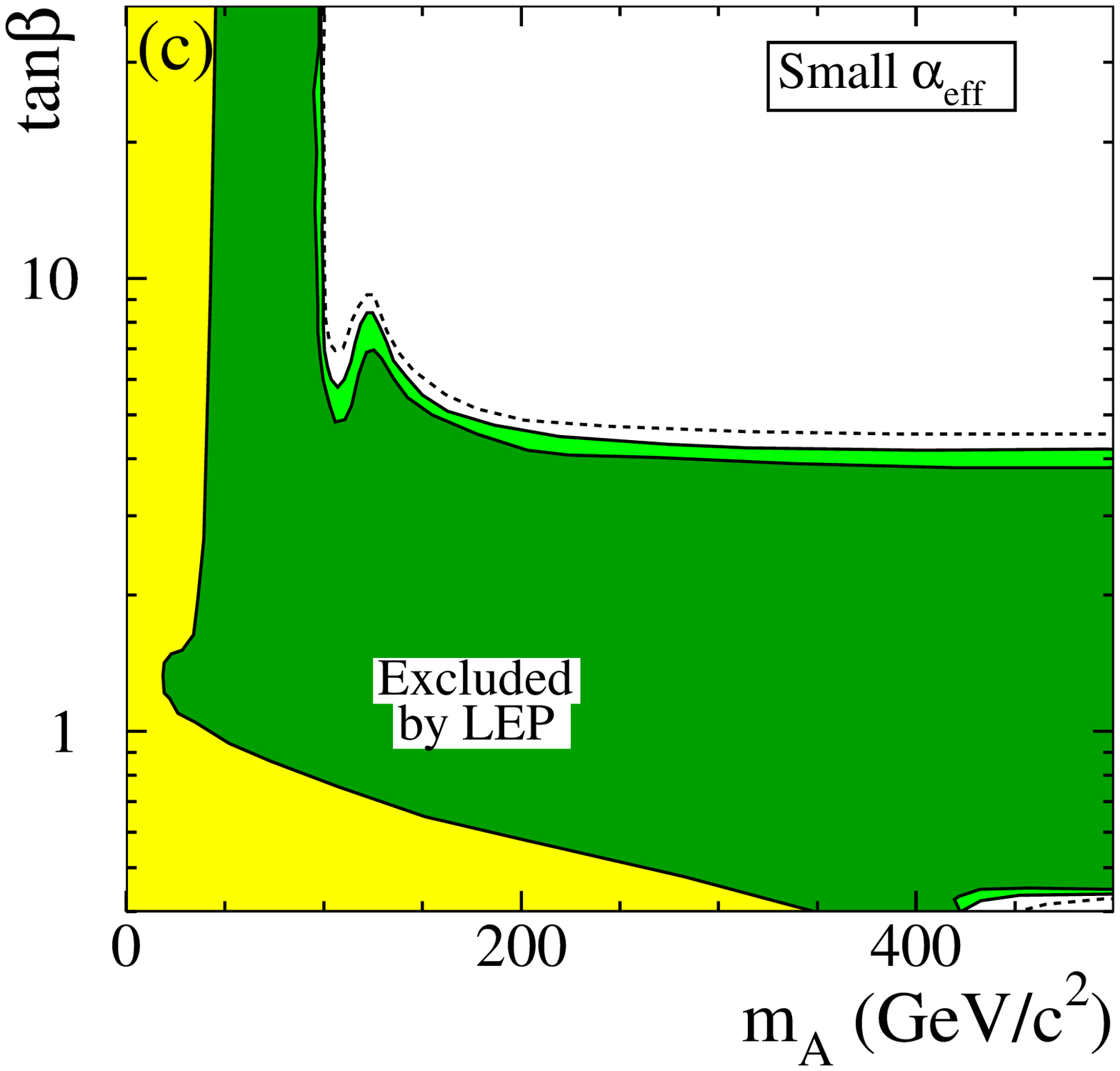} \hfill
\includegraphics[width=0.24\textwidth]{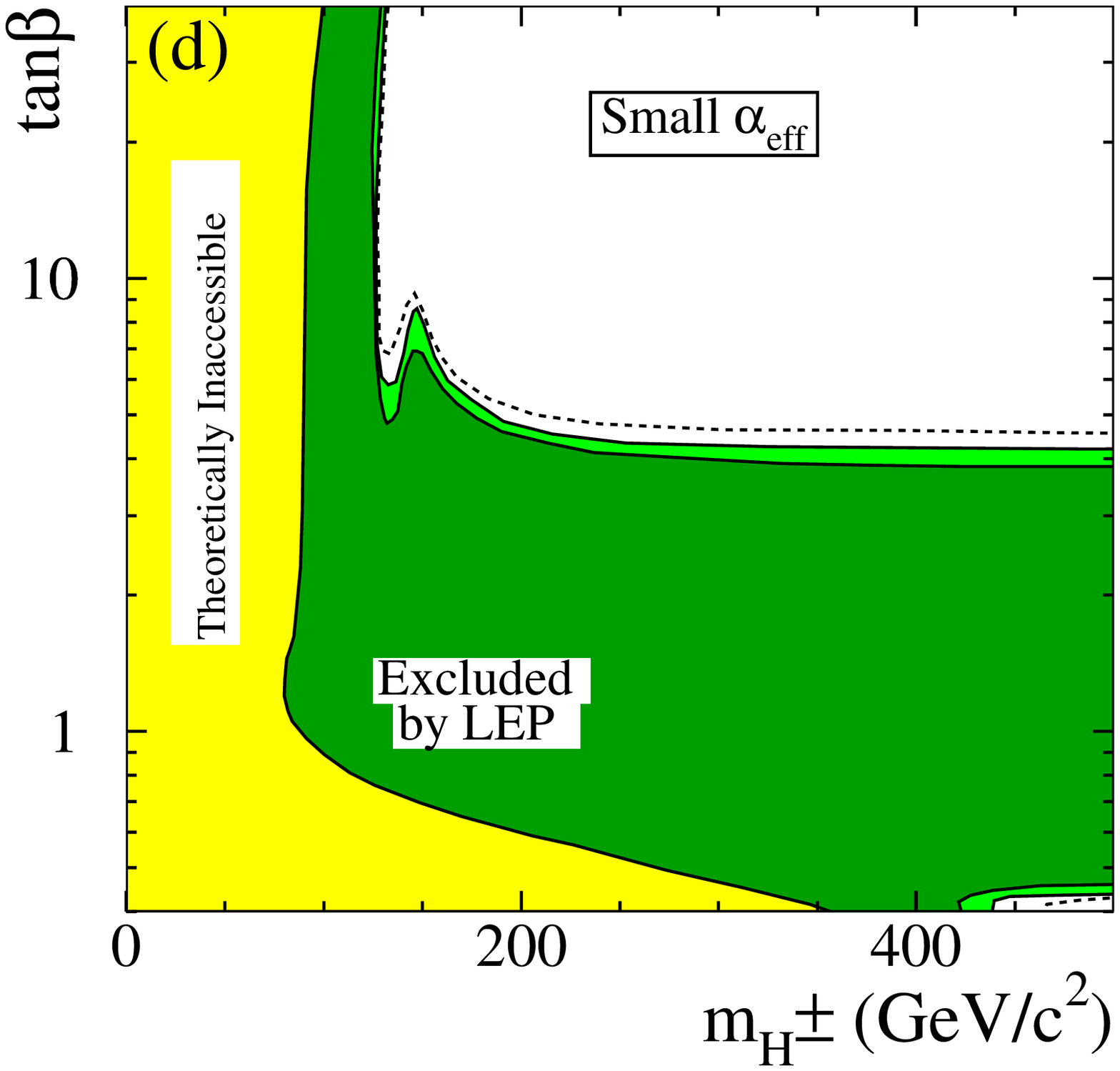}
\end{center}
\vspace*{-8mm}
\caption{Excluded regions in the small-$\alpha_{\rm eff}$ scenario at 95\% CL
(light-green) and at 99.7\% CL (dark-green).
Expected limits in the absence of a signal are indicated with a 
dotted line.
}
\label{fig:alpha}
\end{figure}

The background-only hypothesis is tested in the h-max scenario and expressed in Fig.~\ref{fig:clb_hmax}
as $1-CL_{\rm b}$ regions for the $1\sigma$, 1-2$\sigma$, and $>$$2\sigma$ contours (An interpretation
including this $>$$2\sigma$ excess near $m_{\rm h}=100$~GeV/$c^2$ as a Three-Higgs-Boson hypothesis was 
given~\cite{three_higgs}.)
The corresponding exclusion contours are shown in Fig.~\ref{fig:cls_hmax} for 
$m_{\rm t} = 174.3$~GeV/$c^2$~\cite{top}.
The variation of the $\tan\beta$ limit is shown in Fig.~\ref{fig:clb_hmax} (right plot) as a function
of $m_{\rm t}$.
The results in the no-mixing scenario are given in Fig.~\ref{fig:cls_hmin}.
The large-$\mu$ scenario, depending on the top mass value, is extirely excluded at 95\% CL, 
as shown in Fig.~\ref{fig:large_mu}.
The results for the gluophobic scenario are given in Fig.~\ref{fig:gluophobic}
and for the small-$\alpha_{\rm eff}$ scenario in Fig.~\ref{fig:alpha}.
The CP-conserving MSSM limits are summarized in Table~\ref{tab:cp}.

\clearpage

\clearpage
\begin{table}[h!]
\caption{Summary of CP-conserving MSSM limits at 95\% CL.
}
\label{tab:cp}
\begin{center}
\begin{tabular}{lc|cccc}
Benchmark~~           &~~$m_{\rm t}$ (GeV$/c^2$) ~~&~~  $m_{\rm h}$ (GeV/$c^2$) ~~&~~  $m_{\rm A}$ (GeV/$c^2$)    ~~&~~  exclusion of  \\
scenario            &                        &                          &                             &  $\tan\beta$   \\ \hline
$m_{\rm h}$-max     &  169.3     & 92.9 (94.8)&93.4 (95.1)&0.6-2.6 (0.6-2.7) \\
                    &  174.3     & 92.8 (94.9)&93.4 (95.2)&0.7-2.0 (0.7-2.1) \\
                    &  179.3     & 92.9 (94.8)&93.4 (95.2)&0.9-1.5 (0.9-1.6) \\
                    &  183.0     & 92.8 (94.8)&93.5 (95.2)&~~no excl. (no excl.)~~ \\   \hline
no-mixing           &  169.3     & excl. (excl.)&excl. (excl.)&excl. (excl.) \\
                    &  174.3     & 93.6 (96.0)&93.6 (96.4)&0.4-10.2 (0.4-19.4) \\
                    &  179.3     & 93.3 (95.0)&93.4 (95.0)&0.4-5.5 (0.4-6.5) \\
                    &  183.0     & 92.9 (95.0)&93.1 (95.0)&0.4-4.4 (0.4-4.9) \\   \hline
large-$\mu$         &  169.3     & excl. (excl.)&excl. (excl.)&excl. (excl.) \\
                    &  174.3     & excl. (excl.)&excl. (excl.)&excl. (excl.) \\
                    &  179.3     & 109.2 (109.2)&225.0 (225.0)&0.7-43 (0.7-43)\\
                    &  183.0     &  95.6 (95.6)&98.9 (98.9)& 0.7-11.5 (0.7-11.5) \\  \hline
gluophobic          &  169.3     & 90.6 (93.2) &95.7 (98.2)& 0.4-10.3 (0.4-21.5) \\
                    &  174.3     & 90.5 (92.3) &96.3 (98.0)& 0.4-5.4 (0.4-6.4) \\
                    &  179.3     & 90.0 (91.8) &96.5 (98.2)& 0.4-3.9 (0.4-4.2) \\
                    &  183.0     & 89.8 (91.5) &96.8 (98.7)& 0.5-3.3 (0.5-3.6) \\   \hline
small-$\alpha_{\rm eff}$& 169.3  & 88.2 (90.0)&98.2 (99.6)& 0.4-6.1 (0.4-7.4) \\
                    &  174.3     & 87.3 (89.0)&98.8 (100.0)& 0.4-4.2 (0.4-4.5) \\
                    &  179.3     & 86.6 (88.0)&99.8 (100.7)& 0.5-3.2 (0.5-3.4) \\
                    &  183.0     & 85.6 (87.5)&101.0 (101.3)& 0.6-2.7 (0.5-2.9) 
\end{tabular}
\end{center}
\end{table}

\section{CP-Violating Interpretations}
\begin{figure}[h!]
\begin{center}
\includegraphics[width=0.32\textwidth]{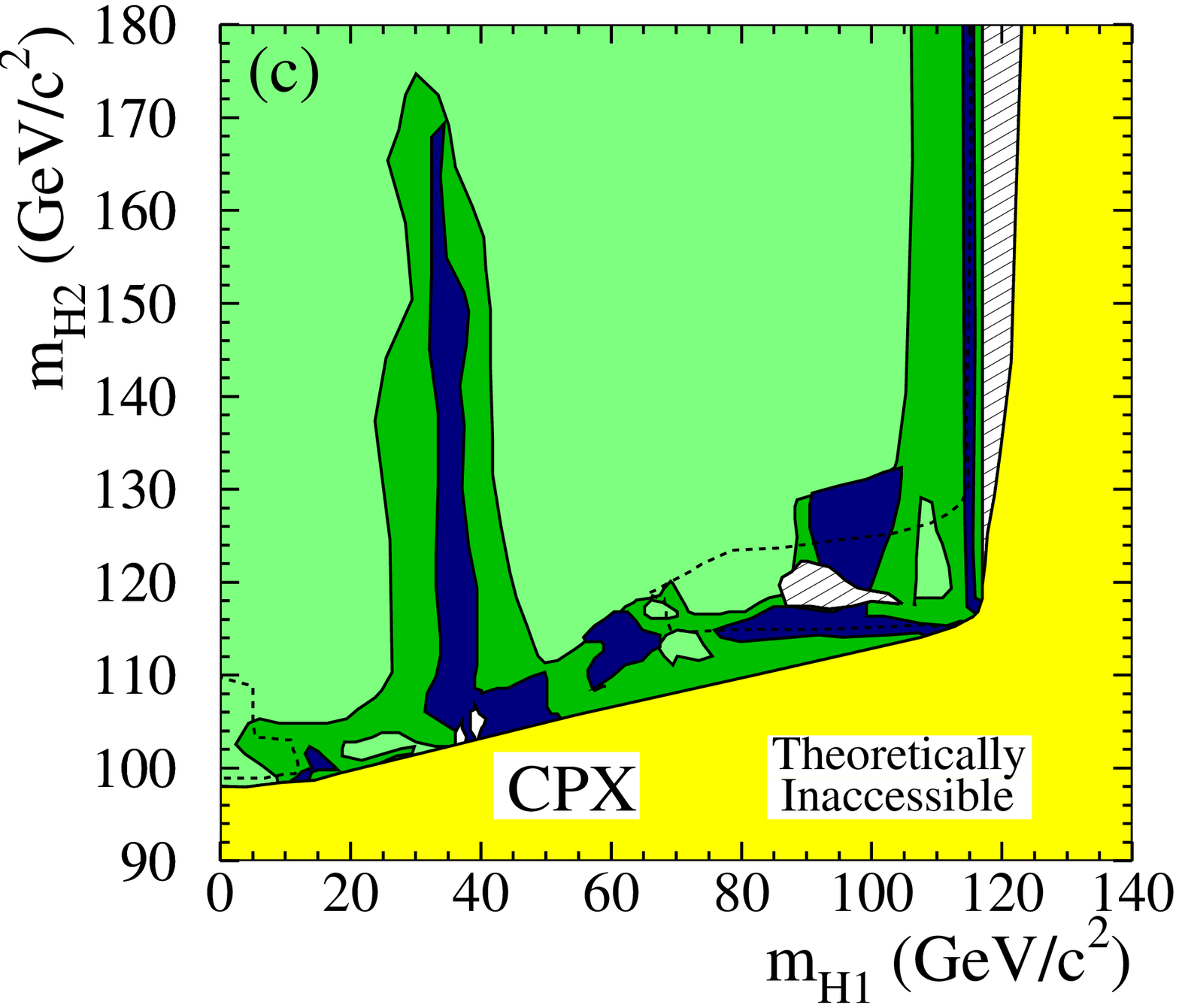} \hfill
\includegraphics[width=0.32\textwidth]{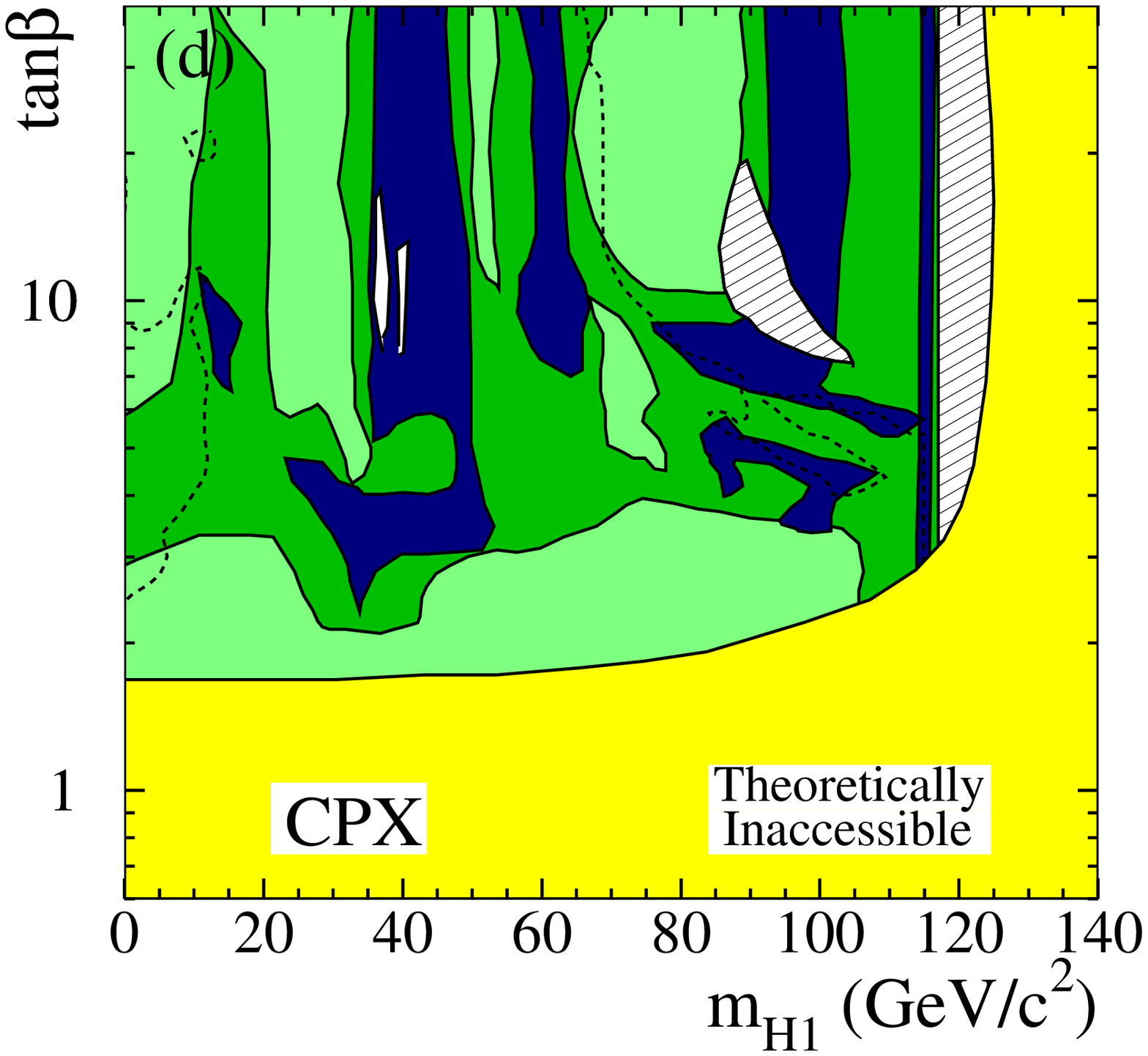}
\end{center}
\vspace*{-8mm}
\caption{Contour regions of $1-CL_{\rm b}$ in the CPX scenario.
Light-green: $1\sigma$, dark-green: 1-2$\sigma$, blue: $>$$2\sigma$.
Expected limits in the absence of a signal are indicated with a 
dotted line.
}
\label{fig:clb_cpx}
\end{figure}

\begin{figure}[tp]
\vspace*{-1mm}
\begin{center}
\includegraphics[width=0.24\textwidth]{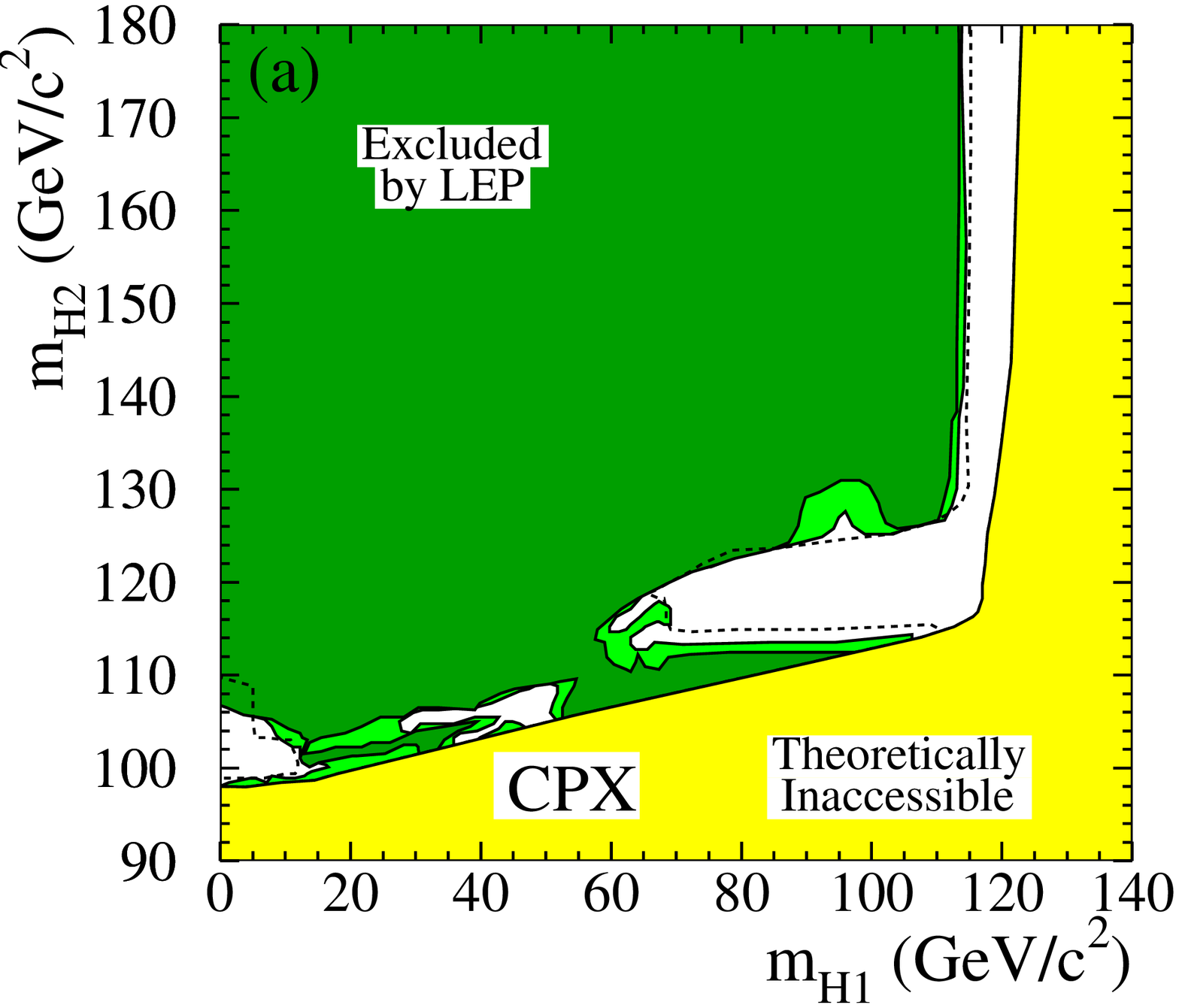} \hfill
\includegraphics[width=0.24\textwidth]{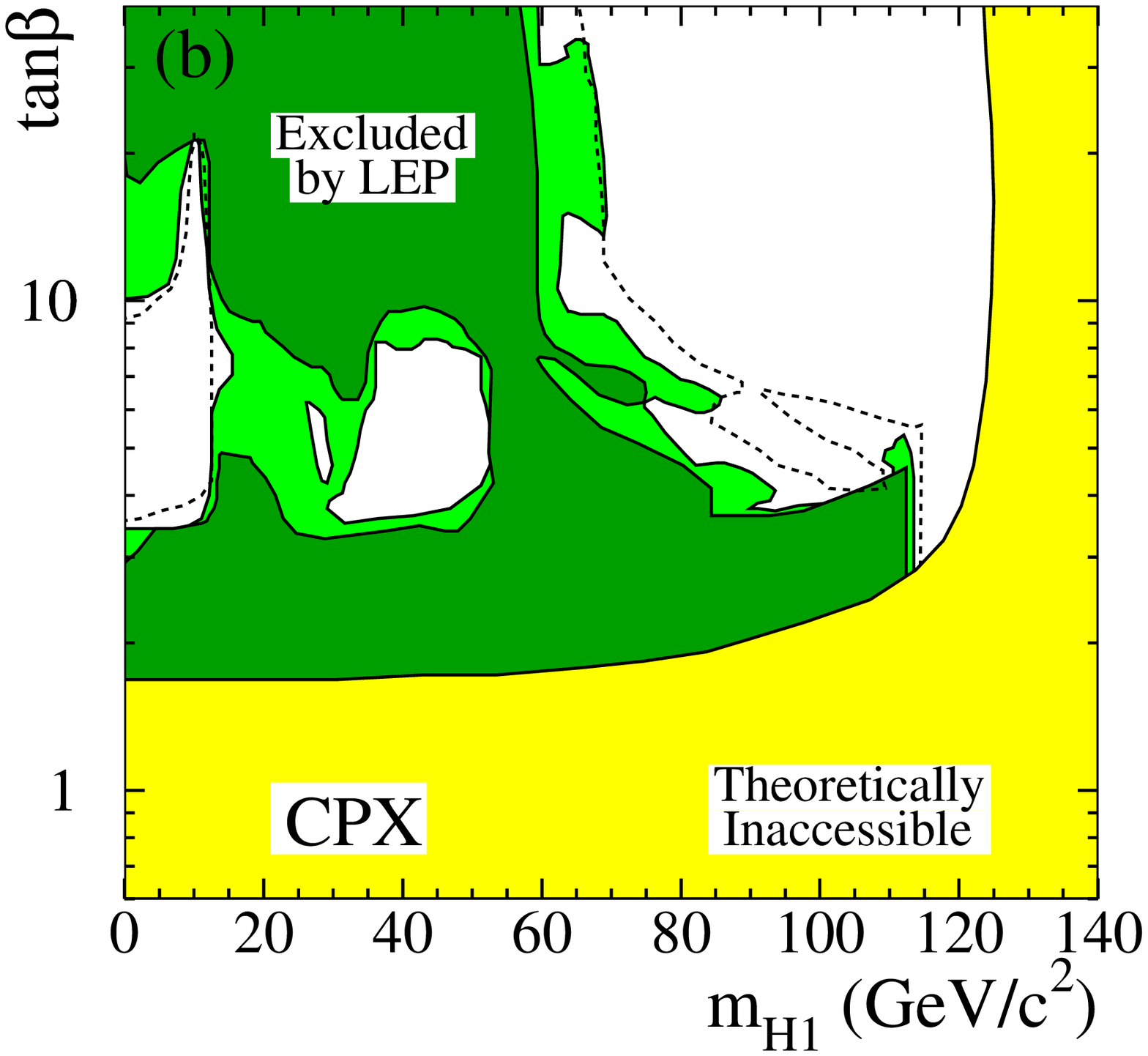} \hfill
\includegraphics[width=0.24\textwidth]{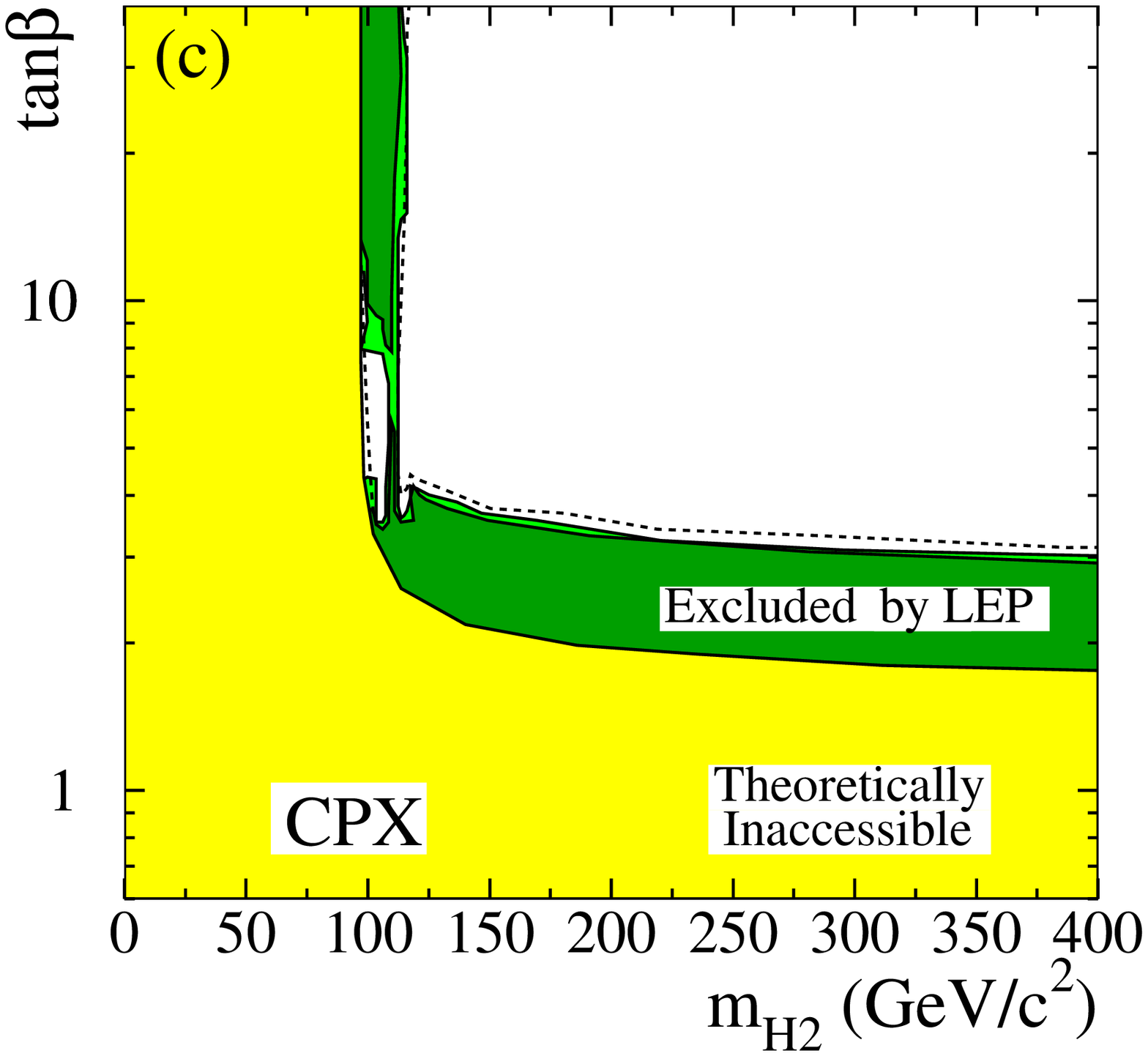} \hfill
\includegraphics[width=0.24\textwidth]{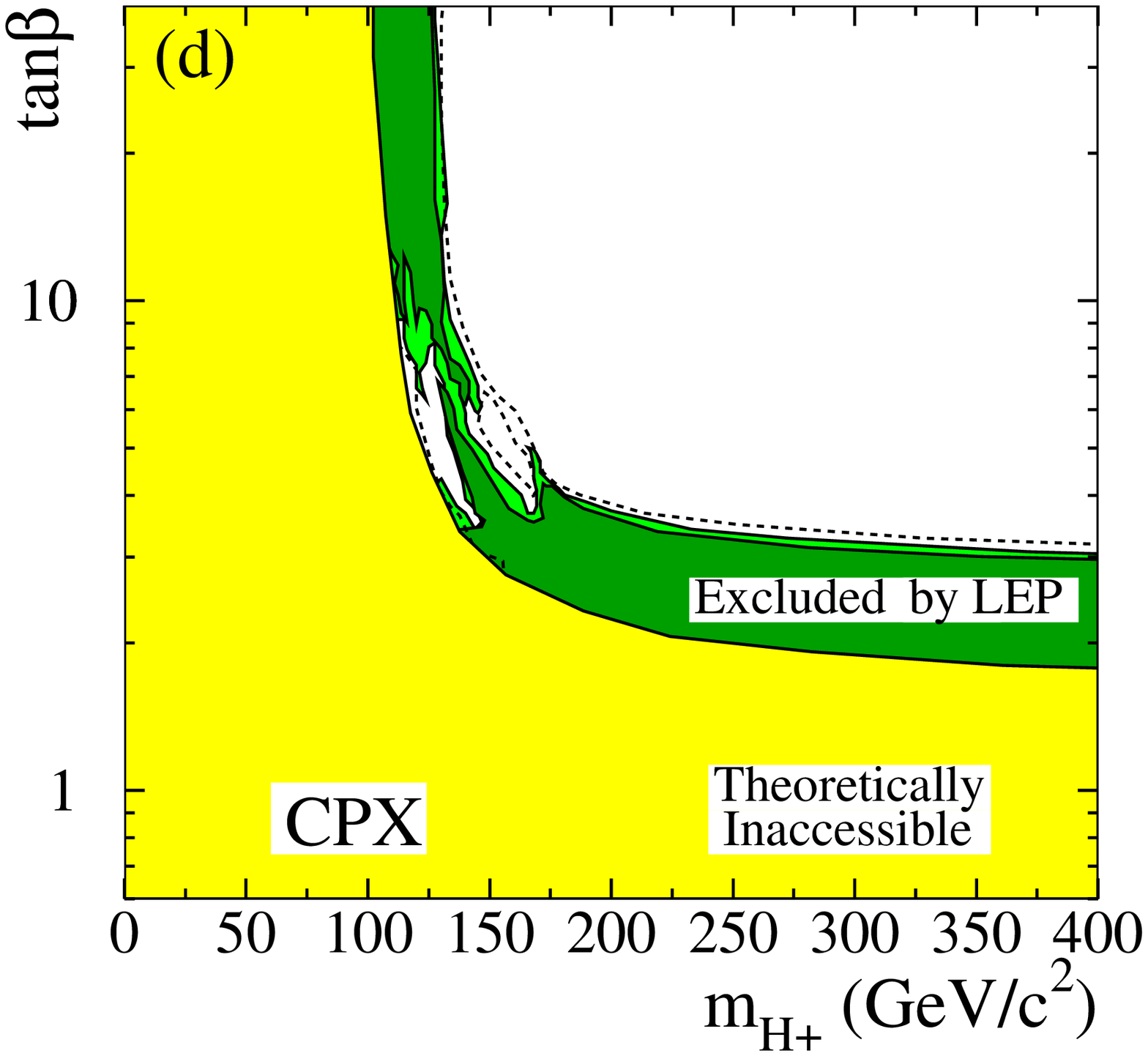}
\end{center}
\vspace*{-8mm}
\caption{Excluded regions in the CPX scenario at 95\% CL
(light-green) and at 99.7\% CL (dark-green).
Expected limits in the absence of a signal are indicated with a 
dotted line.
}
\label{fig:cls_cpx}
\vspace*{-4mm}
\end{figure}

\begin{figure}[h!]
\begin{center}
\includegraphics[width=0.24\textwidth]{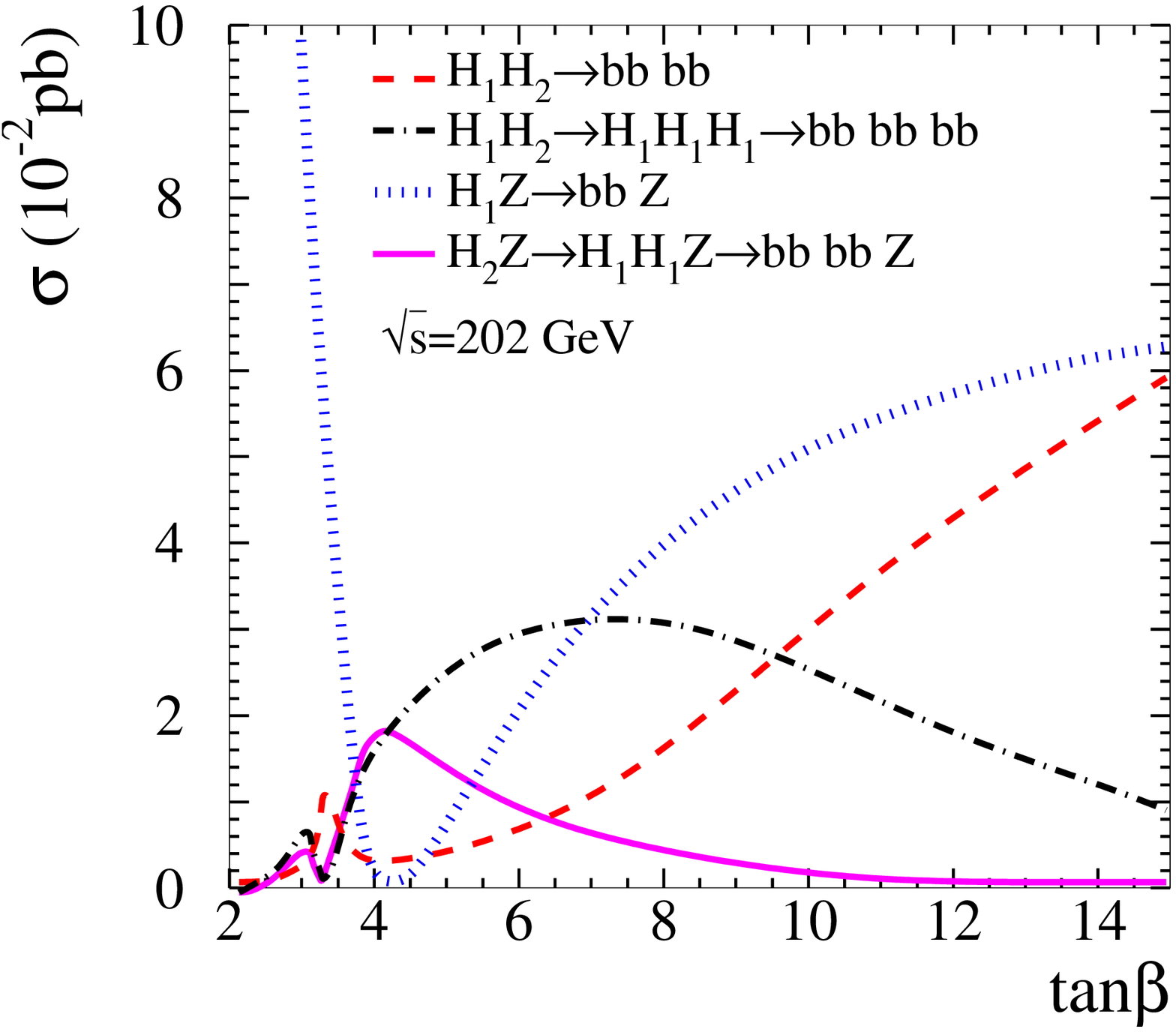} \hfill 
\includegraphics[width=0.24\textwidth]{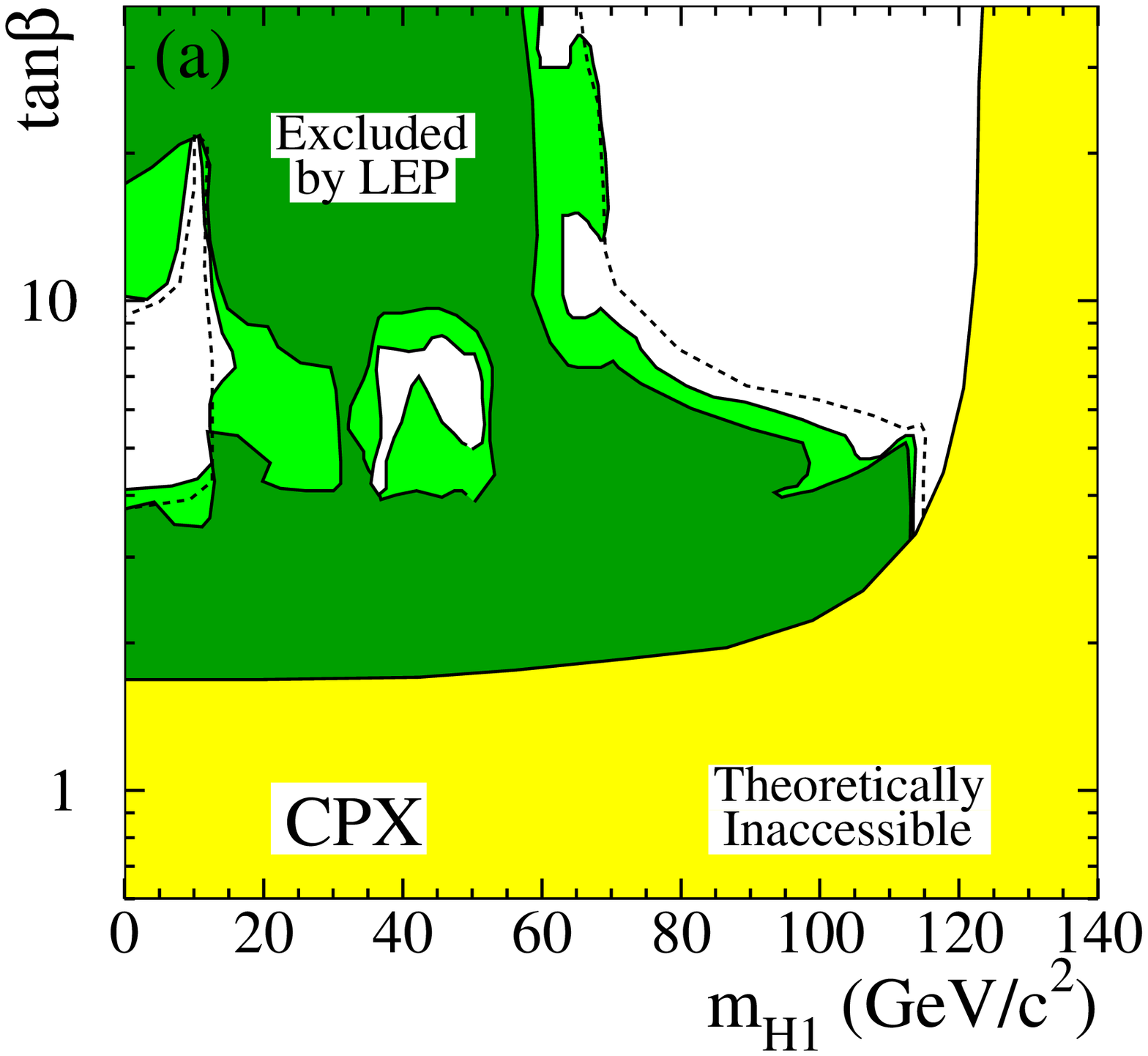} \hfill
\includegraphics[width=0.24\textwidth]{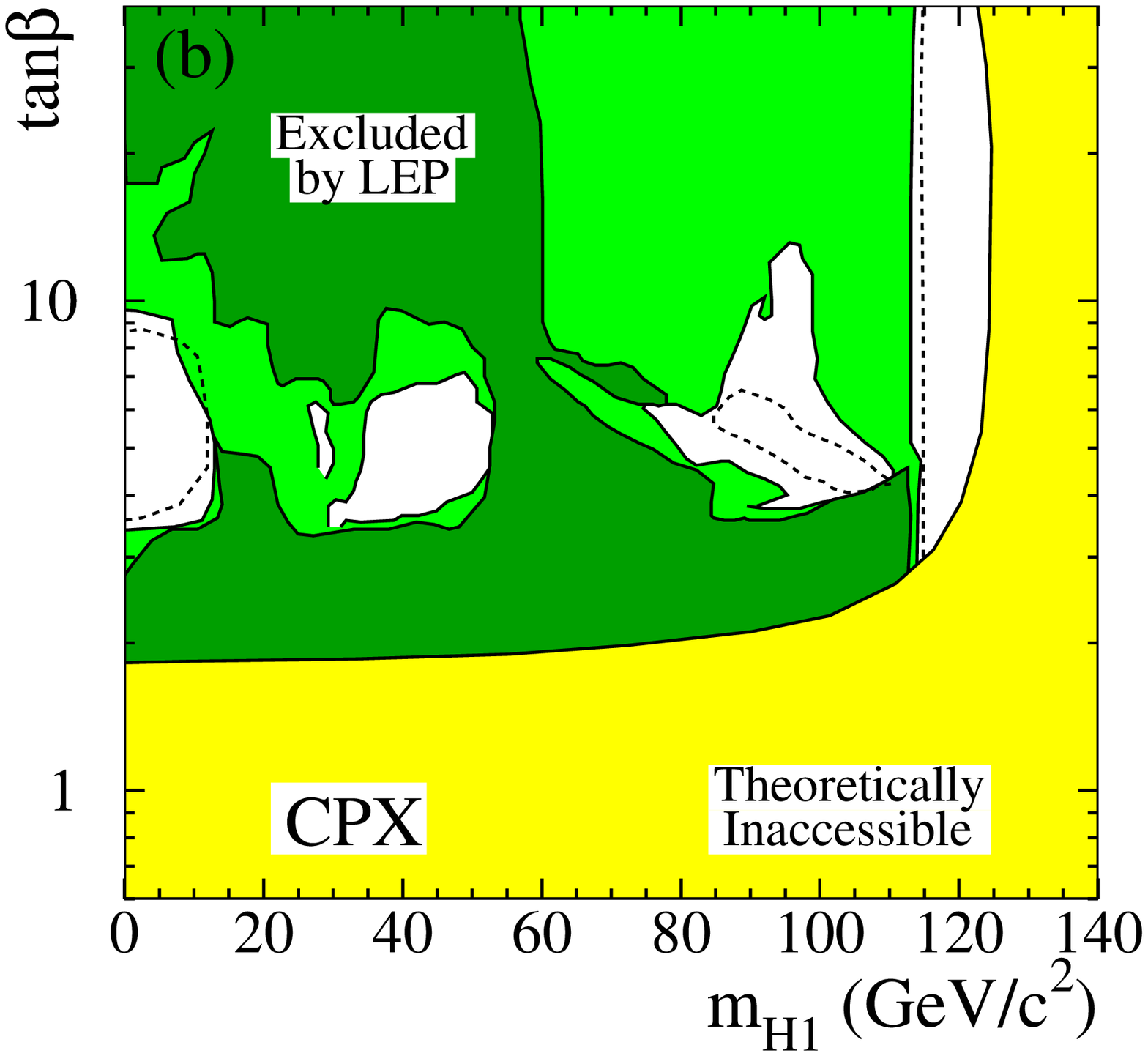} \hfill
\includegraphics[width=0.24\textwidth]{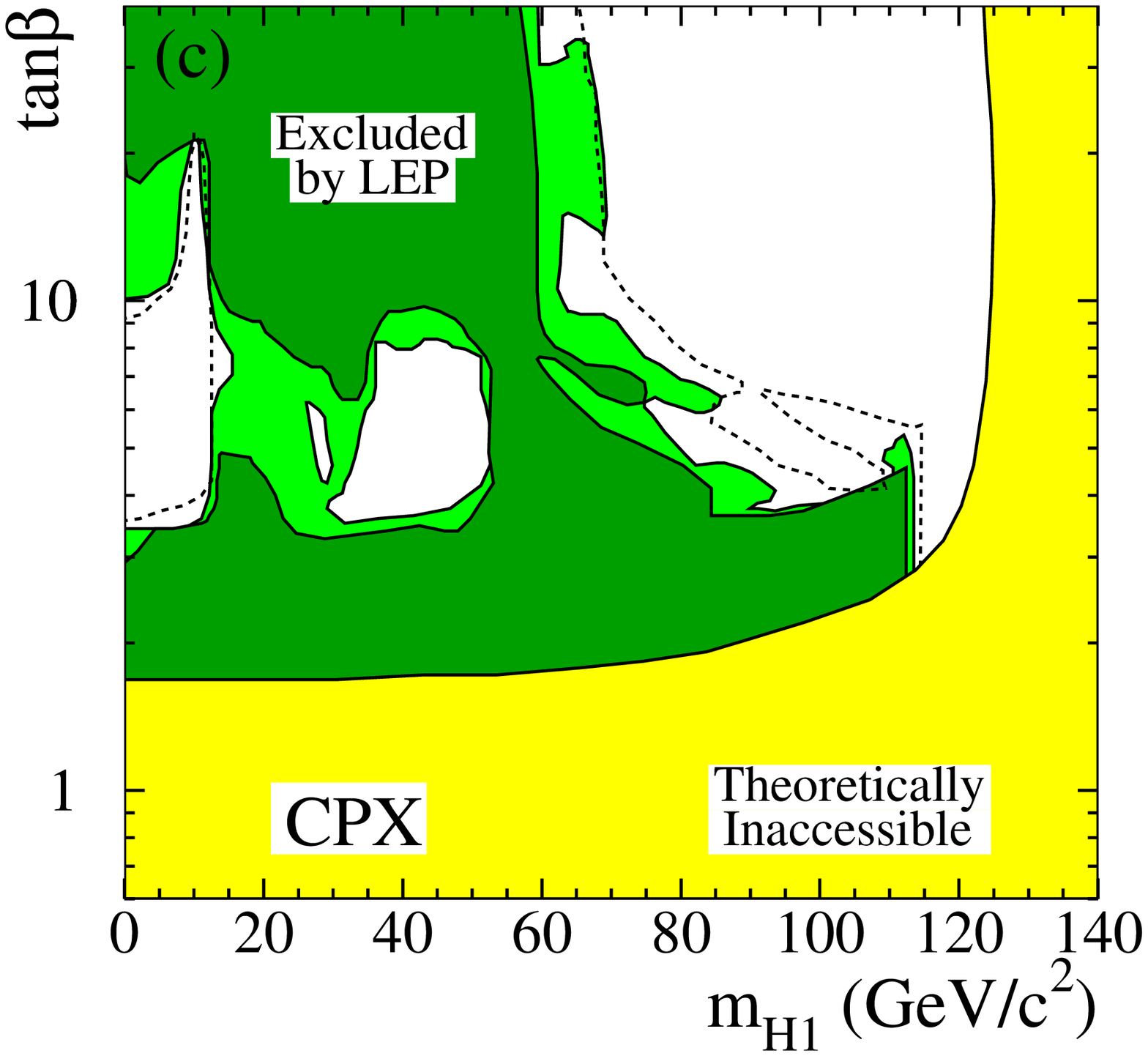}
\end{center}
\vspace*{-8mm}
\caption{Left: expected production cross sections in the CPX scenario.
Center left: CPH calculation.
Center right: FeynHiggs calculation.
Right: logic-OR of CPH and FeynHiggs calculations.
}
\label{fig:xsec}
\vspace*{-4mm}
\end{figure}

\begin{figure}[h!]
\begin{center}
\includegraphics[width=0.24\textwidth]{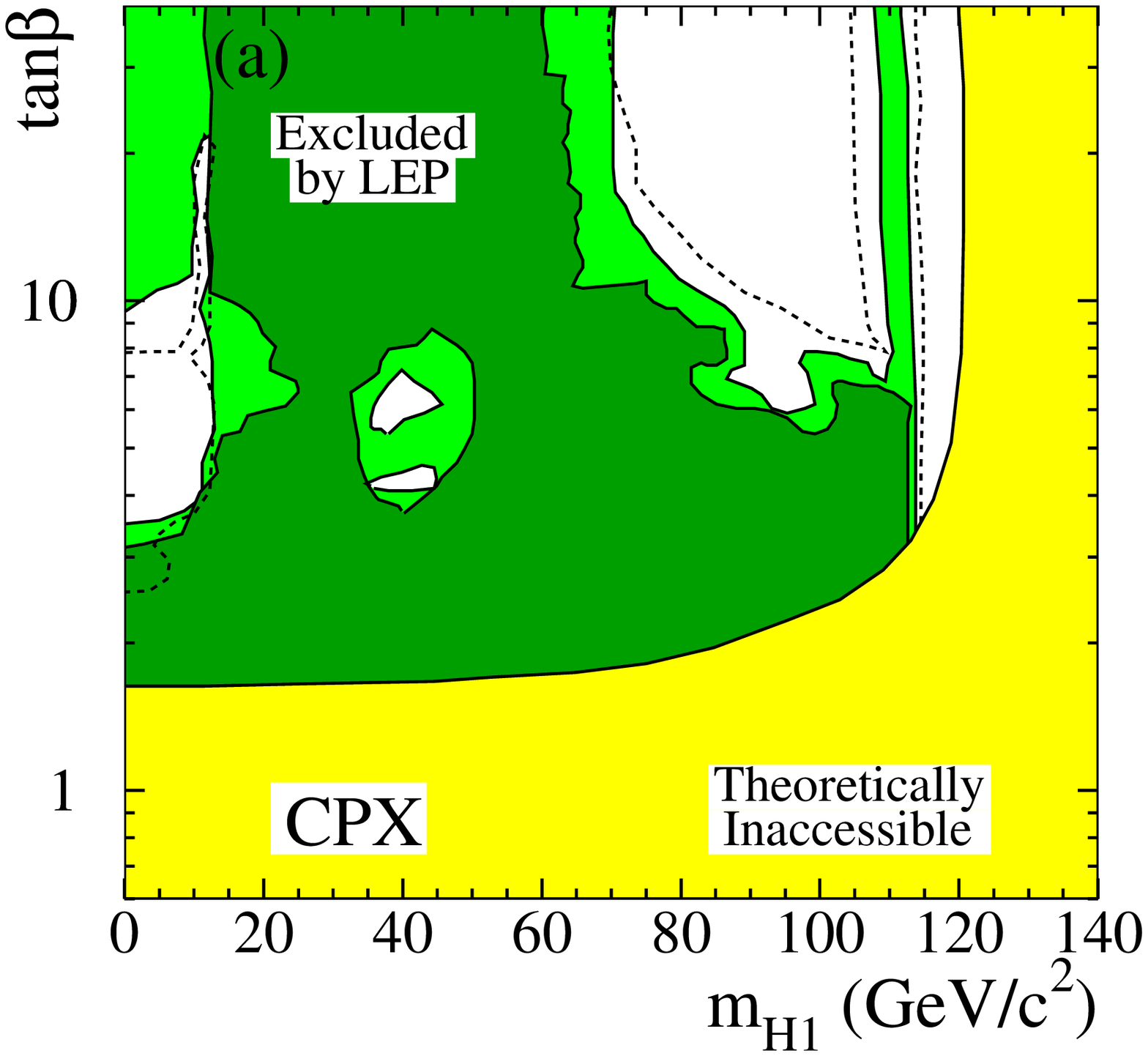} \hfill
\includegraphics[width=0.24\textwidth]{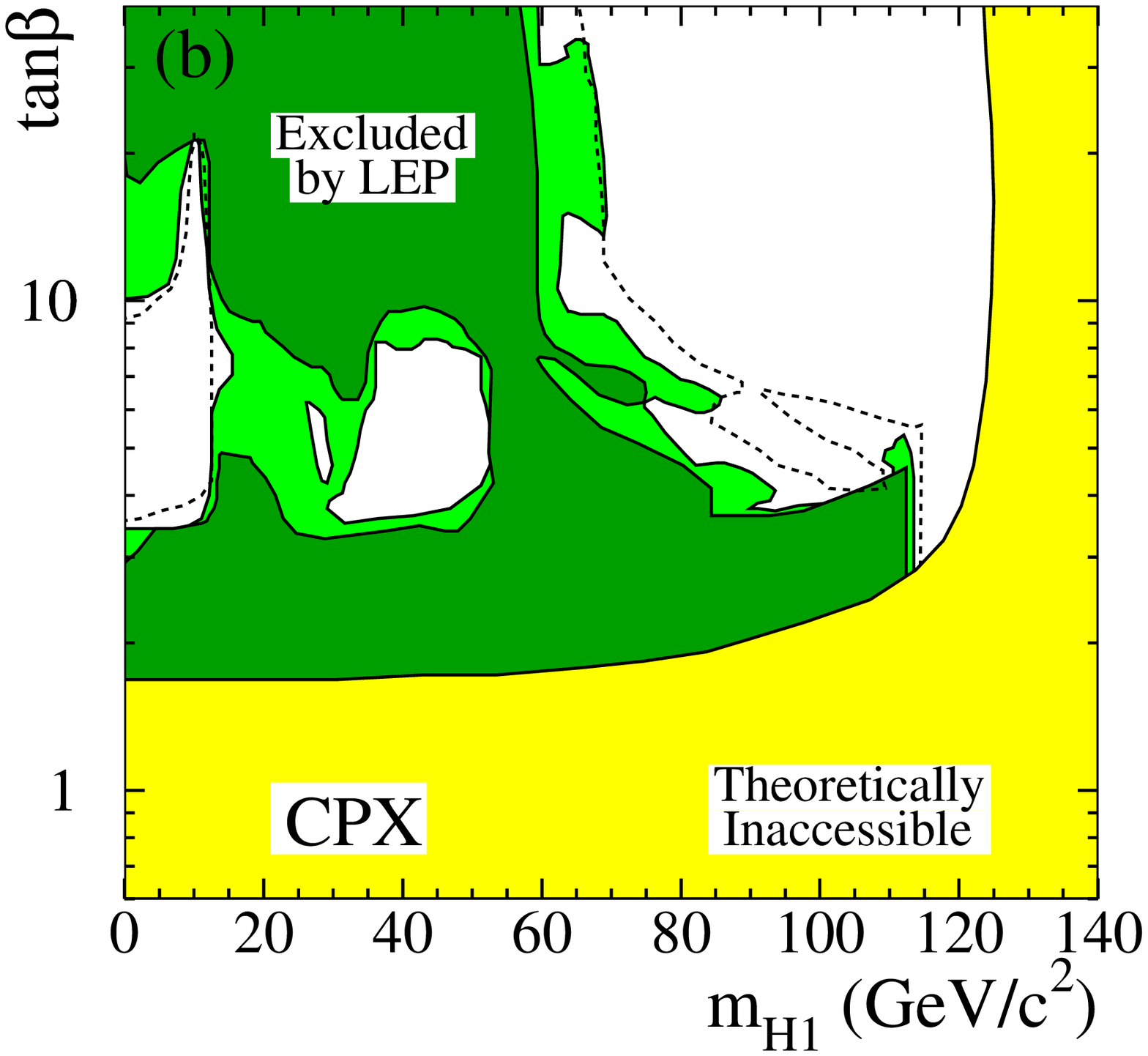} \hfill
\includegraphics[width=0.24\textwidth]{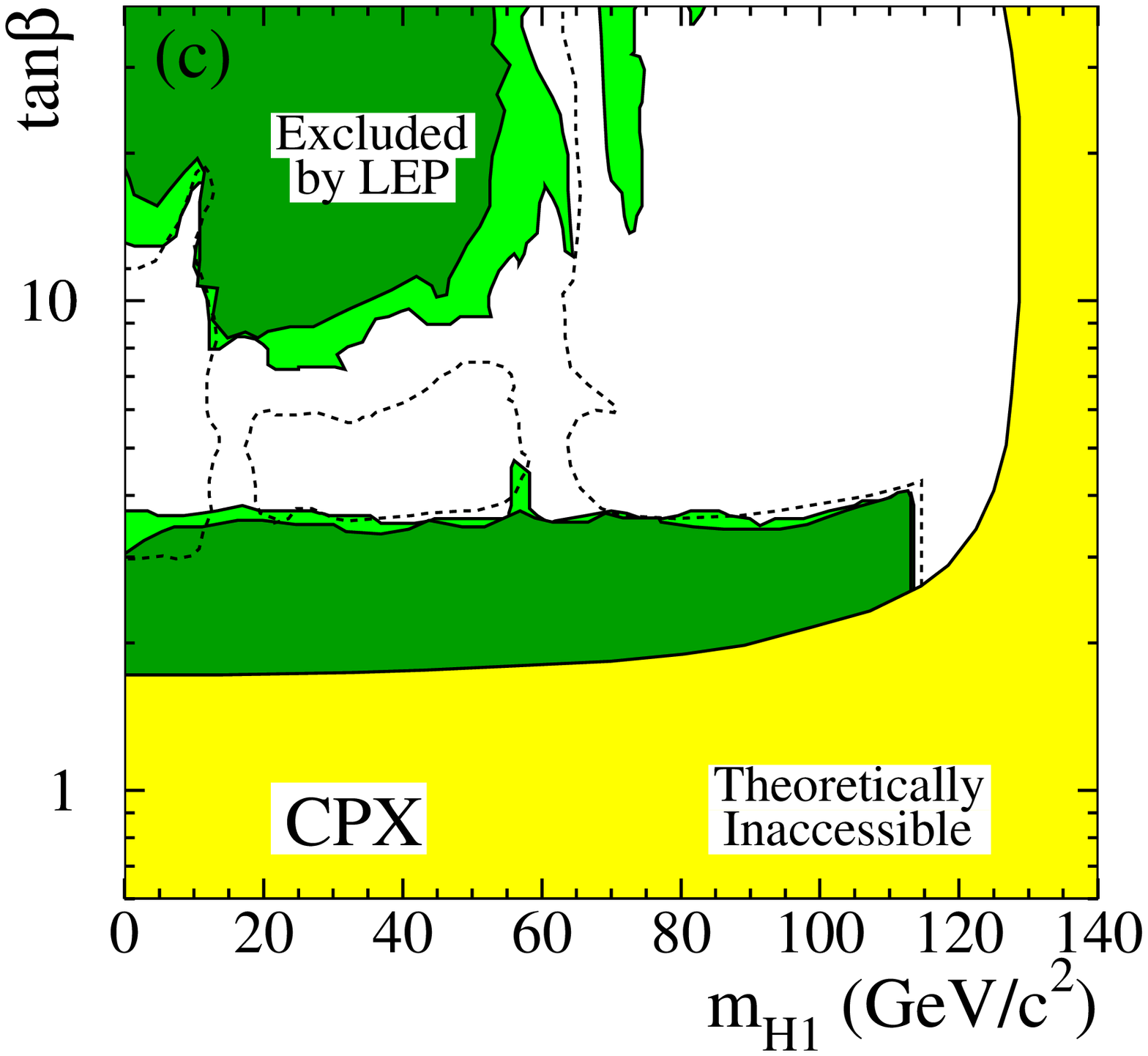}  \hfill
\includegraphics[width=0.24\textwidth]{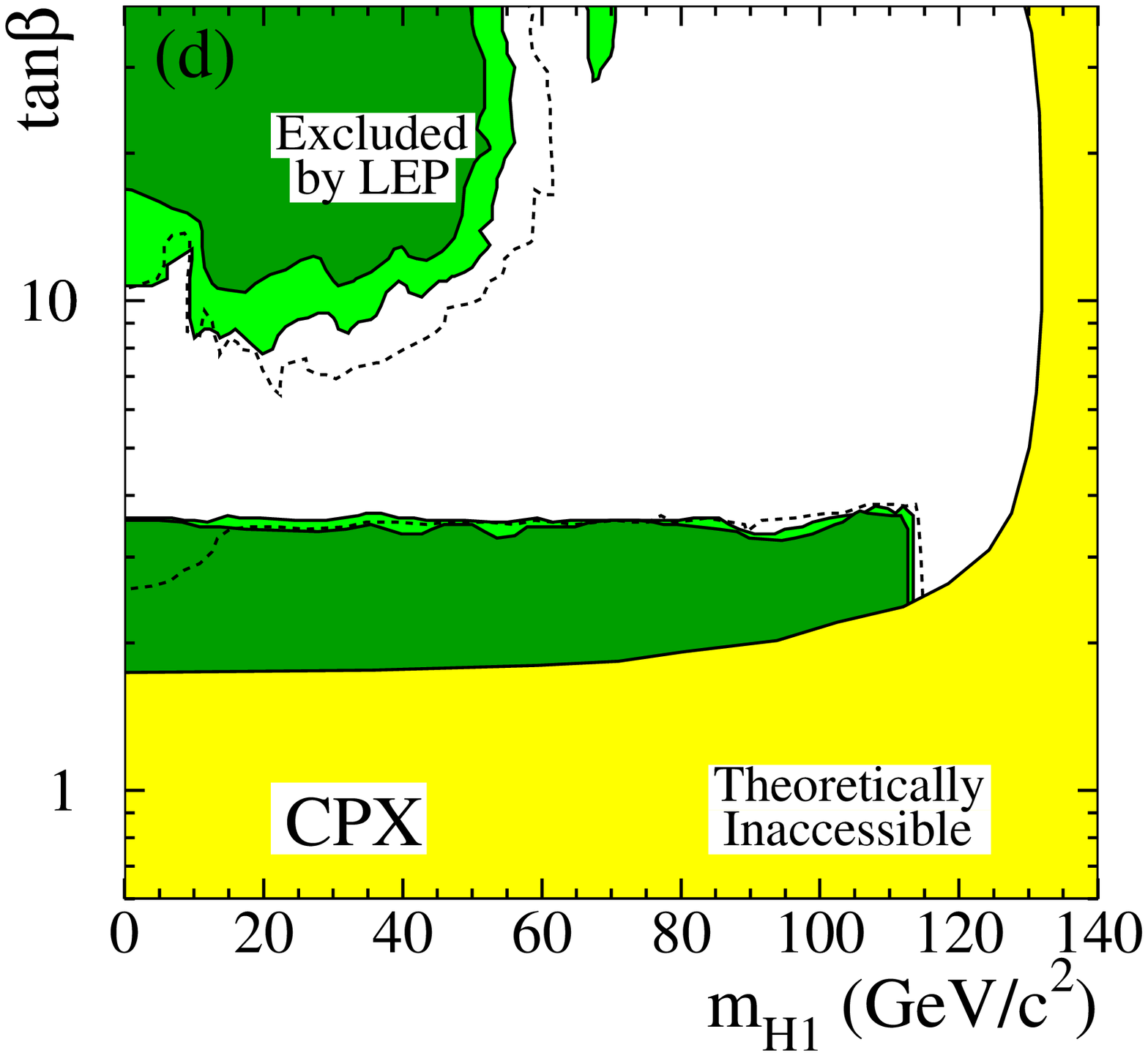} 
\end{center}
\vspace*{-8mm}
\caption{Excluded regions in the CPX scenario at 95\% CL
(light-green) and at 99.7\% CL (dark-green) for
$m_{\rm t} = 169.3,~174.3,~179.3,~183.3$~GeV/$c^2$ in plots (a) to (d), respectively.
Expected limits in the absence of a signal are indicated with a 
dotted line.
}
\label{fig:top}
\vspace*{-5mm}
\end{figure}

\begin{figure}[h!]
\begin{center}
\includegraphics[scale=0.22]{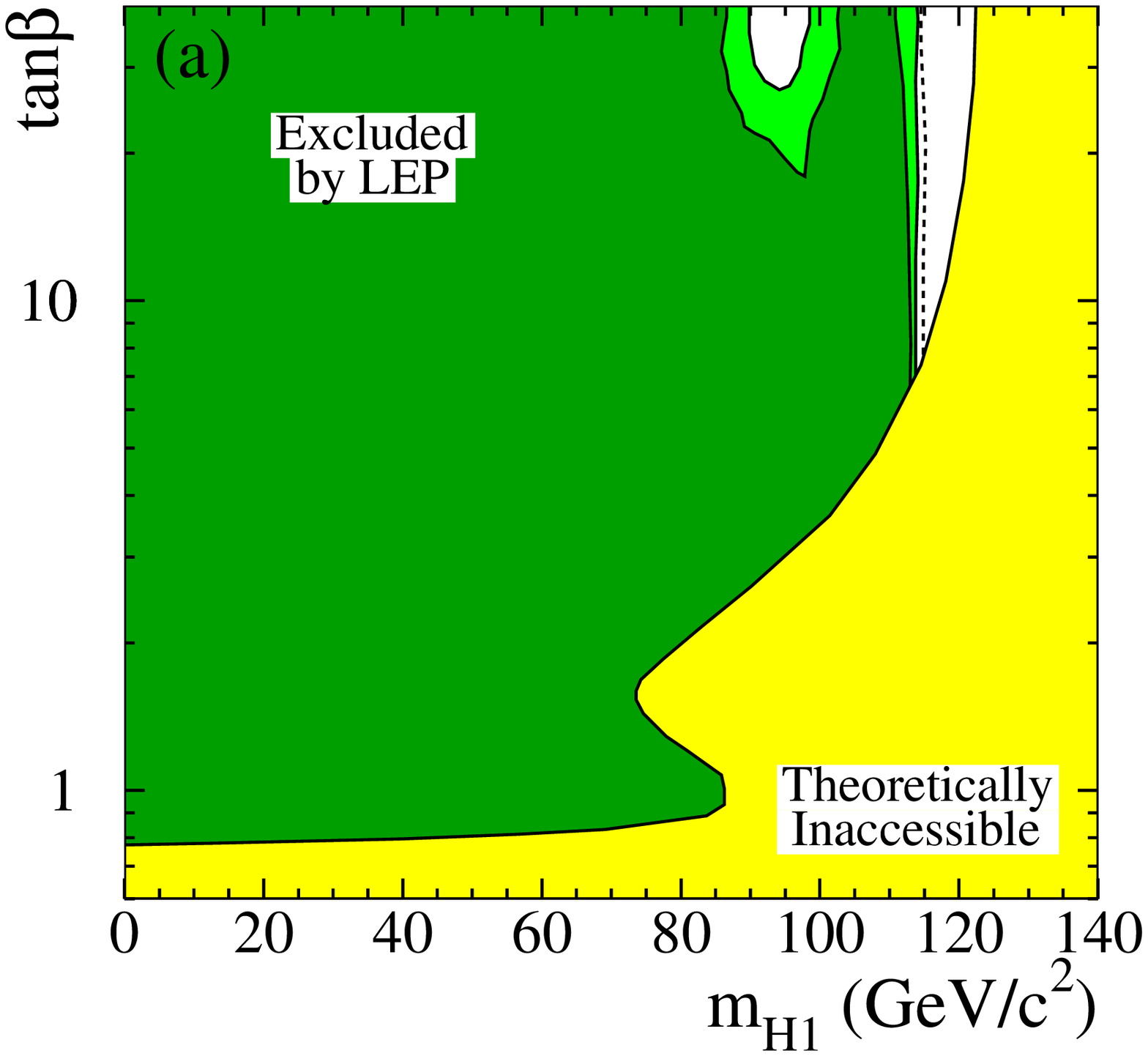} 
\includegraphics[scale=0.22]{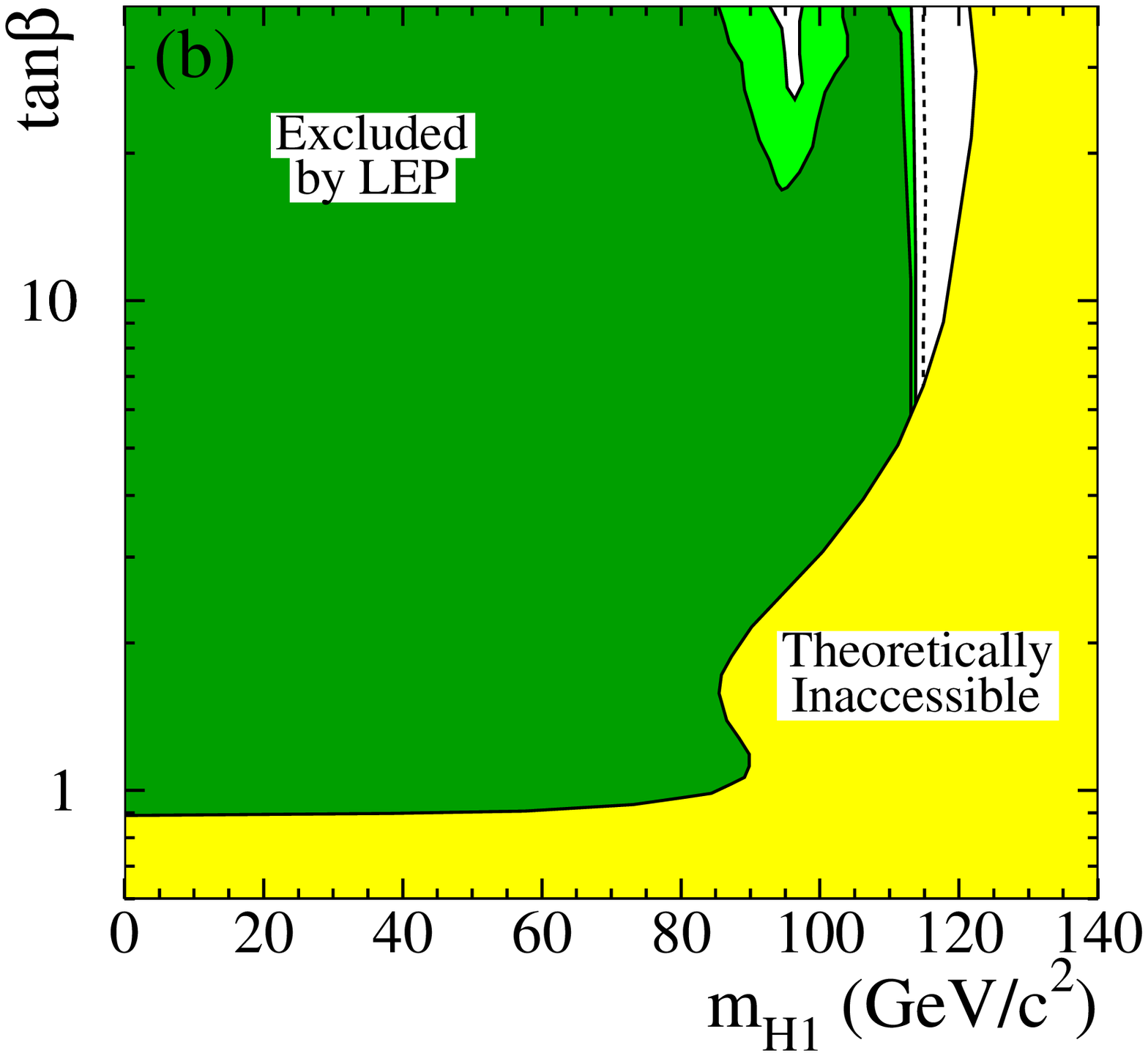} 
\includegraphics[scale=0.22]{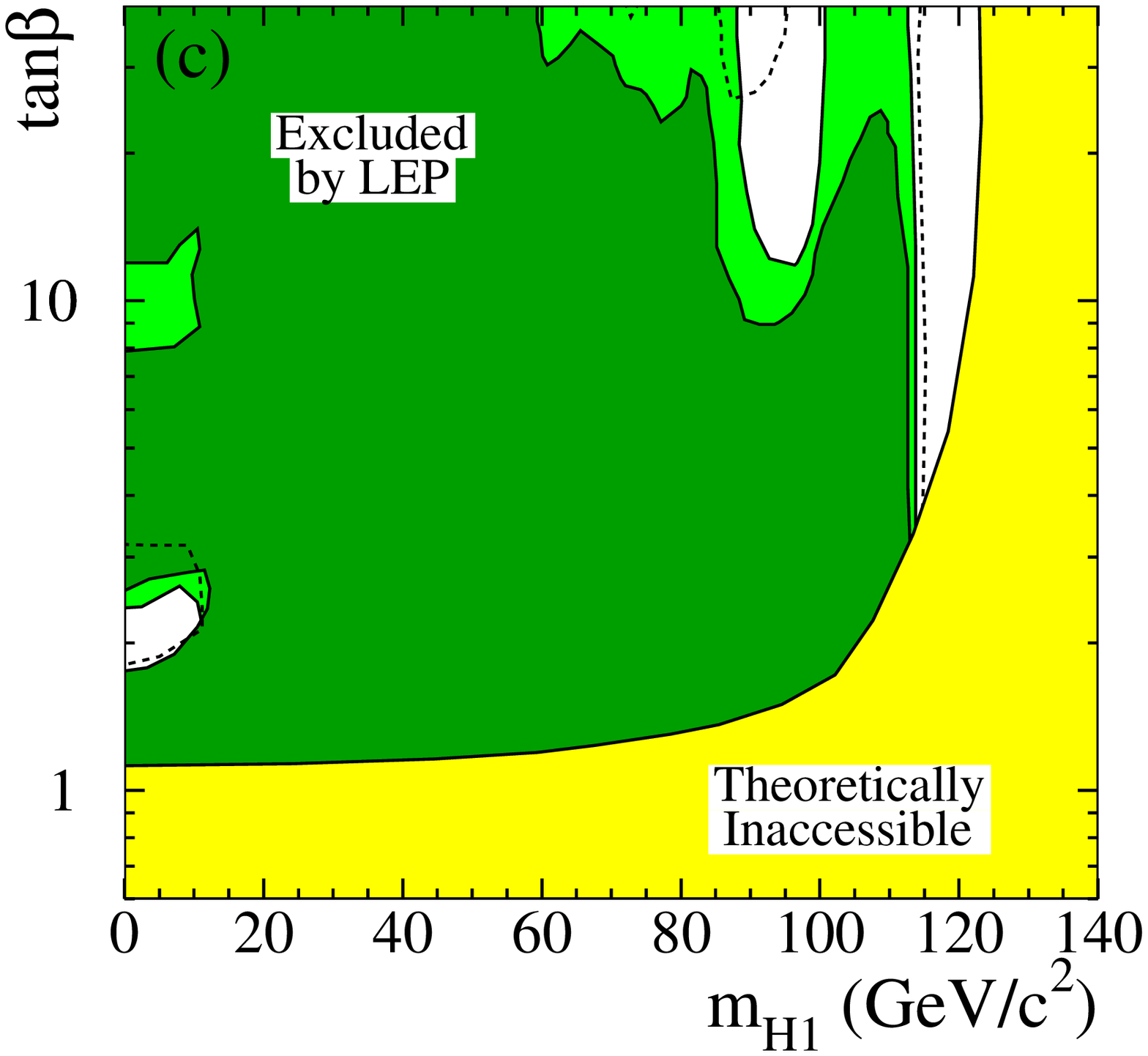} \\
\includegraphics[scale=0.22]{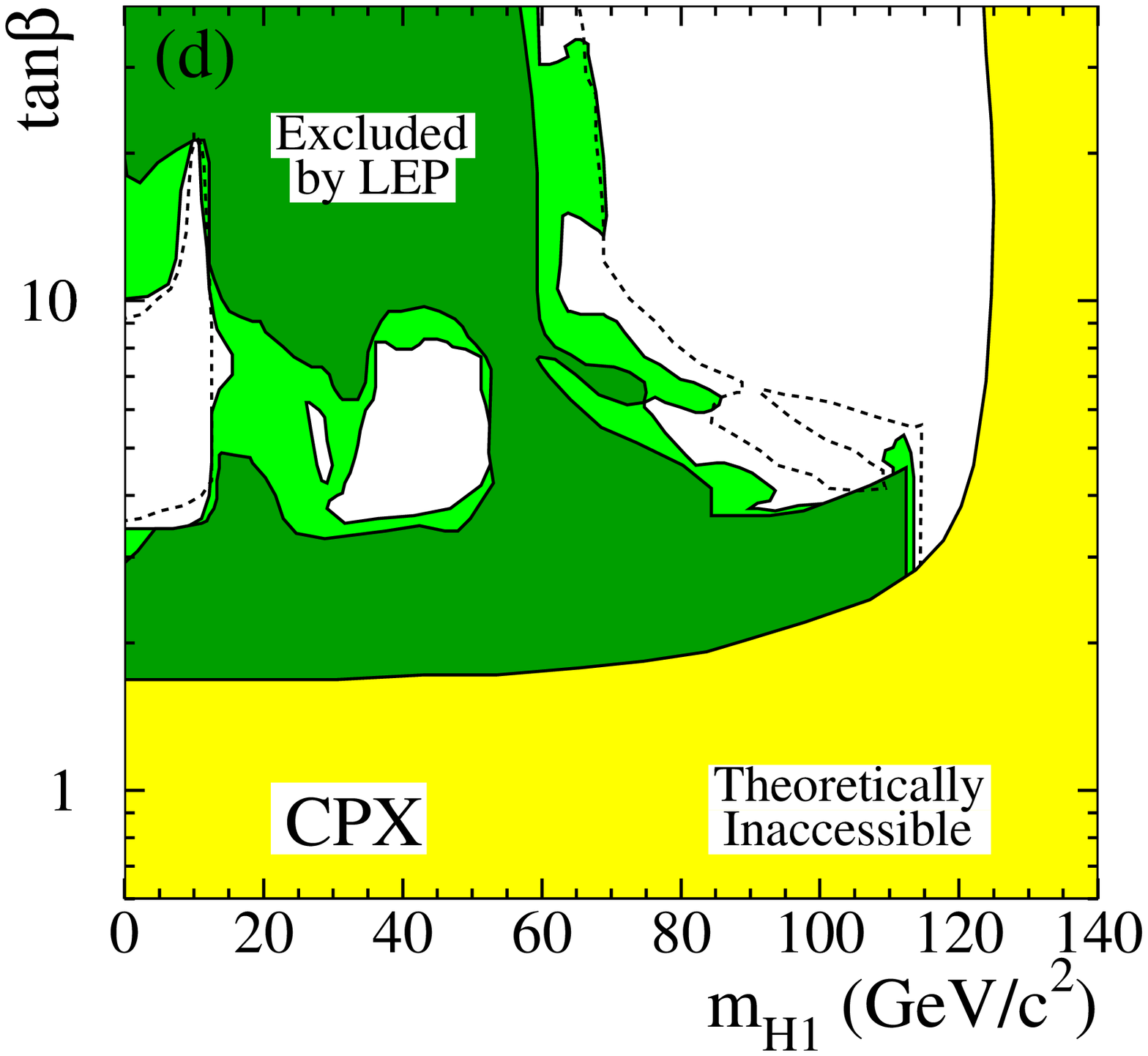} \hspace{-0.2cm}
\includegraphics[scale=0.22]{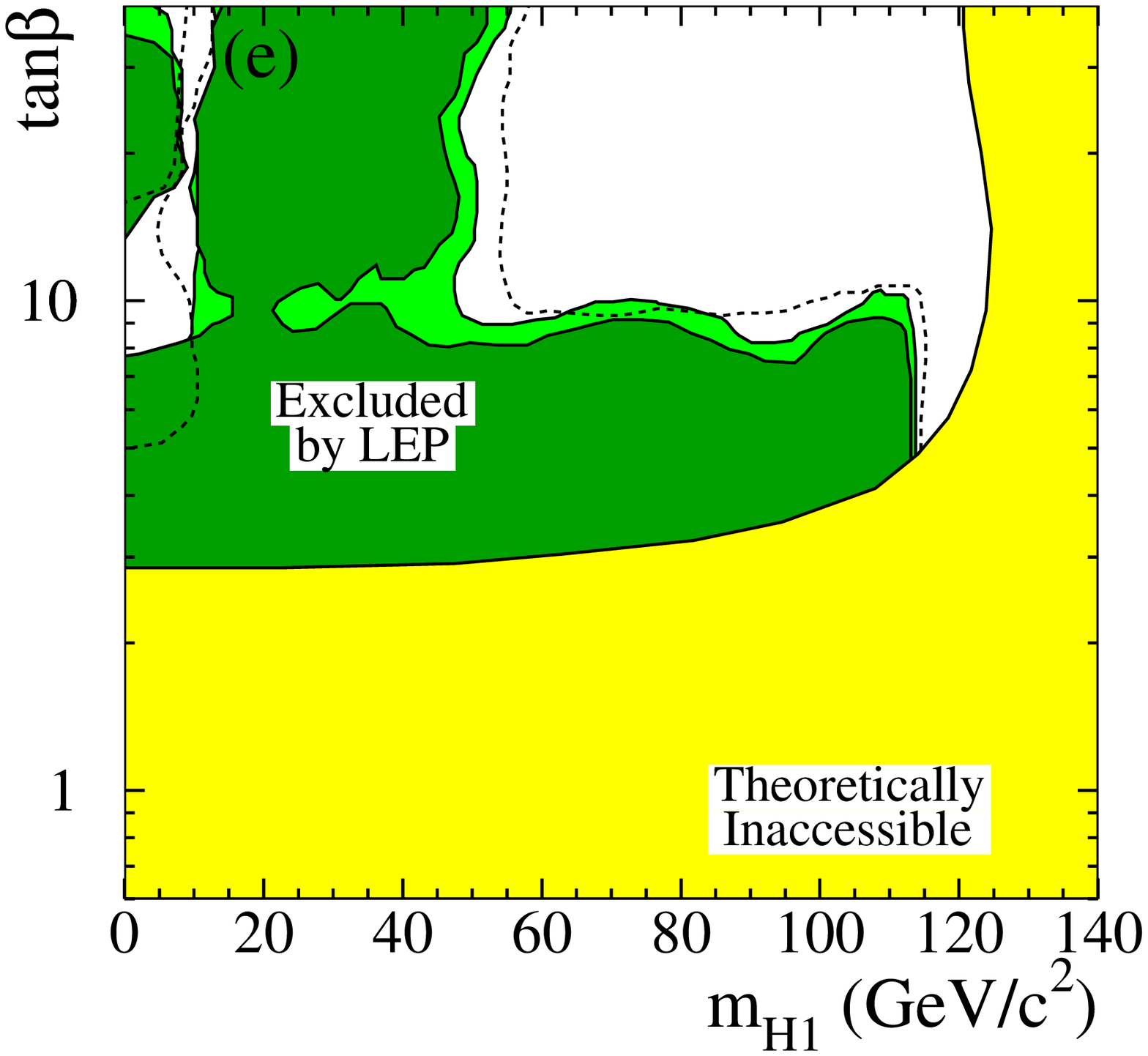} \hspace{-0.2cm}
\includegraphics[scale=0.22]{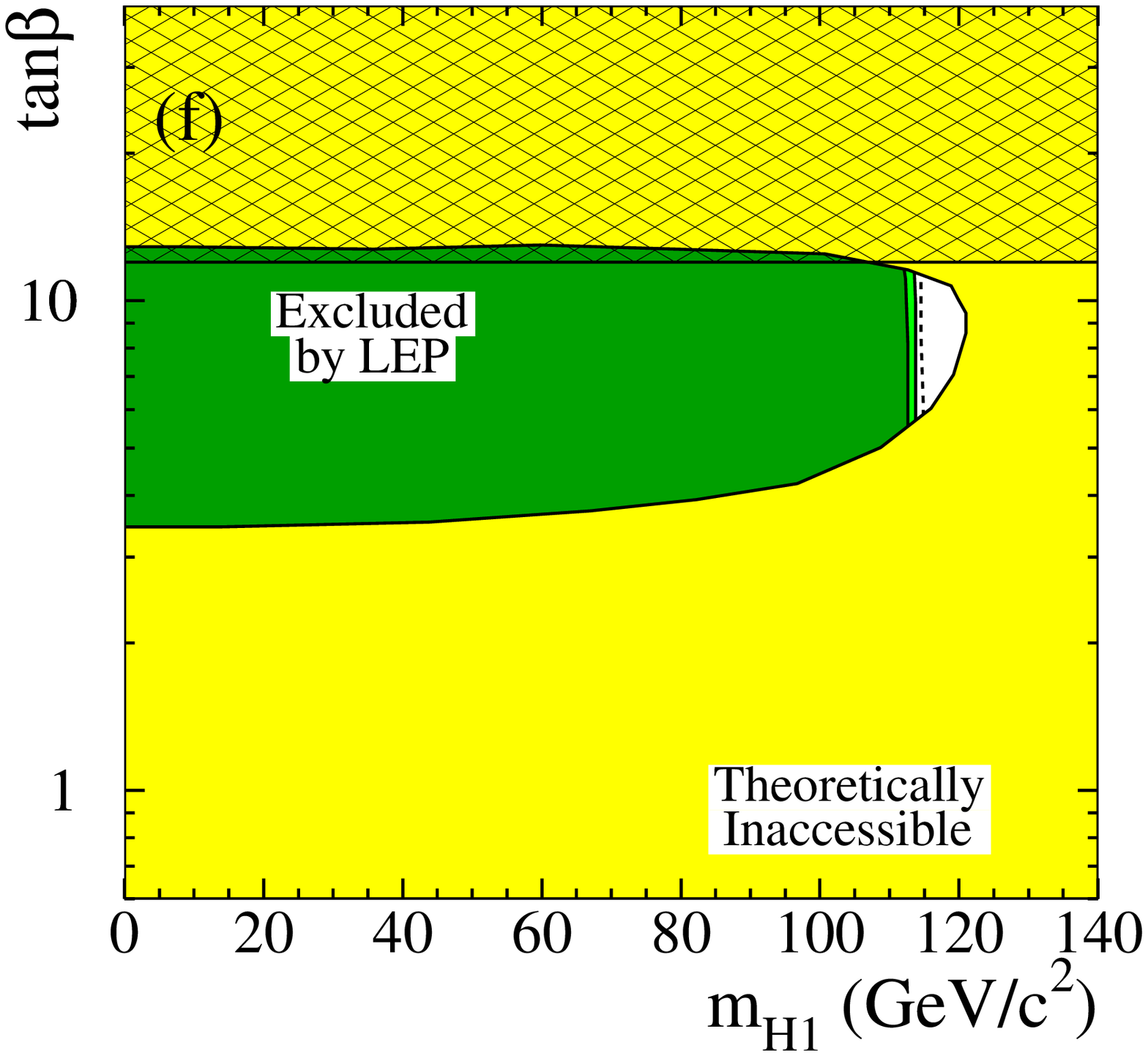} \hspace{-0.2cm}
\end{center}
\vspace*{-8mm}
\caption{Excluded regions in the CPX scenario at 95\% CL
(light-green) and at 99.7\% CL (dark-green) for
$\rm arg(A)= 0^\circ,~30^\circ,~60^\circ,~90^\circ,~135^\circ,~180^\circ$
  in plots (a) to (d), respectively.
Expected limits in the absence of a signal are indicated with a 
dotted line.
}
\label{fig:phase}
\end{figure}

\begin{figure}[h!]
\begin{center}
\includegraphics[width=0.24\textwidth]{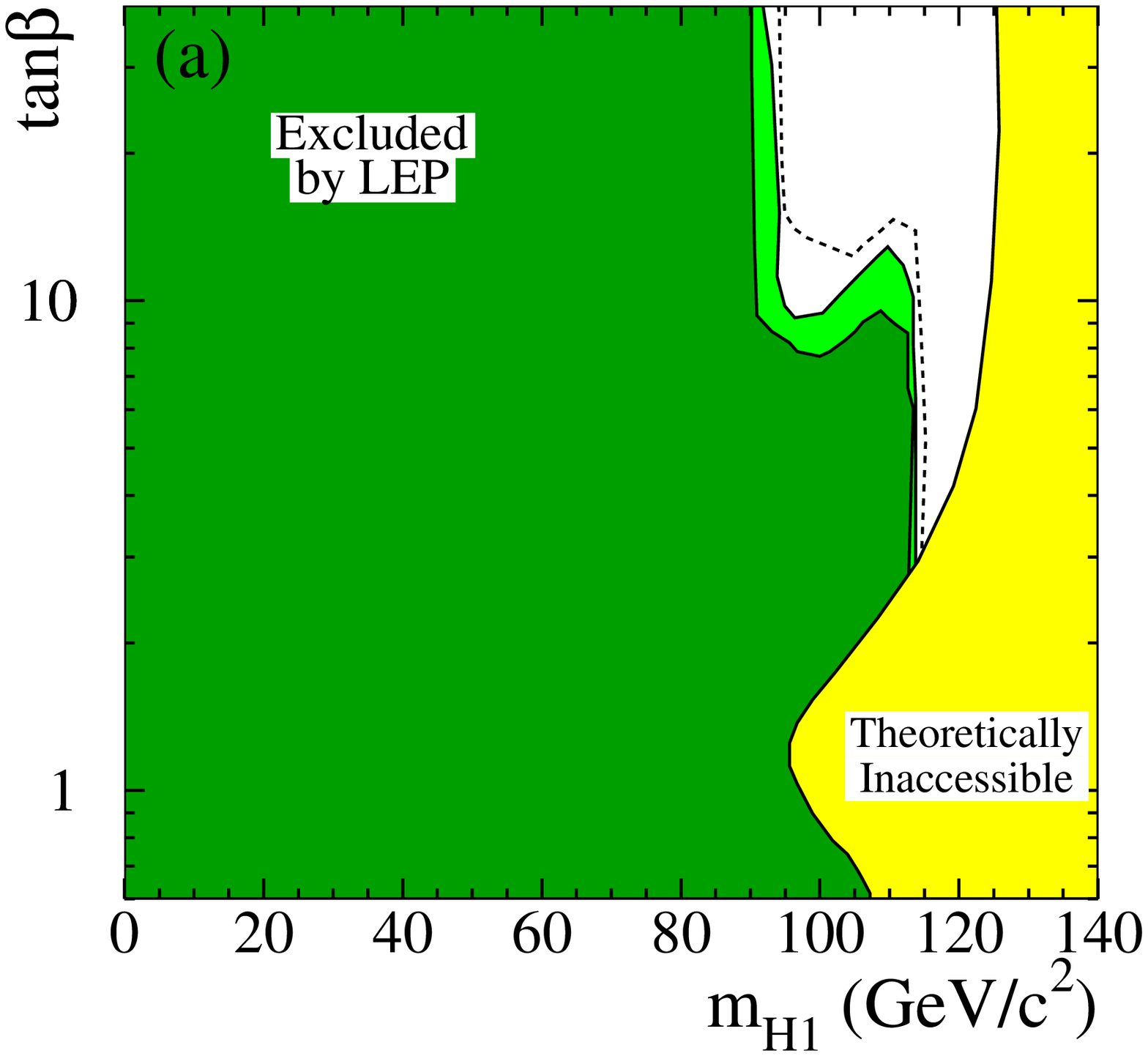} \hfill
\includegraphics[width=0.24\textwidth]{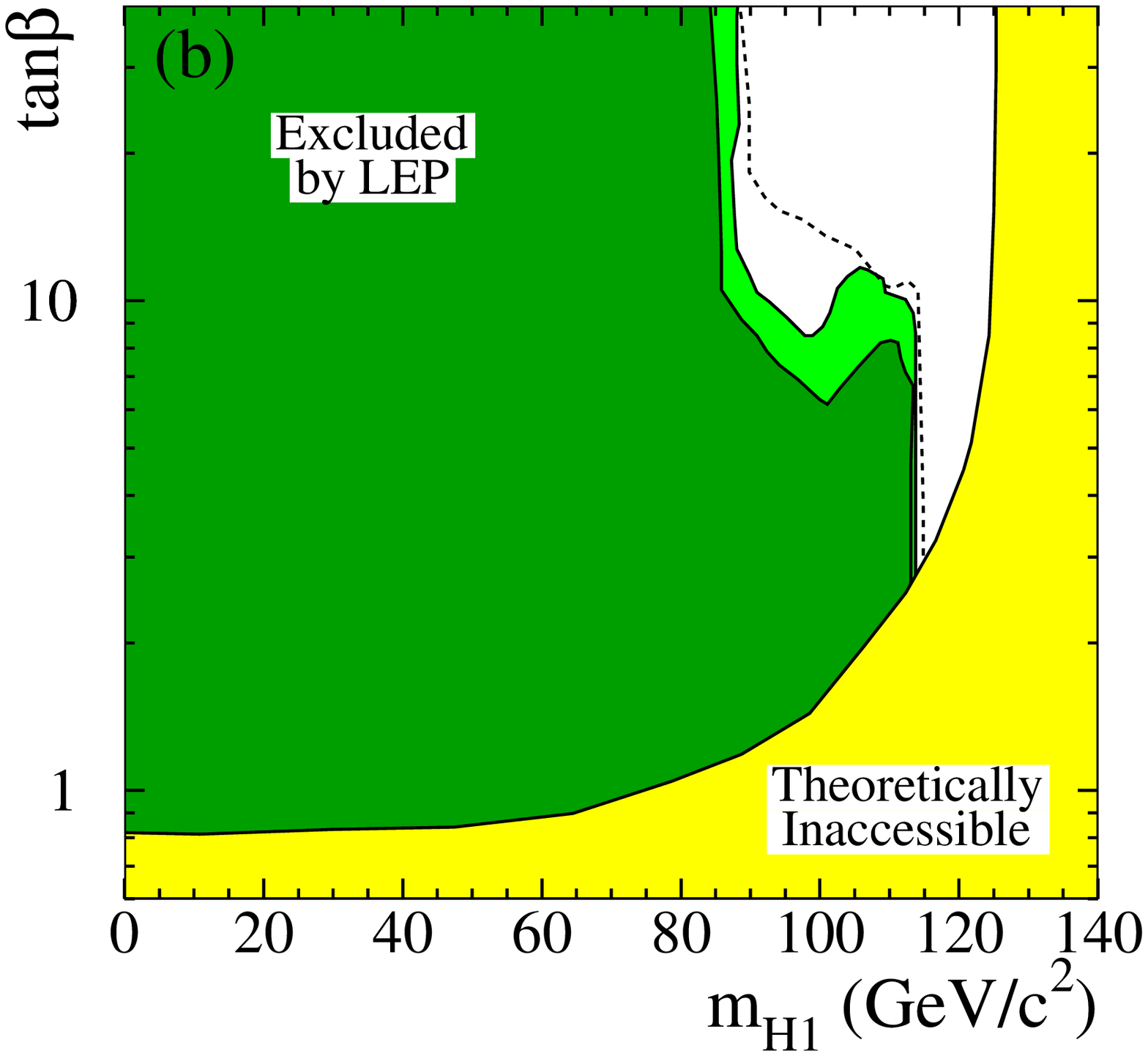} \hfill
\includegraphics[width=0.24\textwidth]{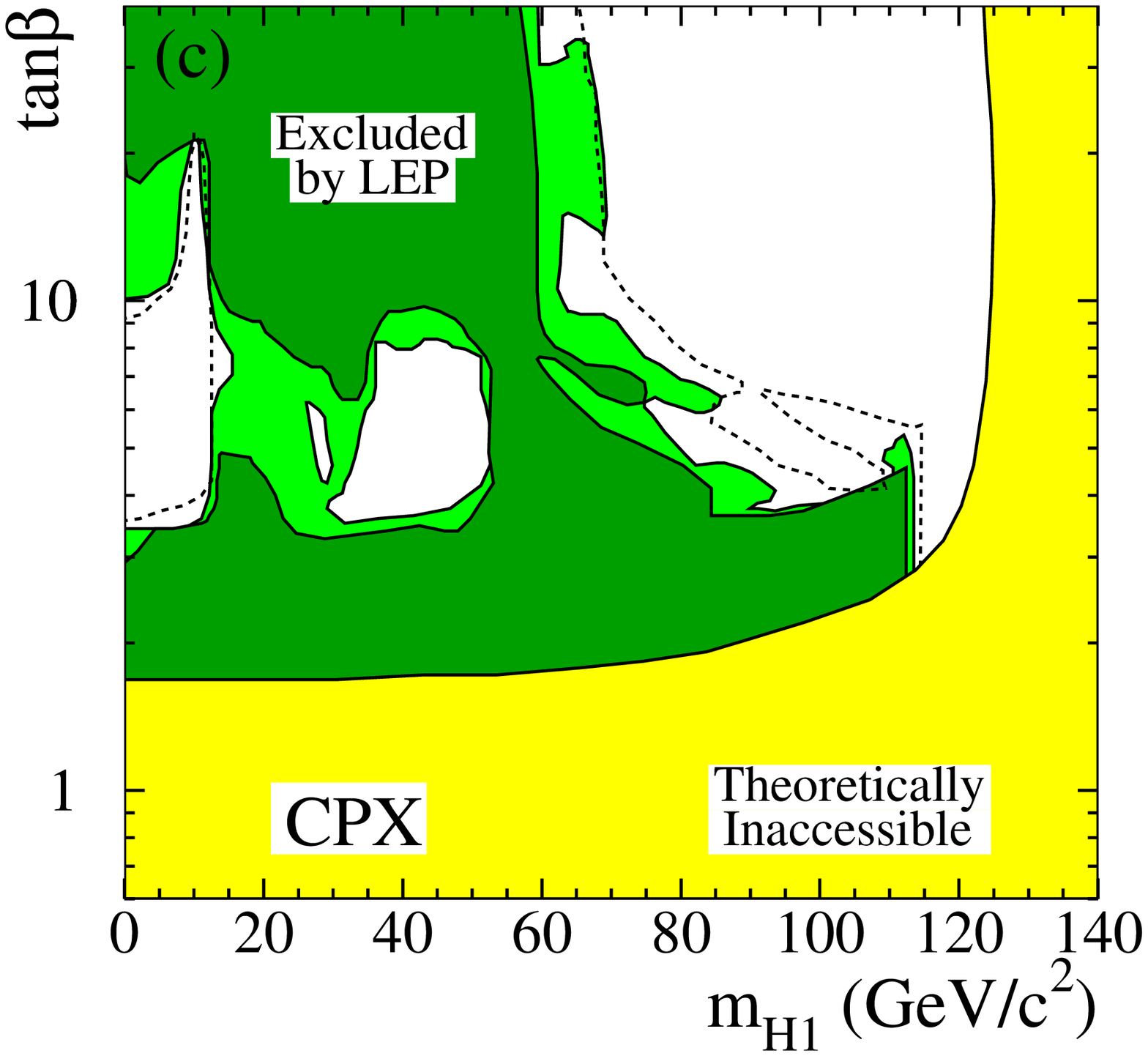} \hfill
\includegraphics[width=0.24\textwidth]{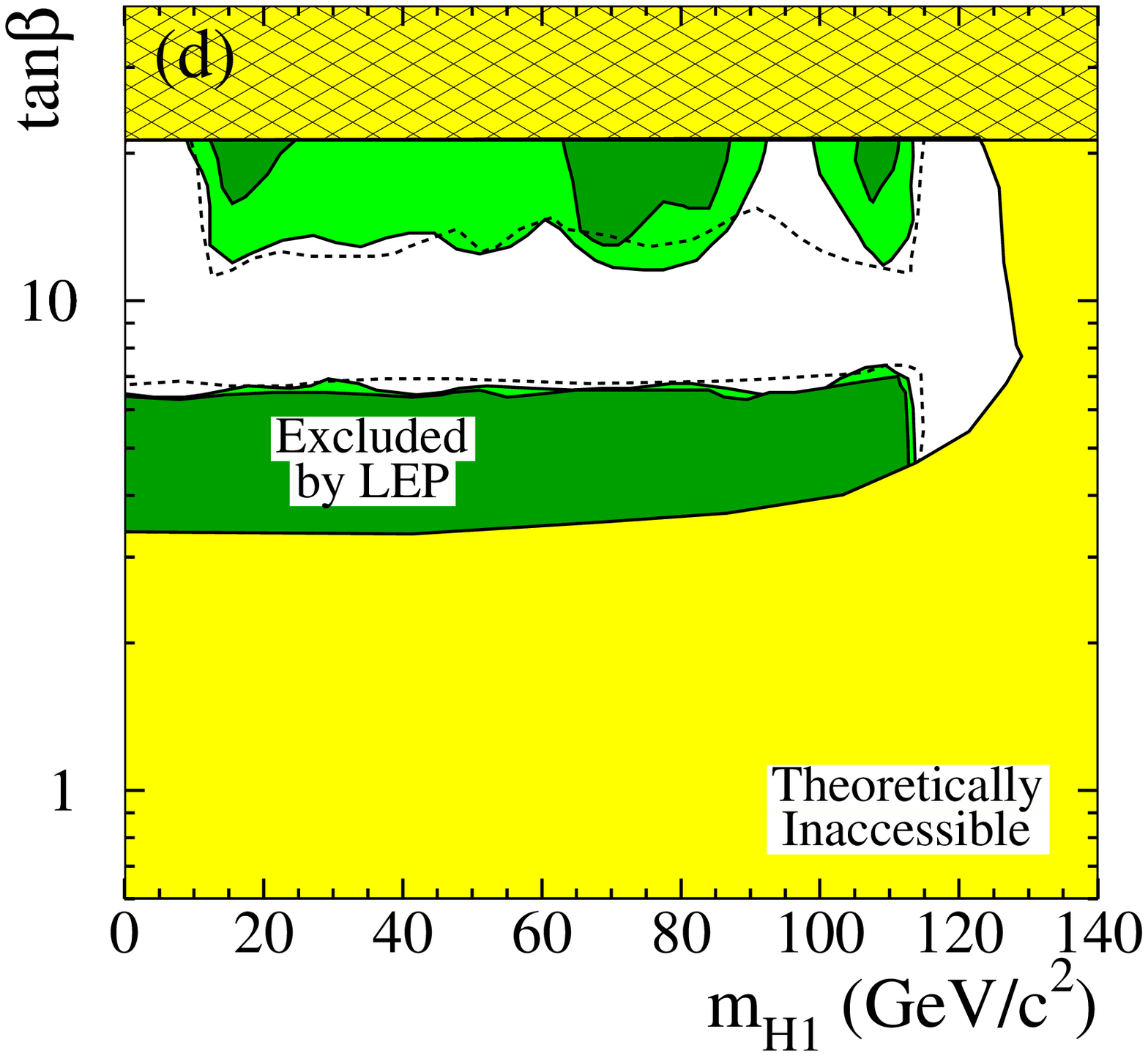} 
\end{center}
\vspace*{-8mm}
\caption{Excluded regions in the CPX scenario at 95\% CL
(light-green) and at 99.7\% CL (dark-green) for
$\mu=500,~1000,~2000,~4000$ GeV in plots (a) to (d), respectively.
Expected limits in the absence of a signal are indicated with a 
dotted line.
}
\label{fig:mu}
\vspace*{-5mm}
\end{figure}

\begin{figure}[h!]
\begin{center}
\includegraphics[width=0.24\textwidth]{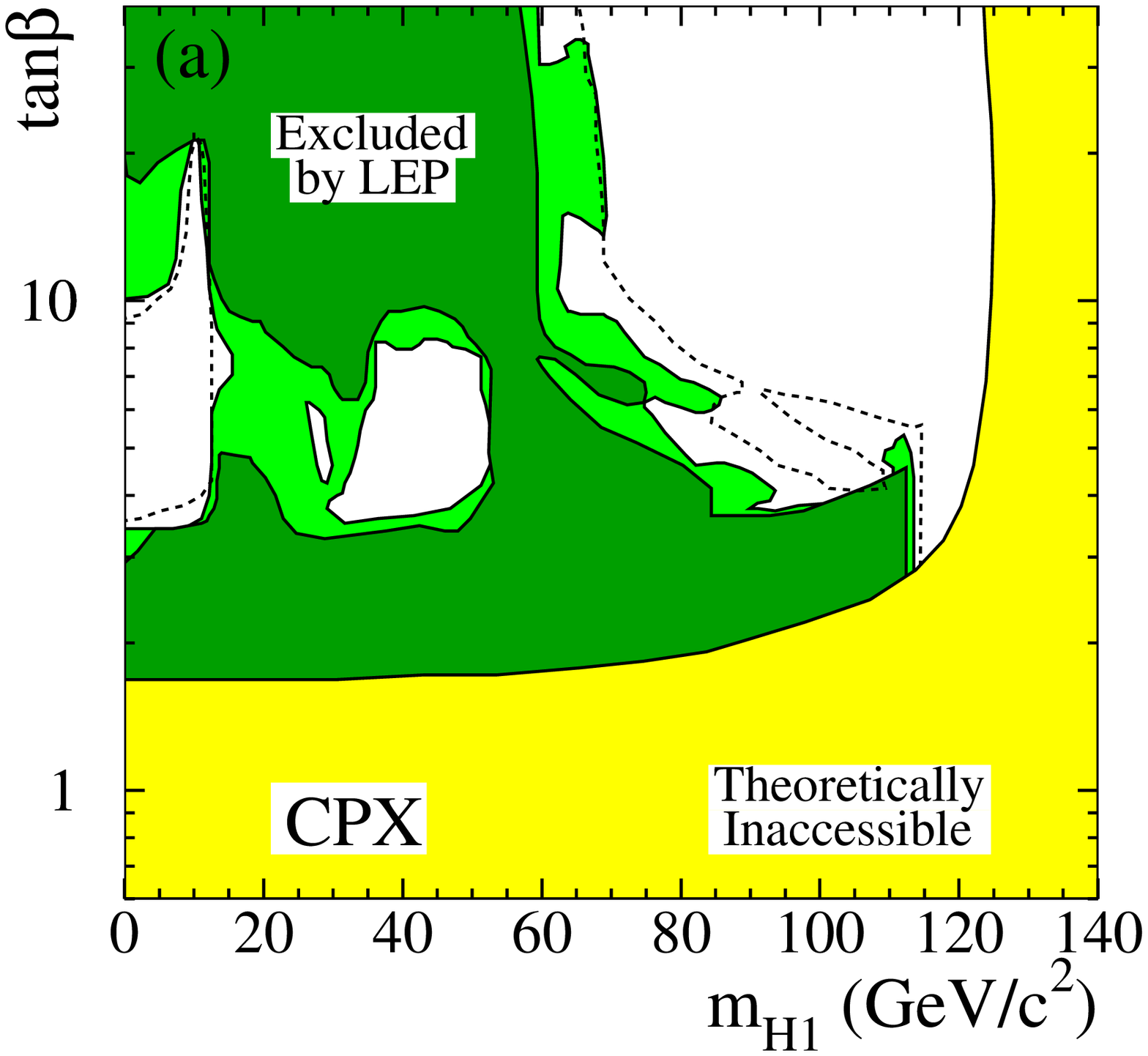} \hfill
\includegraphics[width=0.24\textwidth]{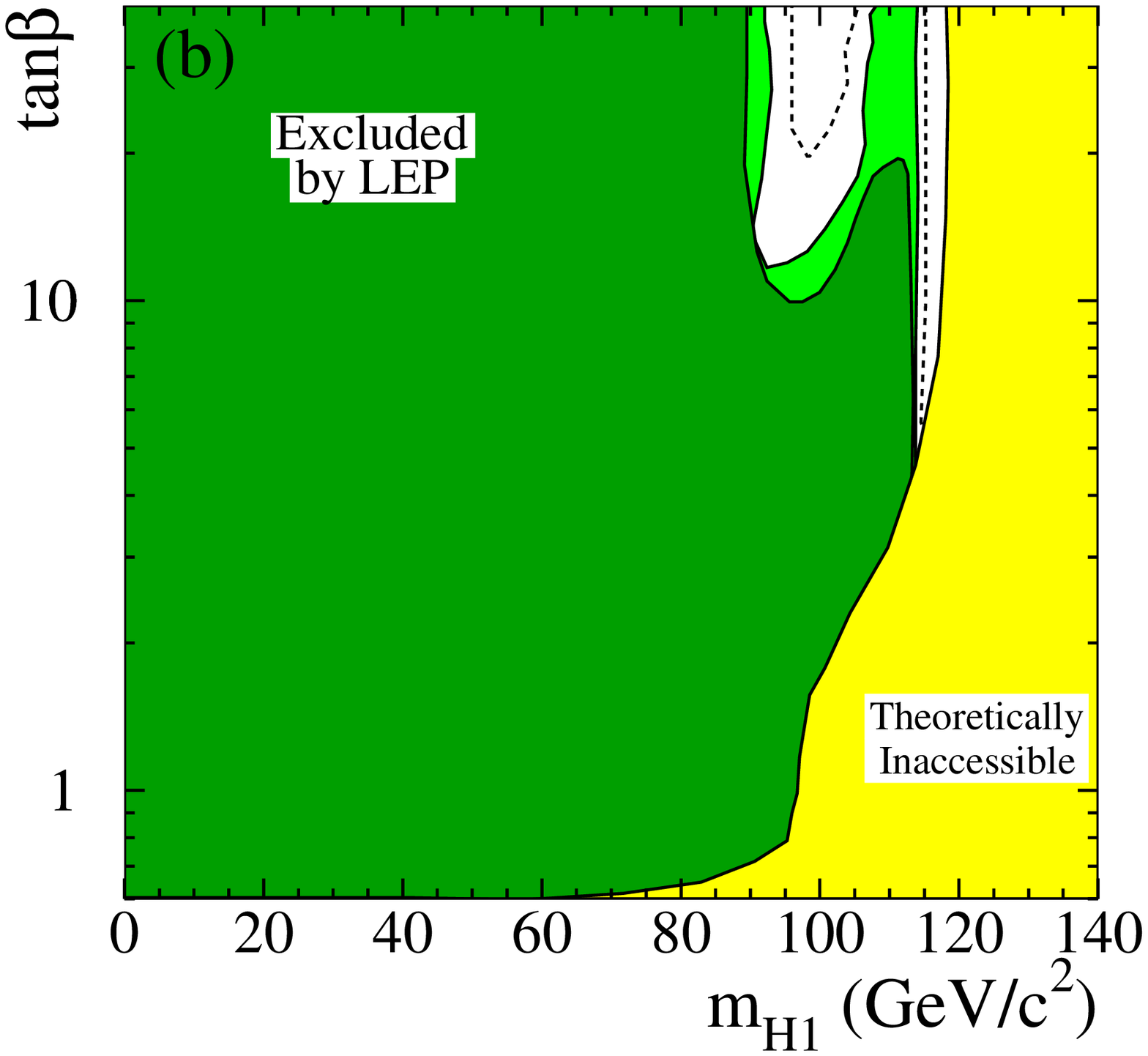}
\end{center}
\vspace*{-8mm}
\caption{Excluded regions in the CPX scenario at 95\% CL
(light-green) and at 99.7\% CL (dark-green) for
$M_{\rm SUSY} = 500,~1000$~GeV in plots (a) and (b), respectively.
Expected limits in the absence of a signal are indicated with a 
dotted line.
}
\label{fig:msusy}
\vspace*{-0.1cm}
\end{figure}

The background-only hypothesis is tested in the CPX scenario
and expressed in Fig.~\ref{fig:clb_cpx} as $1-CL_{\rm b}$ region for the 
$1\sigma$, 1-2$\sigma$, and $>$$2\sigma$ contours.
The corresponding exclusion contours are shown in Fig.~\ref{fig:cls_cpx} 
for $m_{\rm t} = 174.3$~GeV/$c^2$.
The unexcluded region near $\tan\beta=4$ can be understood from the expected
production region when the $\rm H_1\to b\bar b$ mode is suppressed, as 
shown in Fig.~\ref{fig:xsec}. The figure compares also the results from
different cross section programs (CPH and FeynHiggs) and the result
from using the logic-OR of the unexcluded region. A large dependence of the 
unexcluded region on the top quark mass is observed, as shown in Fig.~\ref{fig:top}.
Furthermore, the dependence of the excluded regions on the CP-violating phase
$\rm arg(A)$, and on the $\mu$ and $M_{\rm SUSY}$ parameters are shown in 
Figs.~\ref{fig:phase},~\ref{fig:mu}, and~\ref{fig:msusy}.

\clearpage

\clearpage

\section{Conclusions}
The LEP collaborations have finalized their Higgs boson searches and interpretations in the MSSM.
The new results from the combination of the LEP data have been presented. 
Stringent model-independent limits from Higgsstrahlung and pair-production 
      with b-quarks, $\tau$-leptons or flavour-independent searches have been set.
Limits on the scaling factor $S_{95}$ also constrain general Higgs boson scenarios.
Additional experimental constraints are applied from $\Delta\Gamma_{\rm Z}$,
decay-mode independent $\rm H_1Z$ searches, and searches for the Yukawa processes 
which contribute mostly for very light $\rm H_1$ masses.
The combined LEP data interpretations are given for h-max, no-mixing, large-$\mu$, 
gluophobic, small-$\alpha_{\rm eff}$ and CPX benchmark MSSM parameter sets.
The CP-conserving limits depend mostly on the benchmark scenario and $m_{\rm t}$. 
The limit ranges for $m_{\rm h}$ between 85 GeV/$c^2$ and fully excluded, and for
$m_{\rm A}$ between 93 GeV/$c^2$ and fully excluded.
The corresponding limits on $\tan\beta$ range between no-exclusion and fully excluded.
In the CP-violating scenario much weaker limits compared to the CP-conserving limits are set
depending on $m_{\rm t}$ and the general MSSM parameters.
The Higgs boson searches in LEP data during 1989 and 2006, and many collaborations
with theorists have much advanced the search techniques and interpretations. Much knowledge 
has also been gained for new searches at the current Tevatron collider and the LHC.

\vspace*{-2mm}
\begin{acknowledgments}
I would like to thank the organizers of the SUSY'05 conference for their kind hospitality, and 
the LEP Collaborations who entrusted me to give this presentation.
\end{acknowledgments}

\end{document}